\documentclass[11pt]{article}

\usepackage[letterpaper,top=1.2in,bottom=1.15in,left=1.35in,right=1.35in,
            headheight=15pt,headsep=28pt]{geometry}

\usepackage{amsmath}
\usepackage{unicode-math}
\usepackage[scale=0.92]{librebaskerville}   
\usepackage{graphicx}
\usepackage{xcolor}
\definecolor{accent}{HTML}{000000}     
\definecolor{accentdark}{HTML}{000000}
\usepackage{microtype}                 

\usepackage{booktabs}
\usepackage{array}
\usepackage{makecell}
\usepackage{threeparttable}
\usepackage{longtable}
\usepackage[figuresright]{rotating}
\usepackage{tabularx}
\usepackage{etoolbox}

\setcellgapes{0pt}
\AtBeginEnvironment{table}{\linespread{1}\selectfont}
\AtBeginEnvironment{sidewaystable}{\linespread{1}\selectfont}
\AtBeginEnvironment{figure}{\linespread{1}\selectfont}
\AtBeginEnvironment{tablenotes}{\setlength{\itemsep}{0pt}\setlength{\parsep}{0pt}\linespread{1.05}\selectfont}
\renewcommand{\thetable}{\Roman{table}}   
\usepackage{float}                        

\usepackage{placeins}   
\usepackage{enumitem}
\usepackage{setspace}

\usepackage{caption}
\usepackage{titlesec}
\renewcommand{\thesection}{\Roman{section}}
\renewcommand{\thesubsection}{\Alph{subsection}}
\titleformat{\section}[block]{\filcenter\normalfont\large\bfseries}{\thesection.}{0.6em}{}
\titleformat{\subsection}{\normalfont\bfseries}{\thesubsection.}{0.5em}{}
\titlespacing*{\section}{0pt}{1.9ex plus 0.5ex minus .2ex}{1.1ex}
\titlespacing*{\subsection}{0pt}{1.4ex plus 0.4ex minus .2ex}{0.7ex}

\usepackage[round,authoryear]{natbib}
\newlength{\bodyparindent}

\usepackage[hidelinks,bookmarks=true]{hyperref}
\hypersetup{ pdftitle  = {The Buffett Effect}, pdfauthor = {Pawel Robert Smolinski and Siddharth Saridevi} }

\newcommand{\spacedcaps}[1]{{\addfontfeature{LetterSpace=4.0}\MakeUppercase{#1}}}
\newcommand\blfootnote[1]{%
  \begingroup\renewcommand\thefootnote{}%
  \footnote{\leftskip=1.8em \rightskip=1.8em #1}%
  \addtocounter{footnote}{-1}\endgroup}

\usepackage{fancyhdr}
\usepackage{ifthen}
\fancypagestyle{plain}{\fancyhf{}\fancyfoot[C]{\small\thepage}}

\begin{document}
\setlength{\bodyparindent}{\parindent}
\thispagestyle{plain}
\vspace*{0.25in}
\begin{center}
  {\fontsize{27}{33}\selectfont\bfseries The Buffett Effect}\\[30pt]
  {\large\spacedcaps{Pawel Robert Smolinski \quad and \quad Siddharth Saridevi}\textsuperscript{$\dagger$}%
\blfootnote{$^{\dagger}$\,Pawel Robert Smolinski is at the University of Gda\'nsk, Department of Economics (\href{mailto:pawel.smolinski@ug.edu.pl}{\nolinkurl{pawel.smolinski@ug.edu.pl}}). Siddharth Saridevi is at the Marshall School of Business, University of Southern California. We thank David Hirshleifer and Norbert Schwarz for their help in the preparation of the manuscript. All errors are our own.}}\\[18pt]
  {\small This version: \today\quad\textperiodcentered\quad Working Paper}
\end{center}
\vspace{14pt}
\begin{center}
\begin{minipage}{0.86\textwidth}
  \small
  \begin{center}{\bfseries\spacedcaps{Abstract}}\end{center}
  \vspace{2pt}
  \noindent
  We study whether stock disclosures by recognizable investors generate abnormal market reactions beyond the reactions produced by comparable but non-recognizable managers. We define the Buffett effect as the incremental abnormal return to a disclosed new position that is attributable to the public recognition of the disclosing investor. We construct a recognition index from Wikipedia pageviews and ProQuest news mentions. Using data from 2000 to 2025 we compare market reaction to disclosures of 24 recognizable discretionary funds with disclosures of 24 matched control funds. New-position disclosures by recognizable managers generate a positive disclosure premium over the filing date and the following trading days, and the premium increases with investor recognition. We construct calendar-time portfolios that mechanically copy each group's disclosed purchases and find that copying the most recognized investors earns a significant alpha that persists for months and does not reverse. The evidence is most consistent with an entangled attention-information channel: investor recognition directs scarce investor attention toward selected disclosures, and the information revealed in those disclosures is then capitalized into prices.
\end{minipage}
\end{center}
\vspace{6pt}
\noindent\begin{center}\begin{minipage}{0.86\textwidth}\small
\end{minipage}\end{center}
\vspace{10pt}
\section{Introduction}

Warren Buffett is, arguably, the world's most recognized investor. The long-term performance of his small, concentrated equity portfolio has delivered extraordinary market outperformance and earned him the title of the Oracle of Omaha, as if he possessed the unusual ability to predict the future of the stock market. It is hardly surprising that investors have long spoken of a so-called Buffett effect: the notion that when Berkshire Hathaway, Buffett's firm, discloses its trades, his followers flock to copy the moves of the infallible Oracle. The influx of imitative demand, the story goes, moves the price of the disclosed stocks simply because Buffett was seen to trade them.

Buffett is only one example of a recognizable investor. Many other recognizable investors exist today, and many have existed throughout the history of the market. Some of them are recognizable enough to command a substantial following of their own. A new position taken by a skilled investor can be interpreted as a market signal that the disclosed stock will appreciate. However, this is not true only of recognizable investors. The disclosures of all discretionary funds presumably carry some information about the disclosed stocks, and that information is a tradable signal. Copying these disclosures has been shown to generate abnormal returns in the long horizon, even for ordinary funds (Frank et al., 2004; Cao et al., 2021). The Buffett effect postulates that this signal translates into an immediate market reaction only when the disclosing investor is recognizable, and that the reaction grows with the level of recognition. We therefore define the Buffett effect as the incremental abnormal return a stock earns when a new position in it is disclosed by a recognizable manager, relative to the same disclosure by an otherwise similar but non-recognizable manager.

The correlation between the recognizability of an investor and the market reaction to their disclosures has not yet been tested. A small number of studies look at a single investor and document some form of market reaction. Martin and Puthenpurackal (2008) and Zhang (2022) show that mimicking Buffett's disclosed holdings has historically been profitable. Related evidence shows that investors react more strongly to recommendations from celebrity analysts (Bonner et al., 2007). None of these studies, however, proves the general correlation between the recognizability of an investor and the strength of the market reaction to their disclosures. Nor has anyone tried to study the plausible mechanisms explaining this reaction.

This study provides the first systematic investigation of the anecdotal Buffett effect. In our view, the effect can reflect three explanations/mechanisms: 1) the information in the disclosures of recognizable investors reaches more people; 2) it is more persuasive because of the credibility of the source; or 3) their disclosures elicit mindless imitation trades. In the following sections, we develop the theoretical background for these mechanisms, drawing on rational inattention, social finance, and evidence that investors trade on the actions and statements of prominent figures. We then describe the data and methodology used to test for the Buffett effect in the disclosures of recognizable American investors, present the main results and robustness checks, and use trade-level data to identify which of the proposed mechanism are the most plausible.

We find that the Buffett effect is concentrated among investors with the highest level of recognition. The abnormal returns strengthen after the initial disclosure and do not revert over the following month. The effect is strongest in small, illiquid, and scarcely covered stocks, but it is not completely absent in large, institution-heavy, and widely researched stocks. The trade-level data show that the disclosures of the most recognizable investors are followed by imitation trades of retail investors, and that the disclosures lead to more trades rather than bigger trades. Imitation alone, however, does not fully explain the premium. The price effect is largest in the stocks where information is hardest to obtain, and it leads to lasting outperformance. The evidence points to the following mechanism: the disclosures of recognizable fund managers provide valuable information about the disclosed stocks. Investors are far more attentive when the manager is highly recognizable. They look at the disclosed stocks, find information in them that the market has not yet priced, and begin adding them to their portfolios. This buying raises the price of the disclosed stocks.

\section{Theoretical Background}

We define the Buffett effect as follows. A discretionary fund manager discloses a new position in a stock in a quarterly 13F filing. The disclosed stock earns an abnormal return around the filing date. The Buffett effect is the part of that abnormal return which is attributable to the public recognition of the manager: the incremental abnormal return a stock earns when a new position in it is disclosed by a recognizable manager, relative to the same disclosure by an otherwise similar but non-recognizable manager. The definition has two parts. The counterfactual is the disclosure of a comparable manager who has no significant public recognition. The treatment is recognition itself, so the effect should grow with the level of recognition of the manager and should be strongest for the most highly recognizable managers.

The Buffett effect can only exist among discretionary managers. The goal of a discretionary manager is to identify market mispricing and to profit from it. Such a manager holds a limited number of high-conviction positions and takes a directional view on every stock in their book. By contrast, a quantitative fund holding hundreds or thousands of stocks with substantial quarterly rebalancing carries almost no signal about the individual stocks it discloses, and the same is true of endowments, pension books, sovereign portfolios, and other institutions whose 13F carries no directional view on any disclosed stock. The disclosure of a discretionary manager is therefore the only kind of disclosure that can be interpreted as a market signal that the disclosed stock will appreciate.

Information acquisition is costly (Grossman and Stiglitz, 1980). Discretionary managers do the costly research required to identify undervalued stocks, and once their holdings are disclosed, other market participants can update on that research for free. The evidence shows that this research has value to those who copy it. Copycat portfolios built on the mandated disclosures of ordinary funds earn returns similar to, or even better than, the original funds after expenses (Frank et al., 2004). Institutions systematically view and imitate their peers' filings on the SEC EDGAR website, and the copycatted trades outperform the other trades disclosed by the same funds (Cao et al., 2021). All disclosures of discretionary managers, recognizable or not, therefore carry, at least in theory, some tradable signal.

However, the reaction defined by the Buffett effect happens only to the stocks disclosed by recognizable managers. If all disclosures reveal something about the disclosed stocks, the immediate market reaction should not depend on who made the disclosure. The Buffett effect postulates that it does. There are three possible explanations for why this is the case. We describe each of them as a separate mechanism, and for each mechanism we describe the trace it could leave in the market data.

\textbf{\textit{Mechanism 1}} assumes that all discretionary managers reveal information, with the same level of influence on each individual investor, however the information reveled by the recognizable investors has more market impact because it reaches a larger number of recipients. This is a pure attention mechanism. Attention is scarce, and investors cannot feasibly pay attention to all disclosures. Rational inattention theory treats attention as a resource that agents allocate optimally across signals (Sims, 2003; Maćkowiak et al., 2023), and experimental evidence shows that people respond to attention costs and incentives in ways broadly consistent with this framework (Dean and Neligh, 2023; Loewenstein and Wojtowicz, 2025). Rationally inattentive investors therefore direct their attention only to those disclosures that are easily available to them. The recognition of the manager contributes to that. The higher the recognition, the easier it is to pay attention to the disclosure, and the lower the attention cost for a rationally inattentive investor. We assume that number of trades can be used as a proxy for attention, as more people decide to act on the information reveled in the disclosure. Mechanism 1 predicts then that the number of trades in the disclosed stock will differ significantly between the disclosures of recognizable and non-recognizable managers.

\textbf{\textit{Mechanism 2}} assumes that recognizable managers have information of better quality than non-recognizable managers. The information in the disclosures of recognizable managers is then more persuasive because of its credibility. Recognition typically originates in a history of market outperformance, so investors have reason to grant the disclosures of recognizable managers more weight. Under Mechanism 2, investors trade on all disclosures, but the disclosures of recognizable managers get more weight because of the credibility premium. The abnormal returns of the Buffett effect can then be attributed to the difference in the size of the trades between the disclosures of recognizable and non-recognizable managers. This mechanisms assumes that the size of the trades increases and the number of trades does not differ much. It is possible, however, that the number of trades differs together with the size of the trades, since Mechanism 1 and Mechanism 2 can work together.

One remark applies to both Mechanism 1 and Mechanism 2. Both mechanisms transmit information about the disclosed stock, so both should be more pronounced when the information is costly. The cost of information can be a characteristic of the stock. Small stocks, illiquid stocks, stocks with scarce analyst coverage, and stocks under-held by institutions are the stocks about which the market has less information, and a disclosure is most likely to reveal something the market did not already know about exactly these stocks (Fang and Peress, 2009). For a large stock with substantial institutional ownership and analyst coverage, a disclosure should add less. If the Buffett effect works through Mechanism 1 or Mechanism 2, it should therefore be stronger among the stocks characterized by high information cost.

\textbf{\textit{Mechanism 3}} is a pure imitation mechanism. It works similarly to Mechanism 1, since it also relies on the attention that recognizable managers attract, but in contrast to Mechanism 1 it is detached from the information in the disclosure. Once the disclosure is made, a form of herding takes place. Investors imitate the trade of the recognizable manager regardless of the informational value of that trade. There is ample evidence that this mechanism exists. The on-air recommendations of Jim Cramer produce sharp overnight price increases that subsequently reverse (Engelberg et al., 2012). The cryptocurrency tweets of Elon Musk trigger abnormal returns and trading-volume responses (Ante, 2023), and the tweets of Donald Trump increase trading activity but produce no lasting price effects (Machus et al., 2022). Investors imitate visible peers, and imitation propagates through social networks rather than through fundamentals (Hirshleifer, 2020; Kuchler and Stroebel, 2021). Mechanism 3 leaves a different trace than the first two mechanisms. The increase in the number and size of trades should happen not only in the informationally scarce stocks, where getting the information is costly, but also among large and highly covered stocks, where the disclosure reveals little that the market should not already know. However, more importantly, the effect should reverse once the market realizes that the price increase carries no information and represents mindless copying. A lasting premium should therefore not be possible under Mechanism 3.

A second remark concerns the investors rather than the stocks. The cost of information differs not only across stocks but also across the investors who might act on a disclosure. Institutional investors usually already have the research capacity, the information, and the strategies needed to form their investment decisions. They also monitor the filings of other institutions as a matter of course (Cao et al., 2021), so the disclosure of a recognizable manager adds little to what they already know and little to how much attention they pay. Retail investors are at the other end. They face the highest cost of gathering information and of devising an investment strategy, so a disclosure by a recognizable manager is, for them, a cheap source of investment information. Attention-driven trading is also documented most clearly among retail investors (Barber et al., 2022). The Buffett effect should therefore be driven by the group with the lowest informational access: retail investors.

\section{Data}

\subsection{Defining the universe of discretionary funds}

We restrict attention to discretionary funds, that is, the funds whose disclosures can plausibly carry a market signal about the disclosed stocks. We require a fund to satisfy four criteria over its full 13F history to qualify as a discretionary fund: at least 12 quarterly filings (three years of history), a most-recent filing within 8 quarters of the sample close (the fund is still active), an average of no more than 100 long positions per filing (concentrated breadth), and an average of no more than 20 new-position plus complete-exit events per quarter (low turnover). These cutoffs remove the quant and quasi-quant funds whose disclosures cannot be read as discretionary trades.

The breadth and turnover screens are not sufficient: a fund can be concentrated and low turnover and still not be making a discretionary equity bet. We therefore retain only genuine discretionary equity funds (long/short equity funds, value-oriented funds, concentrated long-only funds, activists) and exclude the structurally non-discretionary institutions whose 13F carries no directional view on any disclosed stock even when it happens to be concentrated: endowments and foundations, insurance and pension asset-liability books, sovereign portfolios, corporate holding companies, funds-of-funds, and single-family offices. The resulting universe consists of discretionary funds: stock pickers and activists whose disclosures can be interpreted as a directional signal that the disclosed stock will appreciate.

\subsection{Recognizable sample}

Having defined the universe of discretionary funds, we turn to the public recognition of the manager behind each fund. Recognition is a property of the manager rather than of the fund, so throughout the paper a fund is labelled by the recognition of the manager who runs it. We start from the English-language Wikipedia ``American hedge fund managers'' category list, which catalogues individuals identified as prominent American hedge fund managers by Wikipedia (accessed May 2026). We process the resulting list manually in two steps. First, we drop entries that are long deceased or inactive (e.g., Jim Cramer) or not asset managers (e.g., Bernie Madoff), together with entries we cannot match to any 13F filing entity: commentators, managed-futures and crypto managers whose holdings are not 13F-reportable, and funds too new or too small to file. In a second step, the entries that do map to a 13F filer are passed through the discretionary-fund screen described above. Funds that fail the breadth and turnover criteria are dropped here. Examples of funds filtered out at this stage include quant funds such as Jim Simons' Renaissance Technologies, multi-strategy complexes like Ken Griffin's Citadel and Ray Dalio's Bridgewater, broad asset managers such as Larry Fink's BlackRock, and vehicles that have converted away from stock-picking, such as Stanley Druckenmiller's Duquesne that is now a family office. What survives is the candidate pool of 24 recognizable funds, ranging from household names like Warren Buffett and Bill Ackman to less universally recognized but still prominent managers such as David Goel and Joon Yun.

To rank recognition within the candidate pool we construct an Investor Recognition Index that combines two complementary measures of public recognition: mean monthly pageviews on the manager's English-language Wikipedia page (2015 to 2025), and total news mentions in the ProQuest news database over 2000 to 2025. Because both measures are heavily right-skewed, we take the natural logarithm of each, standardize it to a z-score within the candidate pool, and average the two z-scores. The two components correlate strongly (Spearman 0.832, Pearson 0.854). Table I reports the full fund characteristics and the recognition ranking.

\begin{sidewaystable}[p]
\centering
\caption{Recognizable funds: characteristics and investor recognition}
\setlength{\tabcolsep}{3pt}
\begin{threeparttable}
\footnotesize
\begin{tabular*}{\linewidth}{@{\extracolsep{\fill}}rllrrrrrrrrr}
\toprule
\# & Fund & Manager & \makecell{AUM\\(\$M)} & \makecell{Pos.\\(\#)} & \makecell{Years\\active} & \makecell{Turn.\\(ev./q)} & \makecell{Rec.\\z} & \makecell{z\\Wiki} & \makecell{z\\ProQ} & \makecell{Wiki\\views/mo} & \makecell{ProQuest\\mentions} \\
\midrule
1 & \textbf{Berkshire} & \textbf{Warren Buffett} & 128,146 & 38.3 & 25.8 & 4.94 & 2.33 & 2.14 & 2.25 & 323,969 & 584,341 \\
2 & \textbf{Icahn} & \textbf{Carl Icahn} & 13,695 & 14.0 & 25.5 & 2.86 & 1.36 & 1.10 & 1.46 & 44,792 & 103,210 \\
3 & \textbf{Scion} & \textbf{Michael Burry} & 79 & 11.3 & 9.7 & 13.00 & 1.29 & 1.61 & 0.82 & 117,613 & 25,800 \\
4 & \textbf{Pershing Square} & \textbf{Bill Ackman} & 7,372 & 7.8 & 20.0 & 2.84 & 1.24 & 1.28 & 1.06 & 63,549 & 43,281 \\
5 & \textbf{Trian} & \textbf{Nelson Peltz} & 5,703 & 9.7 & 19.3 & 3.91 & 1.22 & 1.26 & 1.04 & 60,571 & 41,131 \\
6 & Appaloosa & David Tepper & 3,326 & 34.2 & 25.8 & 16.88 & 0.74 & 0.87 & 0.52 & 28,921 & 13,426 \\
7 & Greenlight & David Einhorn & 2,753 & 33.0 & 25.8 & 13.59 & 0.50 & 0.21 & 0.73 & 8,218 & 21,141 \\
8 & Himalaya & Li Lu & 1,569 & 5.4 & 9.0 & 1.50 & 0.42 & 0.01 & 0.78 & 5,555 & 23,499 \\
9 & Baupost & Seth Klarman & 3,872 & 27.9 & 25.8 & 11.19 & 0.32 & 0.23 & 0.38 & 8,510 & 9,810 \\
10 & Ares & Tony Ressler & 3,463 & 26.8 & 18.0 & 9.11 & 0.30 & 0.77 & $-$0.20 & 23,937 & 2,755 \\
11 & Avenue & Marc Lasry & 438 & 19.2 & 21.7 & 14.72 & 0.19 & 0.30 & 0.06 & 9,733 & 4,911 \\
12 & Alden & Randall D. Smith & 179 & 16.4 & 15.0 & 15.52 & 0.09 & $-$0.29 & 0.47 & 3,102 & 11,900 \\
13 & Pabrai & Mohnish Pabrai & 254 & 7.6 & 21.0 & 3.02 & $-$0.13 & 0.09 & $-$0.34 & 6,455 & 2,024 \\
14 & Altimeter & Brad Gerstner & 3,773 & 22.4 & 14.0 & 8.42 & $-$0.19 & 0.02 & $-$0.38 & 5,611 & 1,861 \\
15 & ValueAct & Jeffrey W. Ubben & 6,056 & 15.1 & 23.8 & 5.16 & $-$0.41 & $-$0.69 & $-$0.09 & 1,466 & 3,530 \\
16 & Steel Partners & Warren Lichtenstein & 501 & 18.1 & 22.0 & 8.00 & $-$0.52 & $-$0.82 & $-$0.15 & 1,127 & 3,077 \\
17 & Fairholme & Bruce R. Berkowitz & 3,140 & 24.5 & 25.8 & 8.01 & $-$0.62 & $-$1.08 & $-$0.09 & 690 & 3,505 \\
18 & Mudrick & Jason Mudrick & 247 & 4.8 & 8.0 & 2.13 & $-$0.64 & $-$0.79 & $-$0.43 & 1,209 & 1,677 \\
19 & Valiant & Chris R. Hansen & 728 & 16.3 & 17.0 & 5.24 & $-$1.05 & $-$0.72 & $-$1.26 & 1,368 & 273 \\
20 & GoldenTree & Steven Tananbaum & 737 & 25.6 & 22.0 & 19.99 & $-$1.13 & $-$0.87 & $-$1.26 & 1,030 & 272 \\
21 & Matrix & David Goel & 2,536 & 18.5 & 23.5 & 10.46 & $-$1.14 & $-$0.80 & $-$1.34 & 1,171 & 228 \\
22 & Palo Alto Investors & Joon Yun & 1,167 & 50.6 & 23.3 & 14.92 & $-$1.21 & $-$1.30 & $-$0.99 & 454 & 492 \\
23 & Cantillon & William von Mueffling & 6,817 & 41.9 & 22.0 & 11.84 & $-$1.33 & $-$1.21 & $-$1.29 & 537 & 252 \\
24 & Dalton & James B. Rosenwald & 184 & 35.1 & 19.0 & 11.39 & $-$1.62 & $-$1.32 & $-$1.74 & 437 & 94 \\
\bottomrule
\end{tabular*}
\begin{tablenotes}[flushleft]\footnotesize
\item \textit{Note.} The 24 recognizable discretionary managers, ordered by investor recognition. AUM is mean reported portfolio value; Pos. is the mean number of distinct holdings; turnover is portfolio events per quarter. The Investor Recognition Index (Rec. z) is the average of two standardized components, the log of mean monthly Wikipedia page views of the manager (z Wiki) and the log of total ProQuest news mentions, 2000--2025 (z ProQ), re-standardized to mean 0 and SD 1 within the 24 recognizable funds; the two components agree strongly (Spearman 0.832, Pearson 0.854). ``Wiki views/mo'' is mean monthly page views of the manager's page and ``ProQuest mentions'' is the total 2000--2025 count. High-recognition funds (Rec. z $\geq$ 1) in bold.
\end{tablenotes}
\end{threeparttable}
\end{sidewaystable}

\subsection{Control sample}

Without a comparison to non-recognizable but otherwise comparable discretionary funds we cannot answer whether the price change after a disclosure is the effect of the manager's recognition or simply the effect of a stock being disclosed by a discretionary manager whose information then gets incorporated into the price. We therefore construct a control sample of 24 discretionary funds: real funds running concentrated, low-turnover books that have never appeared in the Wikipedia list of American hedge fund managers and have no significant public recognition. Each recognizable fund is matched to a unique control by optimal assignment, minimizing the total distance between the two groups on portfolio value, concentration, and longevity. We then verify every selected control, confirming from its recent filings and public sources that it is an active, discretionary equity stock-picker rather than a fund that has wound down, converted to a family office, or otherwise ceased discretionary stock selection (e.g., an endowment-style allocator). Matching algorithm details, the verification log, and balance statistics are reported in the Internet Appendix, which also re-runs the matching with no manual exclusions and shows that the results do not change our interpretation.

The two groups are well-balanced on the three matched characteristics (Internet Appendix Table IA.XI; Table II lists the control funds). Portfolio value, concentration, and longevity do not differ significantly between the recognizable and control groups (Welch p = 0.42, 0.55, and 0.12). The standardized mean differences are +0.09 for log portfolio value, -0.11 for log concentration, and +0.46 for longevity. Portfolio turnover is likewise balanced (about 9 versus 7 events per quarter, p = 0.91). Concentration is the closest match, at a median of 19 positions per filing for the recognizable sample against 21 for the controls; longevity is the least balanced, with the recognizable funds somewhat longer-lived, though the gap is not significant (p = 0.12). We further control for stock-level differences (size, liquidity, analyst coverage, book-to-market, gross profitability, and institutional ownership) in the regression analysis, so the recognizable-versus-control contrast can be read as the effect of recognition holding fund operations and stock characteristics fixed.

\begin{table}[htbp]
\centering
\caption{Control funds: characteristics}
\setlength{\tabcolsep}{4pt}
\begin{threeparttable}
\small
\begin{tabular*}{\linewidth}{@{\extracolsep{\fill}}rlrrrr}
\toprule
\# & Fund & AUM (\$M) & Pos. (\#) & Years active & Turnover (ev./q) \\
\midrule
1 & Sands & 23,608 & 55.3 & 25.8 & 10.92 \\
2 & Southeastern & 13,846 & 39.7 & 25.8 & 7.43 \\
3 & Generation Investment & 8,896 & 32.0 & 19.0 & 7.86 \\
4 & Independent Franchise & 8,632 & 22.2 & 16.0 & 3.03 \\
5 & Akre & 5,154 & 33.0 & 24.5 & 9.14 \\
6 & Sound Shore & 4,498 & 45.6 & 25.8 & 10.32 \\
7 & Caledonia & 4,218 & 11.0 & 13.0 & 3.26 \\
8 & Greenhaven Associates & 3,921 & 28.9 & 25.8 & 7.04 \\
9 & Ownership & 3,448 & 19.8 & 9.7 & 3.06 \\
10 & Steadfast & 3,282 & 47.0 & 23.0 & 26.13 \\
11 & Jana & 2,782 & 36.5 & 23.0 & 30.43 \\
12 & Theleme & 1,650 & 8.8 & 15.0 & 3.63 \\
13 & Cas Investment & 1,078 & 5.3 & 8.0 & 2.09 \\
14 & Oldfield & 1,044 & 22.8 & 15.5 & 3.67 \\
15 & Hawk Ridge & 1,001 & 40.6 & 11.5 & 24.30 \\
16 & Greenbriar & 566 & 17.2 & 18.0 & 4.08 \\
17 & Cyrus & 461 & 14.4 & 20.8 & 10.55 \\
18 & Thorntree & 432 & 18.5 & 9.0 & 8.89 \\
19 & Prime & 354 & 6.0 & 14.8 & 2.61 \\
20 & Mak & 319 & 10.7 & 19.0 & 5.55 \\
21 & Adw & 240 & 5.1 & 7.0 & 1.80 \\
22 & Pacifica & 205 & 17.4 & 15.8 & 2.70 \\
23 & Archon & 198 & 27.7 & 16.0 & 15.02 \\
24 & Tenzing & 175 & 15.6 & 9.0 & 9.65 \\
\bottomrule
\end{tabular*}
\begin{tablenotes}[flushleft]\footnotesize
\item \textit{Note.} The 24 control managers, each matched one-to-one to a recognizable fund by optimal assignment on portfolio value, concentration, and longevity, ordered by AUM. AUM is mean reported portfolio value; Pos. is the mean number of distinct holdings; turnover is portfolio events per quarter.
\end{tablenotes}
\end{threeparttable}
\end{table}

\subsection{Stock level data and event construction}

The remaining data come from standard sources. Quarterly 13F holdings are from the Thomson Reuters institutional holdings database. Filing dates come from the SEC EDGAR full-text filing index, and the EDGAR filing date defines day 0 of every event window in the paper. Because some managers file under more than one CIK over their history (a predecessor and a successor entity, or parallel vehicles), we resolve the complete set of CIKs for each fund so that its full disclosure history is captured. Daily stock returns and other stock-level variables are from CRSP. The daily market, size, value, and momentum factors and the risk-free rate are from the Kenneth French Data Library, consistent with the common-factor framework of Fama and French (1993) and the momentum extension in Carhart (1997). Analyst coverage is the number of one-year-ahead EPS estimates outstanding from I/B/E/S. Book-to-market and gross profitability are from the WRDS Financial Ratios Suite, winsorized at the 1st and 99th percentiles. Institutional ownership is aggregated from the 13F universe and divided by CRSP shares outstanding. The sample period is 2000 to 2025.

Every (manager, stock) pair in a fund's 13F history is classified as either a new position (the stock appears in the fund's filings for the first time) or a complete exit (a previously held stock drops out). The new position is the bullish disclosure and the complete exit serves as its within-fund counterpart. Each event is dated to the EDGAR filing date of the disclosure that revealed it. Around each filing date we compute abnormal returns using standard daily-return event-study logic (Brown and Warner, 1985). Our primary measure is the Carhart four-factor abnormal return (Carhart, 1997): the stock's daily return in excess of the return predicted by a four-factor model (market, size, value, and momentum), with the factor loadings estimated for each event over a pre-disclosure estimation window. As an alternative baseline we also compute market-adjusted abnormal returns, defined as the daily return minus the return on the CRSP value-weighted market portfolio; and as a robustness measure we compute market-model abnormal returns from a single-factor market regression. Cumulating daily abnormal returns over a sub-window yields the event's cumulative abnormal return, CAR, and all CARs are winsorized at the 1st and 99th percentiles. Because factor loadings cannot be estimated for events with very short return histories, those events retain the market-adjusted CAR but not the factor-based ones, so the baseline measure has essentially full coverage. Our primary event windows are [0,1], [0,3], and [0,5], capturing the immediate market reaction over the filing day and the next one, three, and five trading days. Post-disclosure windows out to [0,30] and pre-disclosure windows back to [-10,-1] are used in the identification and robustness analyses.

The resulting analysis sample contains 48 funds (24 recognizable and 24 control) and 32,785 new-position and complete-exit events: 17,218 from the recognizable funds (8,835 new positions and 8,383 exits) and 15,567 from the controls (8,085 new positions and 7,482 exits). Carhart four-factor abnormal returns are available for 29,664 of these events (90.5\%). The full specification regressions that add stock and fund controls run on 19,747 events.

\subsection{Trade-level data}

To test the proposed mechanisms we collect the NYSE Daily TAQ trade records, which report every trade in US-listed stocks with its price and size. Daily TAQ is available to us from 2015, so the trade-level analysis covers the events filed from 2015 to 2025: 13,514 events (6,838 new positions and 6,676 exits).

Retail trades are identified with the methodology of Boehmer et al. (2021). From the identified trades we compute, for each event stock and trading day, the retail order imbalance (retail buy minus retail sell dollar volume over their sum), the number of retail trades, and the retail share of consolidated volume. From all trades we compute the total trade count, the average trade size, and the share of dollar volume executed in trades of at least \$50,000. Each measure is expressed relative to the stock's own average over days [-15,-6] before the filing and averaged over event windows, and compared between high-recognition and control disclosures separately for purchases and for exits. These regressions use the stock and fund controls of the return regressions, stock and year-quarter fixed effects, and the outcome's own pre-filing value as a control. They do not use fund fixed effects, because the treatment is a characteristic of the fund. The identification captures roughly a third of retail volume and might mis-sign some trades (Barber et al., 2024), so retail magnitudes should be read as lower bounds on retail participation.

\section{Results}

Before presenting the estimates, we fix the outcome and the coefficient of interest. The dependent variable throughout is the cumulative abnormal return around the disclosure date. In every table the coefficient we report is the interaction of the group (or recognition) variable with the new-position indicator: it measures how much more a new position, the bullish disclosure, moves the stock than the same fund's complete exit, its within-fund counterpart, and how much wider that buy-minus-exit gap is for the treated funds than for the controls. We report this difference-in-differences coefficient throughout and refer to it as the disclosure premium.

\subsection{Recognition and the disclosure premium}

Table III presents the market reaction to stock disclosures by the recognizable and the control funds. The disclosure premium for a recognizable fund, measured against a control fund, is +0.253\% over the disclosure day and the next (CAR[0,1], p = 0.026). The effect remains significant in all immediate reaction windows: +0.357\% over [0,3] (p = 0.017) and +0.422\% over [0,5] (p = 0.033). The price of the disclosed stock continues to produce cumulative abnormal returns even 20 days after disclosure (+0.566\% at [0,10], p = 0.024; +0.780\% at [0,20], p = 0.049). The effect loses significance 30 days after disclosure, but the coefficient remains positive (+0.490\%, p = 0.320). Adding the full set of stock-level controls (size, liquidity, analyst coverage, book-to-market, gross profitability, and institutional ownership) and fund-level controls (log portfolio value, log concentration, fund age, and turnover), together with stock, fund, and year-quarter fixed effects and the pre-event drift control, reduces the two-day reaction (+0.200\%, p = 0.159), but the effect still persists in the immediate reaction window (+0.341\% over [0,3], p = 0.043; +0.499\% over [0,5], p = 0.031) and the cumulative abnormal returns stay positive and significant through the post-disclosure windows (+0.684\% at [0,10], p = 0.027; +0.772\% at [0,20], p = 0.099). There is no pre-disclosure run-up: the [-10,-1] window is negative and not significant (-0.431\%, p = 0.244); day-by-day tests of the pre-disclosure window, reported in the Internet Appendix, confirm this at the daily level. Figure 1 traces the average abnormal return path around the disclosure and shows the same pattern: flat before the filing, a jump on and just after filing, and no reversal through day 30. This initial result confirms that stock disclosures of recognizable funds produce an abnormal market reaction that does not reverse after the disclosure and continues to grow.

\begin{table}[htbp]
\centering
\caption{Recognizable versus control funds}
\setlength{\tabcolsep}{3pt}
\begin{threeparttable}
\small
\begin{tabular*}{\linewidth}{@{\extracolsep{\fill}}lcccccc}
\toprule
 & (1) & (2) & (3) & (4) & (5) & (6) \\
 & Raw & \makecell{+ Stock\\ctrls} & \makecell{+ Stock\\FE} & \makecell{+ Fund\\ctrls} & \makecell{+ Fund\\FE} & Full \\
\midrule
\multicolumn{7}{l}{\textit{Panel A. Immediate reaction}}\\
CAR[0,1] & 0.253$^{**}$ & 0.213 & 0.227 & 0.229 & 0.231 & 0.200 \\
 & (0.110) & (0.130) & (0.150) & (0.150) & (0.150) & (0.140) \\[1pt]
CAR[0,3] & 0.357$^{**}$ & 0.310$^{*}$ & 0.340$^{*}$ & 0.343$^{*}$ & 0.343$^{*}$ & 0.341$^{**}$ \\
 & (0.145) & (0.157) & (0.176) & (0.176) & (0.175) & (0.164) \\[1pt]
CAR[0,5] & 0.422$^{**}$ & 0.475$^{**}$ & 0.503$^{**}$ & 0.509$^{**}$ & 0.501$^{**}$ & 0.499$^{**}$ \\
 & (0.192) & (0.219) & (0.245) & (0.245) & (0.246) & (0.224) \\[1pt]
\addlinespace[3pt]
\multicolumn{7}{l}{\textit{Panel B. Post-disclosure}}\\
CAR[0,10] & 0.566$^{**}$ & 0.585$^{**}$ & 0.652$^{**}$ & 0.661$^{**}$ & 0.661$^{**}$ & 0.684$^{**}$ \\
 & (0.243) & (0.276) & (0.305) & (0.307) & (0.307) & (0.300) \\[1pt]
CAR[0,20] & 0.780$^{**}$ & 0.794$^{*}$ & 0.843$^{*}$ & 0.858$^{*}$ & 0.849$^{*}$ & 0.772$^{*}$ \\
 & (0.386) & (0.396) & (0.437) & (0.437) & (0.439) & (0.458) \\[1pt]
CAR[0,30] & 0.490 & 0.477 & 0.511 & 0.532 & 0.516 & 0.435 \\
 & (0.487) & (0.505) & (0.558) & (0.555) & (0.558) & (0.552) \\[1pt]
\addlinespace[3pt]
\multicolumn{7}{l}{\textit{Panel C. Pre-disclosure}}\\
CAR[-10,-1] & $-$0.440 & $-$0.358 & $-$0.426 & $-$0.430 & $-$0.419 & $-$0.431 \\
 & (0.275) & (0.331) & (0.369) & (0.368) & (0.368) & (0.366) \\[1pt]
\midrule
Stock controls & No & Yes & Yes & Yes & Yes & Yes \\
Stock FE & No & No & Yes & Yes & Yes & Yes \\
Fund controls & No & No & No & Yes & Yes & Yes \\
Fund FE & No & No & No & No & Yes & Yes \\
Time FE & No & No & No & No & No & Yes \\
Drift control & No & No & No & No & No & Yes \\
Observations & 29,664 & 19,747 & 19,747 & 19,747 & 19,747 & 19,747 \\
Clusters (G) & 48 & 48 & 48 & 48 & 48 & 48 \\
R$^{2}$ & 0.001 & 0.004 & 0.003 & 0.003 & 0.003 & 0.003 \\
\bottomrule
\end{tabular*}
\begin{tablenotes}[flushleft]\footnotesize
\item \textit{Note.} The dependent variable is the cumulative abnormal return (CAR, in \%, Carhart four-factor adjusted, winsorized at 1/99). Entries are the difference-in-differences interaction treat $\times$ buy (the extra reaction to a treated fund's new purchase relative to its own exit, net of the control funds' buy--exit gap), with standard errors clustered by manager in parentheses. Columns add controls cumulatively: column (1) is raw; stock-level controls, stock fixed effects, fund-level controls, fund fixed effects and a year--quarter fixed effect plus the pre-event drift control enter in turn. The drift control is omitted when the dependent variable is itself a pre-event window. Observations, clusters and R$^{2}$ are reported at CAR[0,1]; the sample is essentially constant across windows. R$^{2}$ is the overall R$^{2}$ in columns (1)--(2) and the within-R$^{2}$ (after absorbing fixed effects) in columns (3)--(6). $^{*}p<.10$, $^{**}p<.05$, $^{***}p<.01$.
\end{tablenotes}
\end{threeparttable}
\end{table}

\begin{figure}[htbp]
\centering
\includegraphics[width=\linewidth]{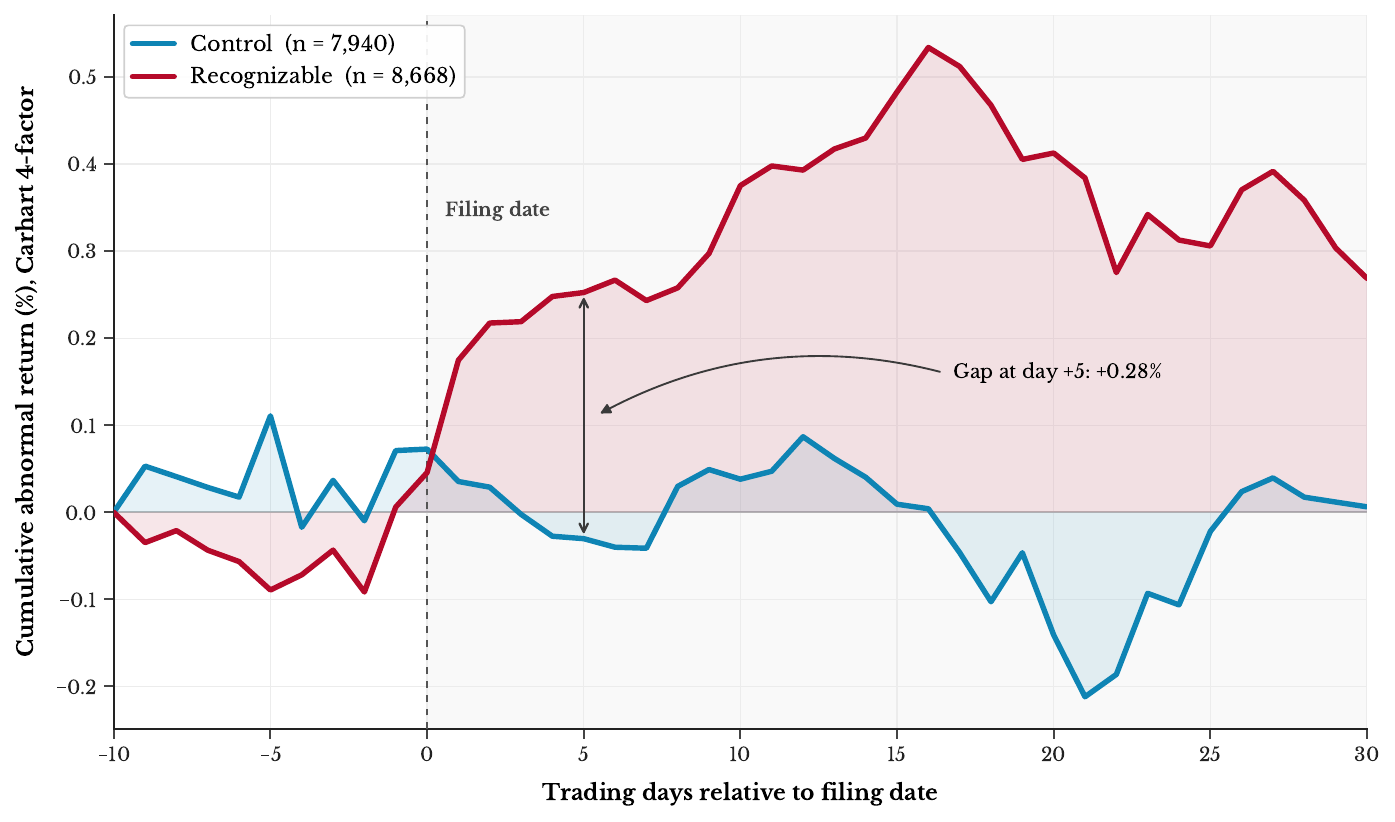}
\caption{Cumulative abnormal return around the disclosure date: recognizable versus control funds}
{\footnotesize\noindent\textit{Note.} Mean cumulative Carhart four-factor abnormal return (\%) on new positions, by group, from day $-$10 to day +30 relative to the EDGAR filing date (day 0). The vertical arrow marks the gap between the two buy-event paths at day +5.\par}
\end{figure}

\subsection{The recognition gradient}

Within the recognizable group there is significant variation in the level of recognition. Some managers are more recognizable than others. Warren Buffett is the most recognizable by far, with 323,969 average monthly Wikipedia views and 584,341 ProQuest mentions over 2000 to 2025. The next two are Carl Icahn and Michael Burry, the former more visible on ProQuest (103,210 ProQuest mentions; 44,792 Wikipedia monthly views), the latter more popular on Wikipedia (25,800 ProQuest mentions; 117,613 Wikipedia monthly views). We expect the level of recognition to correlate with the cumulative abnormal returns generated by a fund's stock disclosure. Figure 2 presents the relationship between fund recognition and the CAR the stock disclosure produces five days after disclosure. Table IV presents the regression coefficients of recognition on the cumulative abnormal returns in each window. Each within-group standard deviation of recognition adds +0.373\% to the two-day CAR (p = 0.005) and +0.427\% over [0,5] (p = 0.047). The correlation is significant in the immediate windows and across all specifications, and it persists through the post-disclosure windows, remaining significant at [0,10] (+0.479\%, p = 0.049) and, marginally, at [0,30] (+0.611\%, p = 0.061).

\begin{figure}[htbp]
\centering
\includegraphics[width=\linewidth]{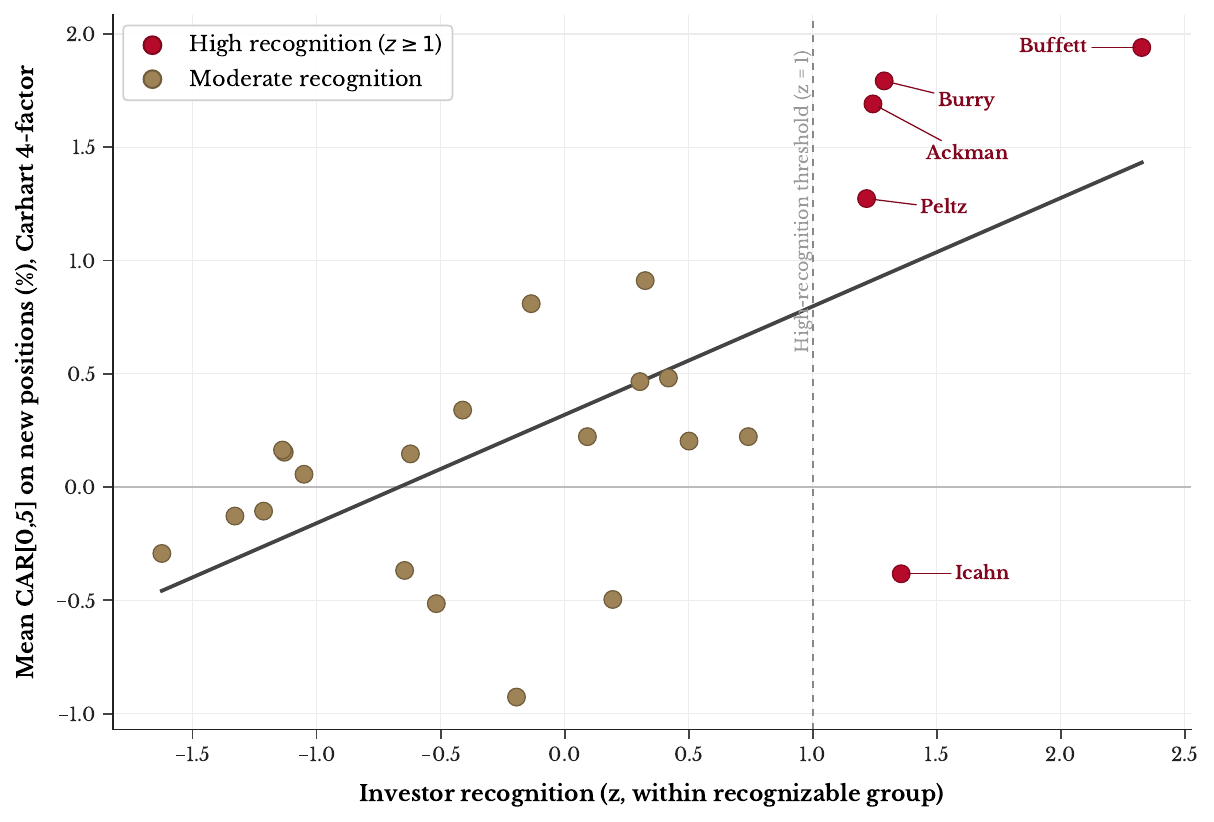}
\caption{Investor recognition and the five-day buy reaction within the recognizable group}
{\footnotesize\noindent\textit{Note.} Each point is one recognizable fund: the Investor Recognition Index (z, standardized within the recognizable group) against the fund's mean CAR[0,5] on new positions (\%). High-recognition funds (z $\geq$ 1) are labeled. The line is the least-squares fit across the 24 funds.\par}
\end{figure}

\begin{table}[htbp]
\centering
\caption{Recognition and the buy reaction within the recognizable funds}
\setlength{\tabcolsep}{3pt}
\begin{threeparttable}
\small
\begin{tabular*}{\linewidth}{@{\extracolsep{\fill}}lcccccc}
\toprule
 & (1) & (2) & (3) & (4) & (5) & (6) \\
 & Raw & \makecell{+ Stock\\ctrls} & \makecell{+ Stock\\FE} & \makecell{+ Fund\\ctrls} & \makecell{+ Fund\\FE} & Full \\
\midrule
\multicolumn{7}{l}{\textit{Panel A. Immediate reaction}}\\
CAR[0,1] & 0.284$^{**}$ & 0.321$^{**}$ & 0.358$^{***}$ & 0.357$^{***}$ & 0.355$^{***}$ & 0.373$^{***}$ \\
 & (0.112) & (0.116) & (0.125) & (0.125) & (0.126) & (0.119) \\[1pt]
CAR[0,3] & 0.404$^{***}$ & 0.363$^{**}$ & 0.407$^{**}$ & 0.405$^{**}$ & 0.402$^{**}$ & 0.431$^{**}$ \\
 & (0.123) & (0.154) & (0.163) & (0.162) & (0.163) & (0.154) \\[1pt]
CAR[0,5] & 0.426$^{***}$ & 0.366$^{*}$ & 0.404$^{*}$ & 0.401$^{*}$ & 0.397$^{*}$ & 0.427$^{**}$ \\
 & (0.149) & (0.185) & (0.210) & (0.209) & (0.211) & (0.204) \\[1pt]
\addlinespace[3pt]
\multicolumn{7}{l}{\textit{Panel B. Post-disclosure}}\\
CAR[0,10] & 0.553$^{***}$ & 0.406$^{*}$ & 0.420$^{*}$ & 0.418$^{*}$ & 0.418$^{*}$ & 0.479$^{**}$ \\
 & (0.163) & (0.215) & (0.227) & (0.227) & (0.229) & (0.231) \\[1pt]
CAR[0,20] & 0.684$^{***}$ & 0.446 & 0.482 & 0.486 & 0.491 & 0.583 \\
 & (0.214) & (0.306) & (0.318) & (0.319) & (0.318) & (0.387) \\[1pt]
CAR[0,30] & 0.754$^{***}$ & 0.579$^{**}$ & 0.573$^{**}$ & 0.564$^{**}$ & 0.579$^{**}$ & 0.611$^{*}$ \\
 & (0.232) & (0.230) & (0.251) & (0.250) & (0.250) & (0.310) \\[1pt]
\addlinespace[3pt]
\multicolumn{7}{l}{\textit{Panel C. Pre-disclosure}}\\
CAR[-10,-1] & 0.357$^{**}$ & 0.300$^{*}$ & 0.373 & 0.365 & 0.364 & 0.404 \\
 & (0.127) & (0.172) & (0.220) & (0.220) & (0.218) & (0.243) \\[1pt]
\midrule
Stock controls & No & Yes & Yes & Yes & Yes & Yes \\
Stock FE & No & No & Yes & Yes & Yes & Yes \\
Fund controls & No & No & No & Yes & Yes & Yes \\
Fund FE & No & No & No & No & Yes & Yes \\
Time FE & No & No & No & No & No & Yes \\
Drift control & No & No & No & No & No & Yes \\
Observations & 15,385 & 9,715 & 9,715 & 9,715 & 9,715 & 9,715 \\
Clusters (G) & 24 & 24 & 24 & 24 & 24 & 24 \\
R$^{2}$ & 0.004 & 0.008 & 0.008 & 0.009 & 0.008 & 0.008 \\
\bottomrule
\end{tabular*}
\begin{tablenotes}[flushleft]\footnotesize
\item \textit{Note.} The dependent variable is the cumulative abnormal return (CAR, in \%, Carhart four-factor adjusted, winsorized at 1/99). Here recognition enters continuously rather than as a discrete tier contrast: the sample is the 24 recognizable funds, and each entry is the change in the buy reaction per one within-recognizable standard deviation of recognition (recognition z $\times$ buy), with standard errors clustered by manager in parentheses. Columns add controls cumulatively in the same sequence used throughout; the drift control is omitted when the dependent variable is itself a pre-event window. Observations, clusters and R$^{2}$ are reported at CAR[0,1]. R$^{2}$ is the overall R$^{2}$ in columns (1)--(2) and the within-R$^{2}$ in columns (3)--(6). $^{*}p<.10$, $^{**}p<.05$, $^{***}p<.01$.
\end{tablenotes}
\end{threeparttable}
\end{table}

Since recognition correlates with the abnormal returns to a fund's disclosed positions, we now split the recognizable group into a top half and a bottom half by the median of the Investor Recognition Index, and compare each half with the control group. Table V shows the results for the top half against control. The top half produces stronger cumulative abnormal returns on stock disclosure than the control group across all windows: +0.485\% over the filing day and the next (p = 0.046), +0.869\% over [0,5] (p = 0.014), and +1.022\% at [0,10] (p = 0.019). There is again no clear reversal, although the effect is no longer significant in the window 30 days after disclosure (+0.983\%, p = 0.178). Table VI collects these contrasts in one place. It shows that the bottom half and the control group do not differ significantly in any post-disclosure window (CAR[0,1] -0.054\%, p = 0.689; [0,5] +0.182\%, p = 0.479). From Table VI we also see that the top half differs from the bottom half in the initial reaction, where the top half significantly outperforms the bottom half of the recognizable group (+0.590\% at [0,1], p = 0.035; +0.643\% at [0,3], p = 0.041).

\begin{table}[htbp]
\centering
\caption{Top half versus control funds}
\setlength{\tabcolsep}{3pt}
\begin{threeparttable}
\small
\begin{tabular*}{\linewidth}{@{\extracolsep{\fill}}lcccccc}
\toprule
 & (1) & (2) & (3) & (4) & (5) & (6) \\
 & Raw & \makecell{+ Stock\\ctrls} & \makecell{+ Stock\\FE} & \makecell{+ Fund\\ctrls} & \makecell{+ Fund\\FE} & Full \\
\midrule
\multicolumn{7}{l}{\textit{Panel A. Immediate reaction}}\\
CAR[0,1] & 0.446$^{**}$ & 0.481$^{**}$ & 0.518$^{**}$ & 0.517$^{**}$ & 0.517$^{**}$ & 0.485$^{**}$ \\
 & (0.194) & (0.216) & (0.252) & (0.252) & (0.251) & (0.234) \\[1pt]
CAR[0,3] & 0.614$^{**}$ & 0.595$^{**}$ & 0.675$^{**}$ & 0.675$^{**}$ & 0.674$^{**}$ & 0.691$^{**}$ \\
 & (0.234) & (0.255) & (0.288) & (0.287) & (0.287) & (0.260) \\[1pt]
CAR[0,5] & 0.686$^{**}$ & 0.790$^{**}$ & 0.847$^{**}$ & 0.853$^{**}$ & 0.846$^{**}$ & 0.869$^{**}$ \\
 & (0.276) & (0.294) & (0.367) & (0.367) & (0.367) & (0.334) \\[1pt]
\addlinespace[3pt]
\multicolumn{7}{l}{\textit{Panel B. Post-disclosure}}\\
CAR[0,10] & 0.950$^{***}$ & 0.915$^{**}$ & 0.965$^{**}$ & 0.973$^{**}$ & 0.971$^{**}$ & 1.022$^{**}$ \\
 & (0.336) & (0.371) & (0.438) & (0.439) & (0.439) & (0.416) \\[1pt]
CAR[0,20] & 1.158$^{**}$ & 1.030$^{*}$ & 1.099$^{*}$ & 1.111$^{*}$ & 1.109$^{*}$ & 1.056$^{*}$ \\
 & (0.542) & (0.520) & (0.609) & (0.610) & (0.612) & (0.624) \\[1pt]
CAR[0,30] & 1.034 & 1.044 & 1.032 & 1.052 & 1.039 & 0.983 \\
 & (0.628) & (0.632) & (0.736) & (0.735) & (0.737) & (0.715) \\[1pt]
\addlinespace[3pt]
\multicolumn{7}{l}{\textit{Panel C. Pre-disclosure}}\\
CAR[-10,-1] & $-$0.193 & $-$0.198 & $-$0.169 & $-$0.178 & $-$0.173 & $-$0.183 \\
 & (0.316) & (0.389) & (0.440) & (0.440) & (0.440) & (0.433) \\[1pt]
\midrule
Stock controls & No & Yes & Yes & Yes & Yes & Yes \\
Stock FE & No & No & Yes & Yes & Yes & Yes \\
Fund controls & No & No & No & Yes & Yes & Yes \\
Fund FE & No & No & No & No & Yes & Yes \\
Time FE & No & No & No & No & No & Yes \\
Drift control & No & No & No & No & No & Yes \\
Observations & 21,862 & 14,789 & 14,789 & 14,789 & 14,789 & 14,789 \\
Clusters (G) & 36 & 36 & 36 & 36 & 36 & 36 \\
R$^{2}$ & 0.002 & 0.006 & 0.006 & 0.006 & 0.005 & 0.006 \\
\bottomrule
\end{tabular*}
\begin{tablenotes}[flushleft]\footnotesize
\item \textit{Note.} The dependent variable is the cumulative abnormal return (CAR, in \%, Carhart four-factor adjusted, winsorized at 1/99). Entries are the difference-in-differences interaction treat $\times$ buy, with standard errors clustered by manager in parentheses. The treated group is the top half of the recognizable funds by recognition; the control group is the matched control sample. Columns add controls cumulatively as throughout; the drift control is omitted when the dependent variable is itself a pre-event window. R$^{2}$ is the overall R$^{2}$ in columns (1)--(2) and the within-R$^{2}$ in columns (3)--(6). $^{*}p<.10$, $^{**}p<.05$, $^{***}p<.01$.
\end{tablenotes}
\end{threeparttable}
\end{table}

\begin{table}[htbp]
\centering
\caption{The recognition gradient (full specification)}
\setlength{\tabcolsep}{3pt}
\begin{threeparttable}
\footnotesize
\begin{tabularx}{\linewidth}{>{\raggedright\arraybackslash\hangindent=1em}Xccccccc}
\toprule
 & \multicolumn{7}{c}{CAR window} \\
\cmidrule(l{3pt}){2-8}
Contrast & [0,1] & [0,3] & [0,5] & [0,10] & [0,20] & [0,30] & [$-$10,$-$1] \\
\midrule
Recognizable $-$ Control & 0.200 & 0.341$^{**}$ & 0.499$^{**}$ & 0.684$^{**}$ & 0.772$^{*}$ & 0.435 & $-$0.431 \\
Top half $-$ Control & 0.485$^{**}$ & 0.691$^{**}$ & 0.869$^{**}$ & 1.022$^{**}$ & 1.056$^{*}$ & 0.983 & $-$0.183 \\
Bottom half $-$ Control & $-$0.054 & 0.033 & 0.182 & 0.377 & 0.525 & $-$0.052 & $-$0.672$^{*}$ \\
Top half $-$ Bottom half & 0.590$^{**}$ & 0.643$^{**}$ & 0.645 & 0.713 & 0.642 & 1.048 & 0.457 \\
High-recognition $-$ Control & 1.422$^{***}$ & 1.869$^{***}$ & 2.309$^{***}$ & 2.547$^{***}$ & 3.091$^{***}$ & 2.319$^{**}$ & 0.392 \\
Moderate-recognition $-$ Control & 0.088 & 0.201 & 0.331 & 0.504$^{*}$ & 0.544 & 0.252 & $-$0.530 \\
High-recognition $-$ Moderate-recognition & 1.246$^{***}$ & 1.456$^{***}$ & 1.719$^{**}$ & 1.773$^{**}$ & 2.563$^{**}$ & 1.856$^{**}$ & 1.119 \\
\bottomrule
\end{tabularx}
\begin{tablenotes}[flushleft]\footnotesize
\item \textit{Note.} Each entry is the full-specification difference-in-differences interaction treat $\times$ buy (\%, Carhart four-factor adjusted) for the indicated contrast, read across the recognition distribution. The pooled and bottom-half/moderate-recognition contrasts are economically small and mostly insignificant, whereas the effect is concentrated in the top half and strongest in the high-recognition tier. $^{*}p<.10$, $^{**}p<.05$, $^{***}p<.01$.
\end{tablenotes}
\end{threeparttable}
\end{table}

\subsection{The high-recognition premium}

A closer look at the cumulative returns generated by each fund (Internet Appendix Tables IA.II and IA.III) shows that the highest abnormal returns in the immediate reaction windows are generated mainly by a small group of managers at the very top of the recognition distribution. These are the managers whose recognition, measured by ProQuest mentions and Wikipedia views, outweighs that of every other manager in the recognizable group. We define the high-recognition group as the managers whose recognition is at least one standard deviation above the mean of the recognition index within the recognizable group (recognition z $\geq$ 1). This criterion yields five funds: Berkshire (Buffett), Icahn (Carl Icahn), Scion (Burry), Pershing Square (Ackman), and Trian (Peltz). We interpret the high-recognition premium as the return associated with that level of recognition and not as a property of any particular fund on that list. Table VII presents the regression comparing the cumulative abnormal returns of these high-recognition funds with the control group. Figure 3 plots the cumulative abnormal returns across all windows. The results show that the high-recognition group produces the strongest disclosure premium, hereafter the high-recognition premium. The immediate reaction to the high-recognition disclosure is +1.422\% over the filing day and the next (p < 0.001) and +2.309\% over [0,5] (p = 0.002), and it remains significant and grows after disclosure, reaching +3.091\% at [0,20] (p = 0.003) and staying significant at [0,30] (+2.319\%, p = 0.016), accounting for all controls. From Table VI we also see that this group significantly outperforms the rest of the recognizable group, the moderate-recognition funds (recognition z < 1), by +1.246\% over the filing day and the next (p < 0.001), a gap that widens across the post-disclosure windows, while the moderate-recognition funds on their own do not differ from the control group.

\begin{table}[htbp]
\centering
\caption{High-recognition versus control funds}
\setlength{\tabcolsep}{3pt}
\begin{threeparttable}
\small
\begin{tabular*}{\linewidth}{@{\extracolsep{\fill}}lcccccc}
\toprule
 & (1) & (2) & (3) & (4) & (5) & (6) \\
 & Raw & \makecell{+ Stock\\ctrls} & \makecell{+ Stock\\FE} & \makecell{+ Fund\\ctrls} & \makecell{+ Fund\\FE} & Full \\
\midrule
\multicolumn{7}{l}{\textit{Panel A. Immediate reaction}}\\
CAR[0,1] & 1.347$^{***}$ & 1.286$^{***}$ & 1.436$^{***}$ & 1.445$^{***}$ & 1.448$^{***}$ & 1.422$^{***}$ \\
 & (0.256) & (0.295) & (0.290) & (0.289) & (0.289) & (0.316) \\[1pt]
CAR[0,3] & 1.758$^{***}$ & 1.690$^{***}$ & 1.910$^{***}$ & 1.941$^{***}$ & 1.938$^{***}$ & 1.869$^{***}$ \\
 & (0.266) & (0.396) & (0.406) & (0.407) & (0.402) & (0.442) \\[1pt]
CAR[0,5] & 1.956$^{***}$ & 1.916$^{***}$ & 2.268$^{***}$ & 2.309$^{***}$ & 2.303$^{***}$ & 2.309$^{***}$ \\
 & (0.459) & (0.601) & (0.661) & (0.671) & (0.662) & (0.668) \\[1pt]
\addlinespace[3pt]
\multicolumn{7}{l}{\textit{Panel B. Post-disclosure}}\\
CAR[0,10] & 2.096$^{***}$ & 2.001$^{**}$ & 2.300$^{***}$ & 2.375$^{***}$ & 2.378$^{***}$ & 2.547$^{***}$ \\
 & (0.565) & (0.783) & (0.790) & (0.802) & (0.792) & (0.808) \\[1pt]
CAR[0,20] & 3.105$^{***}$ & 2.545$^{***}$ & 2.764$^{***}$ & 2.859$^{***}$ & 2.875$^{***}$ & 3.091$^{***}$ \\
 & (0.580) & (0.870) & (0.903) & (0.905) & (0.896) & (0.959) \\[1pt]
CAR[0,30] & 2.722$^{***}$ & 1.700$^{**}$ & 1.917$^{**}$ & 2.022$^{**}$ & 2.022$^{**}$ & 2.319$^{**}$ \\
 & (0.792) & (0.824) & (0.925) & (0.919) & (0.892) & (0.905) \\[1pt]
\addlinespace[3pt]
\multicolumn{7}{l}{\textit{Panel C. Pre-disclosure}}\\
CAR[-10,-1] & 0.406 & 0.595 & 0.472 & 0.487 & 0.459 & 0.392 \\
 & (0.471) & (0.615) & (0.859) & (0.856) & (0.859) & (0.830) \\[1pt]
\midrule
Stock controls & No & Yes & Yes & Yes & Yes & Yes \\
Stock FE & No & No & Yes & Yes & Yes & Yes \\
Fund controls & No & No & No & Yes & Yes & Yes \\
Fund FE & No & No & No & No & Yes & Yes \\
Time FE & No & No & No & No & No & Yes \\
Drift control & No & No & No & No & No & Yes \\
Observations & 15,684 & 10,972 & 10,972 & 10,972 & 10,972 & 10,972 \\
Clusters (G) & 29 & 29 & 29 & 29 & 29 & 29 \\
R$^{2}$ & 0.005 & 0.010 & 0.011 & 0.011 & 0.009 & 0.011 \\
\bottomrule
\end{tabular*}
\begin{tablenotes}[flushleft]\footnotesize
\item \textit{Note.} The dependent variable is the cumulative abnormal return (CAR, in \%, Carhart four-factor adjusted, winsorized at 1/99). Entries are the difference-in-differences interaction treat $\times$ buy, with standard errors clustered by manager in parentheses. The treated group is the five high-recognition funds (recognition z $\geq$ 1); the control group is the matched control sample. Columns add controls cumulatively as throughout; the drift control is omitted when the dependent variable is itself a pre-event window. R$^{2}$ is the overall R$^{2}$ in columns (1)--(2) and the within-R$^{2}$ in columns (3)--(6). $^{*}p<.10$, $^{**}p<.05$, $^{***}p<.01$.
\end{tablenotes}
\end{threeparttable}
\end{table}

\begin{figure}[htbp]
\centering
\includegraphics[width=\linewidth]{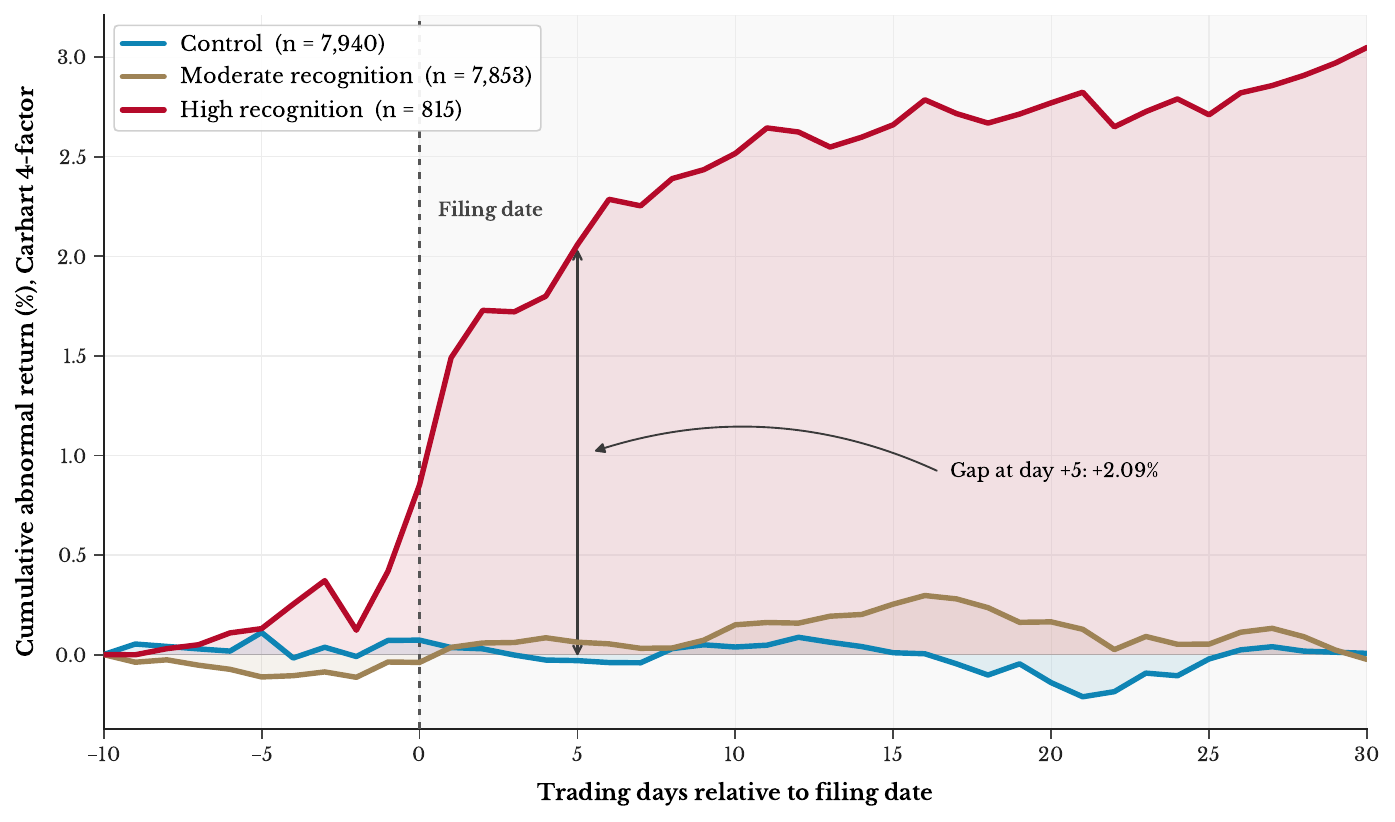}
\caption{Cumulative abnormal return around the disclosure date by recognition tier: control, moderate recognition, and high recognition}
{\footnotesize\noindent\textit{Note.} Mean cumulative Carhart four-factor abnormal return (\%) on new positions, by recognition tier, from day $-$10 to day +30 relative to the EDGAR filing date (day 0). The vertical arrow marks the gap between the high-recognition and the control buy-event paths at day +5.\par}
\end{figure}

Table VIII reports a kink regression with a threshold added at one standard deviation above the mean of the Investor Recognition Index. The base slope, the reaction per standard deviation of recognition, is modest (+0.249\% at [0,1], p = 0.037). The extra slope that switches on above the high-recognition threshold is several times larger (+0.827\% at [0,1], p = 0.014; +1.412\% over [0,5], p = 0.004), so a high-recognition fund faces a total slope of about +1.075\% at [0,1]. Because the tail is identified off the five funds above the threshold, we report small-cluster wild-bootstrap p-values next to the analytic ones; the tail term holds under the unrestricted bootstrap (WCU p = 0.003 at [0,1]) and approaches significance under the restricted version (WCR p = 0.133). The effect of recognition is not linear: it is flat across most of the recognizable group and steep only at the high-recognition end.

\begin{table}[htbp]
\centering
\caption{An extra slope above the high-recognition threshold}
\setlength{\tabcolsep}{4pt}
\begin{threeparttable}
\small
\begin{tabular*}{\linewidth}{@{\extracolsep{\fill}}lcccccc}
\toprule
 & \makecell{Base slope\\(per SD)} & \makecell{Tail (extra)\\(z $\geq$ 1)} & \makecell{Total\\(z $\geq$ 1)} & \makecell{p\\(analytic)} & \makecell{p\\(WCR)} & \makecell{p\\(WCU)} \\
\midrule
\multicolumn{7}{l}{\textit{Panel A. Immediate reaction}}\\
CAR[0,1] & 0.249$^{**}$ & 0.827$^{**}$ & 1.075 & 0.014 & 0.133 & 0.003 \\
CAR[0,3] & 0.265$^{*}$ & 1.101$^{***}$ & 1.366 & <0.001 & 0.043 & <0.001 \\
CAR[0,5] & 0.214 & 1.412$^{***}$ & 1.626 & 0.004 & 0.032 & <0.001 \\
\addlinespace[3pt]
\multicolumn{7}{l}{\textit{Panel B. Post-disclosure}}\\
CAR[0,10] & 0.260 & 1.452$^{**}$ & 1.712 & 0.027 & 0.124 & 0.008 \\
CAR[0,20] & 0.202 & 2.527$^{**}$ & 2.729 & 0.023 & 0.157 & 0.020 \\
CAR[0,30] & 0.469 & 0.942 & 1.411 & 0.419 & 0.419 & 0.357 \\
\addlinespace[3pt]
\multicolumn{7}{l}{\textit{Panel C. Pre-disclosure}}\\
CAR[-10,-1] & 0.186 & 1.445$^{**}$ & 1.631 & 0.036 & 0.281 & 0.051 \\
\bottomrule
\end{tabular*}
\begin{tablenotes}[flushleft]\footnotesize
\item \textit{Note.} Full-specification kink regression on the 24 recognizable funds: CAR $\sim$ FE + controls + buy + (recognition z $\times$ buy) + (max(z-1,0) $\times$ buy). The base slope is the per-SD buy reaction; the tail term is the extra slope that switches on above the high-recognition threshold (z $\geq$ 1); ``Total'' is their sum (the slope faced by high-recognition funds). Stars and the analytic p refer to the tail term (24 clusters). Because identification of the tail rests on the funds above the threshold, the small-cluster-robust wild bootstrap is also reported: WCR (restricted) and WCU (unrestricted), Webb six-point weights, B = 9,999. $^{*}p<.10$, $^{**}p<.05$, $^{***}p<.01$.
\end{tablenotes}
\end{threeparttable}
\end{table}

\section{Robustness}

\subsection{Small-cluster inference}

Table IX reports the small-cluster inference behind the main contrasts. Because the effect is identified off few treated funds (five in the high-recognition tier, twelve in the top half), the cluster-robust standard errors can be unreliable, so we re-test every contrast with the wild-cluster bootstrap (Cameron et al., 2008) in two forms: restricted (WCR), which imposes the null when resampling, and unrestricted (WCU), which does not. We find that the disclosure premium is robust. WCR and WCU stay below the 10\% level across the immediate reaction windows for both the high-recognition-vs-control and top-half-vs-control contrasts. The pooled recognizable-vs-control contrast is borderline at [0,1] (WCR p = 0.126), significant at [0,3] and [0,5], and not significant at [0,30]. The bottom-half contrast fails the bootstrap in every window, and the moderate-recognition contrast passes only at [0,5] (WCR p = 0.089).

\begin{table}[htbp]
\centering
\caption{Small-cluster inference across the recognition gradient}
{\centering\small\textbf{Panel A. Treated tier vs. control}\par\vspace{2pt}}
{\setlength{\tabcolsep}{4pt}\footnotesize
\begin{tabular*}{\linewidth}{@{\extracolsep{\fill}}lcccc}
\toprule
Contrast / window & $\beta$ & Analytic p & WCR p & WCU p \\
\midrule
\multicolumn{5}{l}{\textit{Recognizable $-$ Control}}\\
\quad CAR[0,1] & 0.200 & 0.159 & 0.126 & 0.133 \\
\quad CAR[0,3] & 0.341 & 0.043 & 0.023 & 0.025 \\
\quad CAR[0,5] & 0.499 & 0.031 & 0.018 & 0.018 \\
\quad CAR[0,30] & 0.435 & 0.434 & 0.400 & 0.398 \\
\addlinespace[3pt]
\multicolumn{5}{l}{\textit{Top half $-$ Control}}\\
\quad CAR[0,1] & 0.485 & 0.046 & 0.021 & 0.079 \\
\quad CAR[0,3] & 0.691 & 0.012 & 0.001 & 0.023 \\
\quad CAR[0,5] & 0.869 & 0.014 & 0.008 & 0.008 \\
\quad CAR[0,30] & 0.983 & 0.178 & 0.155 & 0.164 \\
\addlinespace[3pt]
\multicolumn{5}{l}{\textit{Bottom half $-$ Control}}\\
\quad CAR[0,1] & $-$0.054 & 0.689 & 0.678 & 0.679 \\
\quad CAR[0,3] & 0.033 & 0.854 & 0.862 & 0.859 \\
\quad CAR[0,5] & 0.182 & 0.479 & 0.458 & 0.457 \\
\quad CAR[0,30] & $-$0.052 & 0.927 & 0.923 & 0.921 \\
\addlinespace[3pt]
\multicolumn{5}{l}{\textit{High-recognition $-$ Control}}\\
\quad CAR[0,1] & 1.422 & <0.001 & 0.011 & 0.002 \\
\quad CAR[0,3] & 1.869 & <0.001 & 0.045 & 0.004 \\
\quad CAR[0,5] & 2.309 & 0.002 & 0.071 & 0.001 \\
\quad CAR[0,30] & 2.319 & 0.016 & 0.148 & 0.027 \\
\addlinespace[3pt]
\multicolumn{5}{l}{\textit{Moderate-recognition $-$ Control}}\\
\quad CAR[0,1] & 0.088 & 0.481 & 0.455 & 0.458 \\
\quad CAR[0,3] & 0.201 & 0.149 & 0.123 & 0.126 \\
\quad CAR[0,5] & 0.331 & 0.104 & 0.089 & 0.085 \\
\quad CAR[0,30] & 0.252 & 0.651 & 0.617 & 0.623 \\
\bottomrule
\end{tabular*}
\par}
\vspace{4pt}
{\centering\small\textbf{Panel B. Within the recognizable funds}\par\vspace{2pt}}
{\setlength{\tabcolsep}{4pt}\footnotesize
\begin{tabular*}{\linewidth}{@{\extracolsep{\fill}}lccccc}
\toprule
Contrast / window & $\beta$ & Analytic p & WCR p & WCU p & RI p \\
\midrule
\multicolumn{6}{l}{\textit{Top half $-$ Bottom half}}\\
\quad CAR[0,1] & 0.590 & 0.035 & 0.015 & 0.041 & 0.014 \\
\quad CAR[0,3] & 0.643 & 0.041 & 0.019 & 0.022 & 0.021 \\
\quad CAR[0,5] & 0.645 & 0.126 & 0.078 & 0.090 & 0.072 \\
\quad CAR[0,30] & 1.048 & 0.200 & 0.172 & 0.185 & 0.183 \\
\addlinespace[3pt]
\multicolumn{6}{l}{\textit{High-recognition $-$ Moderate-recognition}}\\
\quad CAR[0,1] & 1.246 & <0.001 & 0.027 & 0.002 & <0.001 \\
\quad CAR[0,3] & 1.456 & 0.003 & 0.058 & 0.011 & 0.001 \\
\quad CAR[0,5] & 1.719 & 0.022 & 0.144 & 0.067 & 0.002 \\
\quad CAR[0,30] & 1.856 & 0.040 & 0.155 & 0.031 & 0.063 \\
\bottomrule
\end{tabular*}
\par}
\vspace{4pt}
\begin{minipage}{\linewidth}\footnotesize\linespread{1.05}\selectfont
\textit{Note.} $\beta$ is the full-specification treat $\times$ buy coefficient (\%, Carhart four-factor adjusted). With few treated clusters the analytic (cluster-robust) p is unreliable, so the wild-cluster bootstrap is reported in restricted (WCR) and unrestricted (WCU) form, which bracket the truth (Webb, 2023, six-point weights; B = 9,999). Panel B adds randomization inference (RI), permuting the tier label among the 24 recognizable funds (exact two-sided p; exhaustive over all C(24,5) labelings for the high-recognition tier, 200,000 draws for the top half). RI is valid only within the recognizable group, where the funds are exchangeable under the sharp null that recognition does not affect the buy reaction.
\end{minipage}
\end{table}

Within the recognizable group we also run a randomization inference permutation test. We hold the 24 recognizable funds fixed and permute which of them carry the high-recognition label, recomputing the high-versus-moderate-recognition reaction for every possible relabeling (all 42,504 of them) and reading off the share that match or exceed the observed effect. This tests the sharp null that recognition is unrelated to the reaction among the recognizable funds, under which the labels are exchangeable. We find that the high-versus-moderate-recognition difference is significant in this permutation distribution (p < 0.01 at [0,1] through [0,20]) and marginal at [0,30] (p = 0.063). The top-vs-bottom split also remains significant in the immediate windows (p = 0.014 at [0,1], 0.021 at [0,3]).

Table X reports a leave-one-out jackknife, which re-estimates the high-recognition premium after dropping one high-recognition fund at a time. Removing any single fund, Warren Buffett included, does not meaningfully reduce the premium in the immediate reaction windows, where every re-estimate stays large and significant (the lowest two-day estimate, +1.12\%, comes from dropping Buffett and is still significant at 1\%). Dropping Buffett or Burry costs the 30-day premium its significance, though the coefficient stays positive in both cases. Dropping Icahn raises the premium in every window. Internet Appendix Table IA.II explains why: Icahn is the only high-recognition fund whose own disclosures earn no premium, so taking him out strengthens rather than weakens the high-recognition premium.

\begin{table}[htbp]
\centering
\caption{Leave-one-out jackknife of the high-recognition effect}
\setlength{\tabcolsep}{4pt}
\begin{threeparttable}
\small
\begin{tabular*}{\linewidth}{@{\extracolsep{\fill}}lcccc}
\toprule
 & \multicolumn{4}{c}{CAR window} \\
\cmidrule(l{3pt}){2-5}
Fund removed & [0,1] & [0,3] & [0,5] & [0,30] \\
\midrule
None (full estimate) & 1.422$^{***}$ & 1.869$^{***}$ & 2.309$^{***}$ & 2.319$^{**}$ \\
Buffett (Berkshire) & 1.122$^{***}$ & 1.518$^{**}$ & 2.062$^{*}$ & 1.722 \\
Icahn & 1.598$^{***}$ & 2.180$^{***}$ & 2.722$^{***}$ & 2.749$^{***}$ \\
Burry (Scion) & 1.312$^{***}$ & 1.648$^{**}$ & 1.738$^{**}$ & 2.051 \\
Ackman (Pershing Square) & 1.505$^{***}$ & 1.834$^{***}$ & 2.251$^{***}$ & 2.626$^{***}$ \\
Peltz (Trian) & 1.398$^{***}$ & 1.986$^{***}$ & 2.547$^{***}$ & 2.117$^{*}$ \\
\bottomrule
\end{tabular*}
\begin{tablenotes}[flushleft]\footnotesize
\item \textit{Note.} Leave-one-out jackknife of the headline high-recognition effect (Table VII, column 6). Each row re-estimates the full-specification High-recognition-vs-Control treat $\times$ buy coefficient (\%, Carhart four-factor adjusted) after removing one high-recognition fund from the treated group; the first row retains all five. Standard errors clustered by manager. $^{*}p<.10$, $^{**}p<.05$, $^{***}p<.01$.
\end{tablenotes}
\end{threeparttable}
\end{table}

\subsection{Rolling recognition}

So far we have used the static index of recognition. The static index is measured over 2000 to 2025 as a whole and could in principle classify funds using recognition acquired late in the sample. If we replace the static recognition index with a rolling index built only from the Wikipedia page views available at the filing date (available from 2016), the premium remains at every horizon. It is +1.52\% at CAR[0,1] (p = 0.004) against +1.77\% under the static classification on the same subsample, +1.81\% against +2.05\% at CAR[0,3] (p = 0.006), +2.65\% against +3.07\% at CAR[0,5] (p = 0.013), +3.01\% against +3.39\% at CAR[0,10] (p = 0.036), +3.34\% against +4.29\% at CAR[0,20] (p = 0.062), and +2.64\% against +3.37\% at CAR[0,30] (p = 0.045). The rolling index also predicts the reaction across managers (+0.68\% per standard deviation at CAR[0,1], p = 0.007).

\subsection{Additional checks}

Table XI reports the high-recognition premium across alternative high-recognition-threshold cutoffs (a recognition-z sweep) and across the three abnormal-return measures (Carhart four-factor, market-adjusted, and market-model). The premium rises with the cutoff, from +0.49\% at [0,1] for the top half (z $\geq$ 0) to +1.42\% at z $\geq$ 0.75 or z $\geq$ 1 and +1.87\% for Berkshire alone (z $\geq$ 1.5), and it is similar across the three abnormal-return measures.

\begin{table}[htbp]
\centering
\caption{Robustness of the high-recognition effect}
\setlength{\tabcolsep}{4pt}
\begin{threeparttable}
\small
\begin{tabular*}{\linewidth}{@{\extracolsep{\fill}}lcccc}
\toprule
 & \multicolumn{4}{c}{CAR window} \\
\cmidrule(l{3pt}){2-5}
 & [0,1] & [0,3] & [0,5] & [0,30] \\
\midrule
\multicolumn{5}{l}{\textit{Panel A. High-recognition definition (recognition-z threshold sweep)}}\\
z $\geq$ 0 (n = 12) & 0.485$^{**}$ & 0.691$^{**}$ & 0.869$^{**}$ & 0.983 \\
z $\geq$ 0.25 (n = 10) & 0.524$^{**}$ & 0.818$^{**}$ & 1.106$^{***}$ & 1.664$^{**}$ \\
z $\geq$ 0.5 (n = 7) & 0.601$^{*}$ & 0.831$^{**}$ & 1.205$^{**}$ & 1.384$^{*}$ \\
z $\geq$ 0.75 (n = 5) & 1.422$^{***}$ & 1.869$^{***}$ & 2.309$^{***}$ & 2.319$^{**}$ \\
z $\geq$ 1 (n = 5) & 1.422$^{***}$ & 1.869$^{***}$ & 2.309$^{***}$ & 2.319$^{**}$ \\
z $\geq$ 1.25 (n = 3) & 1.493$^{***}$ & 1.960$^{***}$ & 2.513$^{***}$ & 2.457$^{**}$ \\
z $\geq$ 1.5 (n = 1) & 1.865$^{***}$ & 2.407$^{***}$ & 2.666$^{***}$ & 3.401$^{***}$ \\
\addlinespace[3pt]
\multicolumn{5}{l}{\textit{Panel B. Abnormal-return measure}}\\
Carhart 4-factor & 1.422$^{***}$ & 1.869$^{***}$ & 2.309$^{***}$ & 2.319$^{**}$ \\
Market-adjusted & 1.505$^{***}$ & 1.899$^{***}$ & 2.219$^{***}$ & 1.601$^{*}$ \\
Market-model & 1.476$^{***}$ & 1.917$^{***}$ & 2.309$^{***}$ & 2.116$^{**}$ \\
\bottomrule
\end{tabular*}
\begin{tablenotes}[flushleft]\footnotesize
\item \textit{Note.} High-recognition-vs-Control treat $\times$ buy (full specification). Panel A sweeps the high-recognition cutoff from z $\geq$ 0 (the top half) up to z $\geq$ 1.5 in steps of 0.25, where High-recognition = funds at or above the threshold; n is the resulting number of treated funds (note z $\geq$ 0.75 and z $\geq$ 1 coincide). Panel B re-estimates the headline (z $\geq$ 1) effect under the three abnormal-return measures. $^{*}p<.10$, $^{**}p<.05$, $^{***}p<.01$.
\end{tablenotes}
\end{threeparttable}
\end{table}

Table XII re-estimates the high-recognition premium with the financial-crisis quarters removed. The premium survives every exclusion and, if anything, grows. Dropping both 2008 to 2009 and 2020 increases the five-day cumulative abnormal returns from +2.31\% to +2.76\% and the twenty-day CAR from +3.09\% to +4.07\%, and every estimate through [0,20] remains significant under the restricted wild-cluster bootstrap. The effect is not an artifact of crisis-period volatility.

\begin{table}[htbp]
\centering
\caption{Subperiod stability of the high-recognition effect (crisis exclusions)}
\setlength{\tabcolsep}{3pt}
\begin{threeparttable}
\small
\begin{tabular*}{\linewidth}{@{\extracolsep{\fill}}lcccccc}
\toprule
 & \multicolumn{6}{c}{CAR window} \\
\cmidrule(l{3pt}){2-7}
Sample & [0,1] & [0,3] & [0,5] & [0,10] & [0,20] & [0,30] \\
\midrule
Full sample & 1.422$^{***}$ & 1.869$^{***}$ & 2.309$^{***}$ & 2.547$^{***}$ & 3.091$^{***}$ & 2.319$^{**}$ \\
 & (0.011) & (0.045) & (0.071) & (0.064) & (0.042) & (0.148) \\[1pt]
Drop 2008-09 & 1.386$^{***}$ & 1.792$^{***}$ & 2.480$^{***}$ & 2.859$^{***}$ & 3.551$^{***}$ & 2.893$^{***}$ \\
 & (0.010) & (0.066) & (0.011) & (0.027) & (0.003) & (0.021) \\[1pt]
Drop 2020 & 1.450$^{***}$ & 2.125$^{***}$ & 2.565$^{***}$ & 2.921$^{***}$ & 3.584$^{***}$ & 2.616$^{***}$ \\
 & (0.048) & (0.034) & (0.058) & (0.023) & (0.034) & (0.119) \\[1pt]
Drop 2008-09 \& 2020 & 1.416$^{***}$ & 2.061$^{***}$ & 2.759$^{***}$ & 3.277$^{***}$ & 4.066$^{***}$ & 3.195$^{***}$ \\
 & (0.031) & (0.045) & (0.008) & (0.004) & (0.004) & (0.005) \\[1pt]
\bottomrule
\end{tabular*}
\begin{tablenotes}[flushleft]\footnotesize
\item \textit{Note.} High-recognition-vs-Control treat $\times$ buy (full specification, \%, Carhart four-factor adjusted) after dropping crisis quarters. Each cell reports the point estimate (stars from the analytic cluster-robust p) and, in parentheses, the restricted wild-cluster (WCR) p (Webb six-point weights, B = 9,999). The effect is, if anything, larger once crisis quarters are removed, and it remains significant under the conservative small-cluster bootstrap in every window through [0,20]; at [0,30] the bootstrap p exceeds 0.10 in the full sample and when 2020 is dropped. $^{*}p<.10$, $^{**}p<.05$, $^{***}p<.01$.
\end{tablenotes}
\end{threeparttable}
\end{table}

The Internet Appendix reports further checks. Re-running the control matching with no manual removals returns nine of the excluded funds to the control group and leaves the high-recognition estimates unchanged (Internet Appendix Section IA.5). The Internet Appendix also adds a control for each fund's own past stock-picking record (IA.3) and excludes the purchases that were public before the 13F through the fund's own 13D or 13G filing (IA.6). The immediate premium survives both checks (IA.3).

Overall, the evidence indicates that the premium is a function of recognition, and especially of its highest level.

\section{Mechanism}

We have shown so far that a significant disclosure premium exists among managers with the highest level of recognition. The remaining question is what produces the premium. Section II described three mechanisms. Mechanism 1 and Mechanism 2 are the rational incorporation of information. Investors read the disclosure, treat it as information about the disclosed stock, and trade on it. Under Mechanism 1 the information in the disclosure of a recognizable manager reaches more investors. Under Mechanism 2 the same information is more persuasive to each investor who receives it because of its source. Both mechanisms should work mostly through the investors with the least information of their own, that is retail investors. Mechanism 3 is pure herding. Investors imitate the trade of the manager they follow regardless of its informational value. Price data cannot separate the three mechanisms, because each of them predicts the reaction we have documented. We therefore turn to the trades around high-recognition disclosures.

\subsection{Mechanism 1: the number of trades}

Mechanism 1 implies that the number of trades in the disclosed stock goes up after the disclosure of a recognizable manager. Table XIII reports the effect on the number of trades, for all trades first and then separately for non-retail and retail trades, after purchases (Panel A) and after exits (Panel B). We find that the number of trades goes up. Over the filing day and the next, the total number of trades in a stock bought by a high-recognition fund is 7.2\% higher than in a stock bought by a control fund (p = 0.027). The split by trader type shows that the number of non-retail trades rises by 6.7\% over [0,1] (p = 0.038) and the number of retail trades rises by 15.1\% over [0,1] (p = 0.001), 12.3\% over [0,3] (p = 0.002), 9.0\% over [0,5] (p = 0.012), and is still 6.5\% higher over [0,10] (p = 0.080). Retail investors keep trading the disclosed stock for two weeks after disclosure. The same trading pattern is present after exit disclosures. The number of retail trades in a stock sold by a high-recognition fund is 10.4\% higher than in a stock sold by a control fund over [0,1] (p = 0.010) and 5.3\% higher over [0,10] (p = 0.059), and the total count is 6.4\% higher over [0,1] (p = 0.089). The disclosure attracts trading whichever direction it points. The trading volume increase is concentrated in the high-recognition group. The moderate-recognition funds show no significant change in trading activity (Internet Appendix Table IA.XXI). The Internet Appendix contains additional robustness analyses.

\begin{table}[htbp]
\centering
\caption{Number and size of trades after high-recognition disclosures, by trader type: purchases and exits}
\setlength{\tabcolsep}{4pt}
\begin{threeparttable}
\small
\begin{tabular*}{\linewidth}{@{\extracolsep{\fill}}lccccc}
\toprule
 & \multicolumn{5}{c}{Window} \\
\cmidrule(l{3pt}){2-6}
 & [$-$5,$-$1] & [0,1] & [0,3] & [0,5] & [0,10] \\
\midrule
\multicolumn{6}{l}{\textit{Panel A. High-recognition purchases vs. control purchases}}\\
\quad Total trades & $-$5.97$^{**}$ & 7.24$^{**}$ & 4.50 & 2.64 & 1.76 \\
\quad  & (0.028) & (0.027) & (0.139) & (0.324) & (0.547) \\[1pt]
\quad Non-retail trades & $-$5.97$^{**}$ & 6.71$^{**}$ & 4.00 & 2.22 & 1.42 \\
\quad  & (0.025) & (0.038) & (0.188) & (0.406) & (0.625) \\[1pt]
\quad Retail trades & $-$4.41 & 15.12$^{***}$ & 12.30$^{***}$ & 8.98$^{**}$ & 6.55$^{*}$ \\
\quad  & (0.209) & (0.001) & (0.002) & (0.012) & (0.080) \\[1pt]
\quad Average trade size & $-$0.37 & $-$1.37 & $-$0.60 & 0.38 & 0.65 \\
\quad  & (0.721) & (0.547) & (0.758) & (0.813) & (0.688) \\[1pt]
\quad Average retail trade size & $-$2.65 & $-$1.26 & $-$2.59 & $-$2.68 & $-$0.63 \\
\quad  & (0.218) & (0.697) & (0.353) & (0.246) & (0.772) \\[1pt]
\quad $\geq$\$50k dollar share & $-$0.16 & 0.02 & 0.31 & 0.53 & 0.35 \\
\quad  & (0.687) & (0.971) & (0.532) & (0.110) & (0.424) \\[1pt]
\addlinespace[3pt]
\multicolumn{6}{l}{\textit{Panel B. High-recognition exits vs. control exits}}\\
\quad Total trades & $-$1.55 & 6.36$^{*}$ & 4.23 & 2.19 & 3.73$^{*}$ \\
\quad  & (0.663) & (0.089) & (0.145) & (0.420) & (0.091) \\[1pt]
\quad Non-retail trades & $-$1.33 & 6.02 & 3.95 & 1.97 & 3.63 \\
\quad  & (0.707) & (0.106) & (0.174) & (0.470) & (0.102) \\[1pt]
\quad Retail trades & $-$4.84 & 10.37$^{***}$ & 7.08$^{**}$ & 5.24 & 5.31$^{*}$ \\
\quad  & (0.313) & (0.010) & (0.040) & (0.110) & (0.059) \\[1pt]
\quad Average trade size & $-$0.59 & $-$0.59 & $-$0.77 & $-$0.37 & 0.72 \\
\quad  & (0.683) & (0.852) & (0.734) & (0.859) & (0.707) \\[1pt]
\quad Average retail trade size & 1.10 & $-$4.83 & $-$1.81 & $-$2.73 & $-$0.50 \\
\quad  & (0.630) & (0.163) & (0.444) & (0.161) & (0.698) \\[1pt]
\quad $\geq$\$50k dollar share & $-$0.53 & 0.15 & $-$0.12 & 0.09 & 0.10 \\
\quad  & (0.411) & (0.912) & (0.903) & (0.922) & (0.905) \\[1pt]
\bottomrule
\end{tabular*}
\begin{tablenotes}[flushleft]\footnotesize
\item \textit{Note.} High-recognition minus control, estimated separately for purchases (Panel A) and complete exits (Panel B): the coefficient on the high-recognition indicator in a regression of the outcome on the paper's stock and fund controls, stock and year-quarter fixed effects, and the event's own pre-window value of the outcome, with p-values from manager-clustered standard errors in parentheses. Each outcome is net of the event's own [-15,-6] baseline, averaged over the window, and scaled by 100: trade counts and sizes in log points (approximately percent), the dollar-share row in percentage points. Non-retail trades are all trades minus the retail trades identified by the sub-penny method. Sample 2015 to 2025, the period covered by Daily TAQ. $^{*}p<.10$, $^{**}p<.05$, $^{***}p<.01$.
\end{tablenotes}
\end{threeparttable}
\end{table}

\subsection{Mechanism 2: the size of trades}

Mechanism 2 implies that the size of the trades goes up, while the number of trades need not change much. An investor who finds the disclosure more persuasive because of its source should trade more of the stock. The lower rows of each panel of Table XIII report the effect on the average trade size, on the average retail trade size, and on the share of dollar volume in trades of at least \$50,000. None of the trade size measures differs significantly. After a high-recognition buy disclosure the average trade size changes by $-$1.4\% over [0,1] (p = 0.547) and by +0.4\% over [0,5] (p = 0.813). The average retail trade size changes by $-$1.3\% over [0,1] (p = 0.697) and by $-$2.7\% over [0,5] (p = 0.246). The large-trade share changes by less than one percentage point at every window (p $\geq$ 0.110). The same flat pattern is present for the exit disclosures. Figure 4 plots the count and the size effects side by side. More investors trade after a high-recognition disclosure, and the investors who trade do not trade bigger. The conviction of the investors who trade, as far as the size of their trades shows it, does not differ. We find no trace of Mechanism 2 in the trade data.

\begin{figure}[htbp]
\centering
\includegraphics[width=\linewidth]{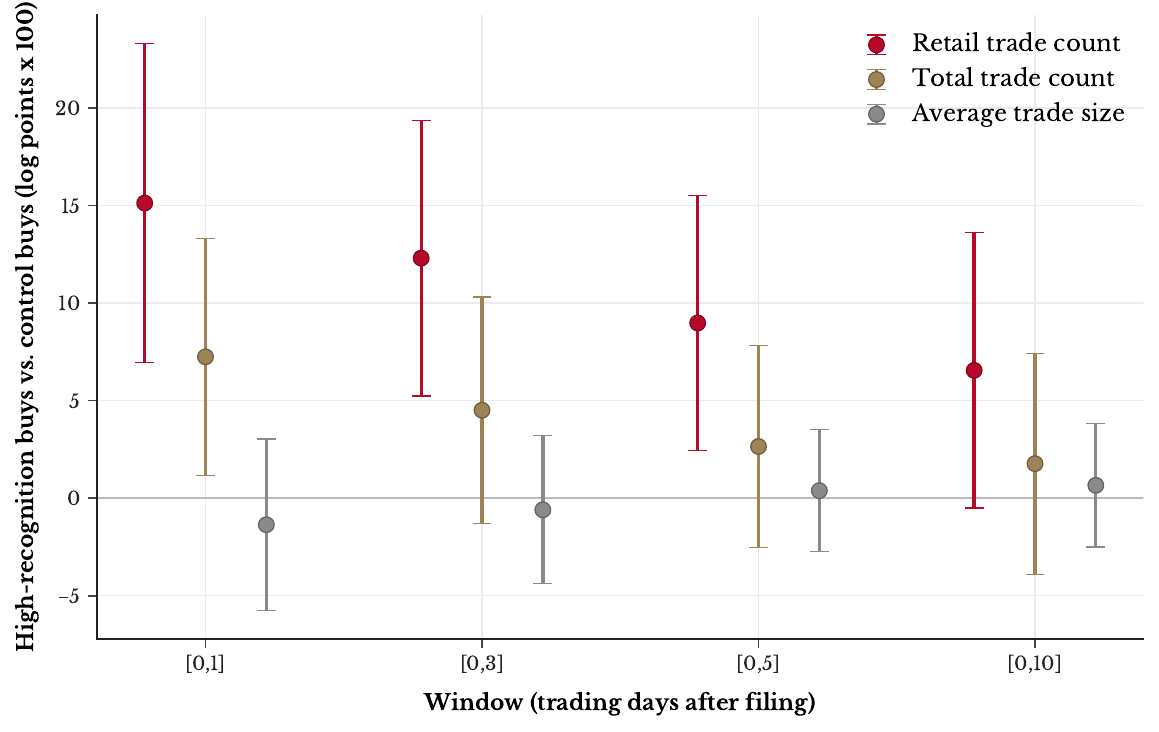}
\caption{More trades, not bigger trades: high-recognition purchases against control purchases}
{\footnotesize\noindent\textit{Note.} High-recognition minus control after purchases, by window, for the abnormal log number of retail trades, the abnormal log number of all trades, and the abnormal log average trade size (log points $\times$ 100), with 95\% confidence intervals from manager-clustered standard errors.\par}
\end{figure}

\subsection{Mechanism 3: imitation}

Mechanism 3 produces the same increase in the number of trades as Mechanism 1. Investors who imitate the manager also trade more. However, Mechanism 1 works through information revelation. The disclosures reveal new positive information about the disclosed stocks, and because more investors pay attention to highly recognizable investors, that information gets incorporated into their decisions. Mechanism 3 ignores the information component and mindlessly copies the disclosed trade. Mechanism 1 transmits information, so it should be strongest where information is costly, in small, illiquid, scarcely covered, and under-held stocks. Mechanism 3 does not depend on the stock, so it should be present just as much in large and widely covered stocks. But more importantly, Mechanism 3 should reverse, because the price increase resulting from mindless copying carries no information about the stock. We test the first prediction in the prices and then in the trades, and the second with calendar-time portfolios in Section VI.D.

Table XIV splits the high-recognition premium across the cross-section of stocks. We split the disclosed stocks at the median of each of five characteristics and re-estimate the high-recognition-vs-control reaction on each half, leaving the splitting characteristic out of the controls. Consistent with Mechanism 1, the premium is larger on the side where the market has less information about the stock. The reaction to a high-recognition disclosure is about twice as large in small stocks as in large ones (+2.21\% versus +1.10\% over the filing day and the next, p = 0.009), larger in stocks with scarce analyst coverage than in widely covered ones (+3.28\% versus +1.72\% over [0,5]; the difference is significant at three days, p = 0.074, and marginal at five, p = 0.101), and larger in illiquid stocks (Amihud, 2002) than in liquid ones (+2.09\% versus +1.10\% over the filing day and the next, p = 0.048). The premium is also larger in stocks under-held by institutions than in widely held ones (for example +3.49\% versus +2.11\% at [0,10]), though here the gap is substantial in size but not statistically significant (p = 0.62); and it is larger in growth stocks than in value stocks at five to twenty days (+2.92\% versus +1.61\% over [0,5], p = 0.090, widening to +3.94\% versus +1.34\% at [0,10], p = 0.001, and p = 0.058 at [0,20]).

\begin{table}[htbp]
\centering
\caption{Mechanism: the high-recognition premium by stock characteristic}
\setlength{\tabcolsep}{3pt}
\begin{threeparttable}
\small
\begin{tabular*}{\linewidth}{@{\extracolsep{\fill}}lcccccc}
\toprule
 & \multicolumn{6}{c}{CAR window} \\
\cmidrule(l{3pt}){2-7}
 & [0,1] & [0,3] & [0,5] & [0,10] & [0,20] & [0,30] \\
\midrule
\multicolumn{7}{l}{\textit{Market capitalization}}\\
\quad Small & 2.211$^{***}$ & 2.741$^{***}$ & 3.343$^{**}$ & 2.934 & 4.318 & 2.827 \\
\quad Large & 1.104$^{**}$ & 1.384$^{***}$ & 1.736$^{***}$ & 2.139$^{***}$ & 2.316$^{***}$ & 1.981$^{***}$ \\
\quad Difference & 1.107 & 1.357 & 1.607 & 0.795 & 2.001 & 0.846 \\
\quad  & (0.009) & (0.030) & (0.116) & (0.291) & (0.194) & (0.598) \\
\addlinespace[3pt]
\multicolumn{7}{l}{\textit{Analyst coverage}}\\
\quad Scarce & 1.641$^{***}$ & 2.544$^{***}$ & 3.280$^{***}$ & 3.056$^{***}$ & 3.551$^{**}$ & 1.327 \\
\quad Broad & 1.242$^{**}$ & 1.420$^{**}$ & 1.715$^{***}$ & 2.172$^{**}$ & 2.692$^{***}$ & 2.547$^{***}$ \\
\quad Difference & 0.400 & 1.124 & 1.565 & 0.884 & 0.859 & $-$1.221 \\
\quad  & (0.391) & (0.074) & (0.101) & (0.135) & (0.243) & (0.417) \\
\addlinespace[3pt]
\multicolumn{7}{l}{\textit{Illiquidity (Amihud)}}\\
\quad Liquid & 1.097$^{**}$ & 1.340$^{***}$ & 1.588$^{***}$ & 2.358$^{***}$ & 2.730$^{***}$ & 2.173$^{***}$ \\
\quad Illiquid & 2.089$^{***}$ & 2.375$^{***}$ & 3.211$^{**}$ & 2.319$^{*}$ & 3.448 & 2.824 \\
\quad Difference & $-$0.993 & $-$1.035 & $-$1.622 & 0.040 & $-$0.718 & $-$0.651 \\
\quad  & (0.048) & (0.074) & (0.121) & (0.665) & (0.382) & (0.513) \\
\addlinespace[3pt]
\multicolumn{7}{l}{\textit{Institutional ownership}}\\
\quad Under-held & 1.457$^{**}$ & 2.315$^{***}$ & 2.836$^{***}$ & 3.489$^{***}$ & 3.487$^{***}$ & 3.830$^{***}$ \\
\quad Widely held & 1.247$^{**}$ & 1.388$^{*}$ & 1.878$^{**}$ & 2.108$^{*}$ & 3.188$^{**}$ & 1.903 \\
\quad Difference & 0.209 & 0.927 & 0.958 & 1.381 & 0.299 & 1.927 \\
\quad  & (0.944) & (0.504) & (0.641) & (0.617) & (0.760) & (0.778) \\
\addlinespace[3pt]
\multicolumn{7}{l}{\textit{Book-to-market}}\\
\quad Growth & 1.501$^{***}$ & 2.216$^{***}$ & 2.924$^{***}$ & 3.938$^{***}$ & 3.681$^{***}$ & 1.845 \\
\quad Value & 1.103$^{**}$ & 1.443$^{**}$ & 1.612$^{**}$ & 1.344$^{*}$ & 1.889$^{**}$ & 1.817$^{*}$ \\
\quad Difference & 0.398 & 0.773 & 1.312 & 2.593 & 1.791 & 0.029 \\
\quad  & (0.904) & (0.526) & (0.090) & ($<$0.001) & (0.058) & (0.992) \\
\bottomrule
\end{tabular*}
\begin{tablenotes}[flushleft]\footnotesize
\item \textit{Note.} Each cell is the high-recognition treat $\times$ buy coefficient (full specification) estimated on the low and high half of the stock characteristic (median split); the splitting characteristic is excluded from the controls. ``Difference'' is low $-$ high, with the analytic p-value of the difference in parentheses beneath it. The high-recognition premium is largest among information-scarce stocks (small, low-coverage). $^{*}p<.10$, $^{**}p<.05$, $^{***}p<.01$.
\end{tablenotes}
\end{threeparttable}
\end{table}

Further, we split the trade data the same way (Internet Appendix Table IA.XXVII). For size and liquidity the pattern points in the direction of Mechanism 1 over Mechanism 3, but the differences are not significant. Over the [0,5] window the number of retail trades after high-recognition purchases rises by 12.7\% in small stocks and 6.9\% in large ones (p = 0.747 for the difference), and by 14.4\% in illiquid stocks and 5.8\% in liquid ones (p = 0.362). The only significant difference runs the other way: more trading in broadly covered stocks than in scarcely covered ones over [0,5] (p = 0.046 for retail trades, p = 0.020 for all trades). The trade data cover ten years and a small number of high-recognition funds, so the splits are noisy. We read them as cautiously in favor of Mechanism 1 over Mechanism 3.

\subsection{The premium as a strategy}

The second prediction that separates Mechanism 3 from Mechanism 1 is reversal. A premium built on imitation should disappear, while a premium built on information should last. So far we have measured the post-disclosure drift in event time, as the cumulative abnormal return over windows anchored at the filing date, and found no reversal within thirty days. To test persistence directly we measure it in calendar time and ask whether a portfolio that buys the stocks a group disclosed would have earned abnormal returns over one, three, and six months. The portfolio copies the group's disclosed purchases mechanically: a stock enters on the first trading day after the filing that disclosed it and is held for exactly H trading days. The portfolio does not rebalance on any calendar schedule and does not hold the recognizable funds' portfolios as such; it holds only their newly disclosed purchases, each for a fixed window. Holdings are equal-weighted and the weights are reset daily across whatever positions are open. We regress the portfolio's daily return in excess of the risk-free rate on the four Carhart factors. The intercept is the daily abnormal return of the copying strategy. If Mechanism 3 were the whole story, this intercept should turn negative at the longer holding periods, as the imitation-driven price increase unwinds. We report holding periods of 21, 63, and 126 trading days.

Table XV reports the alphas. The portfolio of high-recognition disclosures earns +9.8 basis points per day over the 21-day holding period (p < 0.001), +4.4 basis points per day over 63 days (p = 0.003), and a not statistically significant +1.4 basis points over 126 days (p = 0.164). The control portfolio earns nothing at any horizon: +0.3 basis points at 21 days (p = 0.839), +0.5 at 63 days (p = 0.444), and +0.8 at 126 days (p = 0.154).

We further construct the long-short portfolio that buys the high-recognition disclosures and shorts the control disclosures, earns +10.4 basis points per day at the 21-day horizon (p < 0.001) and +4.8 at 63 days (p = 0.003), and is flat at 126 days (+0.8, p = 0.438). A portfolio formed symmetrically from the high-recognition funds' complete exits earns a statistically zero alpha (p = 0.288), so the effect is specific to the bullish disclosure. The calendar-time evidence thus reproduces the event-time pattern: the abnormal returns are strongest in the first month and fade over the subsequent quarter, with no reversal at any horizon.

\begin{table}[htbp]
\centering
\caption{Calendar-time portfolio alphas of disclosed purchases}
\setlength{\tabcolsep}{4pt}
\begin{threeparttable}
\small
\begin{tabular*}{\linewidth}{@{\extracolsep{\fill}}lccc}
\toprule
Portfolio & H = 21 & H = 63 & H = 126 \\
\midrule
Recognizable buys & 1.9 & 2.8$^{***}$ & 2.1$^{***}$ \\
 & (1.7) & (0.8) & (0.7) \\[1pt]
High-recognition buys & 9.8$^{***}$ & 4.4$^{***}$ & 1.4 \\
 & (2.7) & (1.5) & (1.0) \\[1pt]
Moderate-recognition buys & 1.1 & 2.5$^{***}$ & 2.1$^{***}$ \\
 & (1.8) & (0.9) & (0.7) \\[1pt]
Control buys & 0.3 & 0.5 & 0.8 \\
 & (1.3) & (0.7) & (0.6) \\[1pt]
High-recognition $-$ Control & 10.4$^{***}$ & 4.8$^{***}$ & 0.8 \\
 & (2.9) & (1.6) & (1.1) \\[1pt]
\bottomrule
\end{tabular*}
\begin{tablenotes}[flushleft]\footnotesize
\item \textit{Note.} Daily Carhart four-factor alphas, in basis points per day, of calendar-time portfolios that buy every new position a group disclosed. A stock enters the portfolio on the first trading day after the filing that disclosed it, is held for H trading days, and is then dropped; holdings are equal-weighted with weights reset daily, and portfolio days require at least five open positions. Alphas are estimated by regressing the portfolio's daily excess return on the four Carhart factors with HAC standard errors (five lags); standard errors in parentheses, in basis points, and stars from the corresponding p-values. High-recognition $-$ Control is the long-short portfolio, estimated on days where both legs are populated; the daily alpha annualizes at roughly $\times$252 (+10.4 bps per day $\approx$ +26\% per year). Internet Appendix Section IA.7 reports exit portfolios, delayed entry, and buy-and-hold variants. $^{*}p<.10$, $^{**}p<.05$, $^{***}p<.01$.
\end{tablenotes}
\end{threeparttable}
\end{table}

These alphas measure the persistence of the price adjustment and are unlikely to constitute an implementable trading rule. Entering at the first post-filing day credits the strategy with the return earned from the close of the filing day itself, and a copier cannot always trade at that price. Delaying entry by one further day, to a price any follower could obtain in practice, still leaves +3.2 basis points per day at the 63-day horizon (p = 0.029), about three-quarters of the headline figure, and +4.9 basis points at the 21-day horizon (p = 0.064), about half of it. Furthermore, the alphas do not depend on the daily equal-weight reset embedded in the calendar-time convention. The buy-and-hold versions of every portfolio, which never trade within the holding window, produce essentially identical estimates (Internet Appendix Section IA.7). The alphas are moreover gross of transaction costs, and the strategy is costly to run: the high-recognition portfolio holds about ten positions on average, turns its entire book over every holding period, is invested only around the filing seasons at the short horizon, and Section VI.C shows the premium concentrates in precisely the small and illiquid stocks where trading costs are the highest. The annualized magnitudes (the 21-day long-short alpha annualizes to roughly +26\% per year at $\times$252) accordingly assume a full year of such portfolio days and likely overstate what a real copier could capture. However, the alphas do exactly what our reading of the effect requires. A premium driven by temporary copycat pressure should reverse once attention moves on. It does not reverse. On the other hand, a premium that reflects some market information should build and then level off. The results from the calendar-time portfolios are consistent with the information revelation interpretation. The abnormal returns accumulate over one to three months, and never turn negative, so the high-recognition disclosure produces a lasting revaluation of the stock rather than a temporary surge of buying.

\section{Conclusion}

Although our findings largely confirm an old anecdotal effect, we believe they carry implications for modern theories of investor attention and information search. The finding fits naturally with the literature on rational inattention (Sims, 2003; Ma\'{c}kowiak et al., 2023) and with the evidence on attention-driven trading (Barber et al., 2022). The crux of these theories is that attention is a scarce resource that must be allocated like any other: investors searching for investment opportunities must decide which stocks to attend to, and news coverage (Fang and Peress, 2009), salience (Bordalo et al., 2022), peer networks (Kuchler and Stroebel, 2021), and related channels have been shown to direct attention toward particular stocks. The disclosures of recognizable investors constitute another such channel. Moreover, deciding whose disclosures deserve attention is itself costly, so attention concentrates on the filings of the most recognized investors. Because those filings do reveal information that is costly to obtain elsewhere, the attention they attract translates into abnormal returns.

We show that the investors with the highest cost of market research, retail investors, take part in a rational search for market information. Their attention to information sources is limited, however, and they must ration it to the sources that are most available or most salient and hence lowest in cost. Our evidence is most consistent with the Buffett effect being mostly a rational attention mechanism. It may also operate through credibility, the belief that highly recognizable managers are better stock pickers than non-recognizable or less recognizable managers. If that is the case, however, credibility also operates through attention and not through the conviction behind the trades. The size of trade orders does not differ between high-recognition and control disclosures, so it is unlikely that credibility affects how much investors decide to invest. Recognition influences whether investors pay attention, but it does not influence the size of their trades.

The Buffett effect also offers an interesting twist on recent models of reinforcement learning in financial markets, which draw on the older psychology of operant conditioning (Barberis and Jin, 2023). In these models, investors learn from their own past choices and payoffs and adjust their portfolios according to what experience has reinforced. Psychology, however, also offers models of observational learning, in which behavior is reinforced by observing the actions of others, and such social transmission has been shown to shape investor behavior (Hirshleifer, 2020; Han et al., 2022). Mimicking the disclosed trades of the most talented investors is a natural case of such observational learning. Investors observe a highly recognizable investor's success and follow the trades accordingly. One question remains open for data we do not have. Account-level records would separate the number of investors who respond from the amount each of them commits, and hence reach from credibility, more cleanly than trade data allow. We hope that future research on recognition and observational learning in finance can build on our findings and on the mechanism we propose for how the Buffett effect arises.

Lastly, we point out two limitations of the analysis. First, the main effect, the high-recognition premium, is driven by five managers. These managers share one characteristic, being the most recognizable in our data, and the premium survives the robustness checks in Section V, but the effective sample behind the effect is nonetheless small. Second, both samples rest on screening criteria we set, and each recognizable and control fund is verified manually as an active discretionary stock-picker. Re-running the control matching without the manual exclusions leaves the high-recognition estimates unchanged (Internet Appendix Section IA.5). Even so, the matched-sample design is curated, and different screening rules could result in different samples.

\section*{Acknowledgements}

We thank David Hirshleifer and Norbert Schwarz for their help in the preparation of the manuscript.

\section*{References}
\setlength{\parindent}{0pt}

\hangindent=1.5em\hangafter=1\noindent Amihud, Y. (2002). Illiquidity and stock returns: Cross-section and time-series effects. Journal of Financial Markets, 5(1), 31--56.\par\vspace{4pt}

\hangindent=1.5em\hangafter=1\noindent Ante, L. (2023). How Elon Musk's Twitter activity moves cryptocurrency markets. Technological Forecasting and Social Change, 186, Article 122112.\par\vspace{4pt}

\hangindent=1.5em\hangafter=1\noindent Barber, B. M., Huang, X., Jorion, P., Odean, T., \& Schwarz, C. (2024). A (sub)penny for your thoughts: Tracking retail investor activity in TAQ. The Journal of Finance, 79(4), 2403--2427.\par\vspace{4pt}

\hangindent=1.5em\hangafter=1\noindent Barber, B. M., Huang, X., Odean, T., \& Schwarz, C. (2022). Attention-induced trading and returns: Evidence from Robinhood users. The Journal of Finance, 77(6), 3141--3190.\par\vspace{4pt}

\hangindent=1.5em\hangafter=1\noindent Barberis, N. C., \& Jin, L. J. (2023). Model-free and model-based learning as joint drivers of investor behavior (NBER Working Paper No. 31081). National Bureau of Economic Research.\par\vspace{4pt}

\hangindent=1.5em\hangafter=1\noindent Boehmer, E., Jones, C. M., Zhang, X., \& Zhang, X. (2021). Tracking retail investor activity. The Journal of Finance, 76(5), 2249--2305.\par\vspace{4pt}

\hangindent=1.5em\hangafter=1\noindent Bonner, S. E., Hugon, A., \& Walther, B. R. (2007). Investor reaction to celebrity analysts: The case of earnings forecast revisions. Journal of Accounting Research, 45(3), 481--513.\par\vspace{4pt}

\hangindent=1.5em\hangafter=1\noindent Bordalo, P., Gennaioli, N., \& Shleifer, A. (2022). Salience. Annual Review of Economics, 14, 521--544.\par\vspace{4pt}

\hangindent=1.5em\hangafter=1\noindent Brown, S. J., \& Warner, J. B. (1985). Using daily stock returns: The case of event studies. Journal of Financial Economics, 14(1), 3--31.\par\vspace{4pt}

\hangindent=1.5em\hangafter=1\noindent Cameron, A. C., Gelbach, J. B., \& Miller, D. L. (2008). Bootstrap-based improvements for inference with clustered errors. The Review of Economics and Statistics, 90(3), 414--427.\par\vspace{4pt}

\hangindent=1.5em\hangafter=1\noindent Cao, S. S., Du, K., Yang, B., \& Zhang, A. L. (2021). Copycat skills and disclosure costs: Evidence from peer companies' digital footprints. Journal of Accounting Research, 59(4), 1261--1302.\par\vspace{4pt}

\hangindent=1.5em\hangafter=1\noindent Carhart, M. M. (1997). On persistence in mutual fund performance. The Journal of Finance, 52(1), 57--82.\par\vspace{4pt}

\hangindent=1.5em\hangafter=1\noindent Dean, M., \& Neligh, N. (2023). Experimental tests of rational inattention. Journal of Political Economy, 131(12), 3415--3461.\par\vspace{4pt}

\hangindent=1.5em\hangafter=1\noindent Engelberg, J., Sasseville, C., \& Williams, J. (2012). Market madness? The case of Mad Money. Management Science, 58(2), 351--364.\par\vspace{4pt}

\hangindent=1.5em\hangafter=1\noindent Fama, E. F., \& French, K. R. (1993). Common risk factors in the returns on stocks and bonds. Journal of Financial Economics, 33(1), 3--56.\par\vspace{4pt}

\hangindent=1.5em\hangafter=1\noindent Fang, L., \& Peress, J. (2009). Media coverage and the cross-section of stock returns. The Journal of Finance, 64(5), 2023--2052.\par\vspace{4pt}

\hangindent=1.5em\hangafter=1\noindent Frank, M. M., Poterba, J. M., Shackelford, D. A., \& Shoven, J. B. (2004). Copycat funds: Information disclosure regulation and the returns to active management in the mutual fund industry. Journal of Law and Economics, 47(2), 515--541.\par\vspace{4pt}

\hangindent=1.5em\hangafter=1\noindent Grossman, S. J., \& Stiglitz, J. E. (1980). On the impossibility of informationally efficient markets. The American Economic Review, 70(3), 393--408.\par\vspace{4pt}

\hangindent=1.5em\hangafter=1\noindent Han, B., Hirshleifer, D., \& Walden, J. (2022). Social transmission bias and investor behavior. Journal of Financial and Quantitative Analysis, 57(1), 390--412.\par\vspace{4pt}

\hangindent=1.5em\hangafter=1\noindent Hirshleifer, D. (2020). Presidential address: Social transmission bias in economics and finance. The Journal of Finance, 75(4), 1779--1831.\par\vspace{4pt}

\hangindent=1.5em\hangafter=1\noindent Kuchler, T., \& Stroebel, J. (2021). Social finance. Annual Review of Financial Economics, 13, 37--55.\par\vspace{4pt}

\hangindent=1.5em\hangafter=1\noindent Loewenstein, G., \& Wojtowicz, Z. (2025). The economics of attention. Journal of Economic Literature, 63(3), 1038--1089.\par\vspace{4pt}

\hangindent=1.5em\hangafter=1\noindent Machus, T., Mestel, R., \& Theissen, E. (2022). Heroes, just for one day: The impact of Donald Trump's tweets on stock prices. Journal of Behavioral and Experimental Finance, 33, Article 100594.\par\vspace{4pt}

\hangindent=1.5em\hangafter=1\noindent Ma\'{c}kowiak, B., Mat\v{e}jka, F., \& Wiederholt, M. (2023). Rational inattention: A review. Journal of Economic Literature, 61(1), 226--273.\par\vspace{4pt}

\hangindent=1.5em\hangafter=1\noindent Martin, G. S., \& Puthenpurackal, J. (2008). Imitation is the sincerest form of flattery: Warren Buffett and Berkshire Hathaway. SSRN.\par\vspace{4pt}

\hangindent=1.5em\hangafter=1\noindent Sims, C. A. (2003). Implications of rational inattention. Journal of Monetary Economics, 50(3), 665--690.\par\vspace{4pt}

\hangindent=1.5em\hangafter=1\noindent Webb, M. D. (2023). Reworking wild bootstrap-based inference for clustered errors. Canadian Journal of Economics, 56(3), 839--858.\par\vspace{4pt}

\hangindent=1.5em\hangafter=1\noindent Zhang, M. (2022). Warren Buffett anomaly. Finance Research Letters, 46, Article 102473.\par\vspace{4pt}

\FloatBarrier
\clearpage
\setlength{\parindent}{\bodyparindent}
\setcounter{section}{0}
\setcounter{table}{0}
\setcounter{figure}{0}
\renewcommand{\thesection}{IA.\arabic{section}}
\renewcommand{\thetable}{IA.\Roman{table}}
\renewcommand{\thefigure}{IA.\arabic{figure}}
\renewcommand{\theHsection}{IA.\arabic{section}}
\renewcommand{\theHtable}{IA.\arabic{table}}
\renewcommand{\theHfigure}{IA.\arabic{figure}}
\titleformat{\section}[block]{\filcenter\normalfont\large\bfseries}{\thesection}{0.6em}{}
\fancyhead[C]{\footnotesize
  \ifthenelse{\isodd{\value{page}}}{\spacedcaps{Appendix}}{\spacedcaps{The Buffett Effect}}}
\thispagestyle{plain}
\vspace*{0.25in}
\begin{center}
  {\fontsize{27}{33}\selectfont\bfseries Appendix}\\[24pt]
\end{center}
\noindent This appendix reports the supplementary tables and figures referred to in the main text. Section numbers, table numbers, and figure numbers follow the order in which the main text refers to them.

\section{Sample construction}

The main text describes the universe of discretionary funds and the construction of the recognizable group (Sections III.A and III.B). Here we report the filters and the construction stages. Table IA.I supplements Section III.B of the main text.

\subsection{Discretionary-fund filters}

A fund enters the universe if its 13F history satisfies: at least 12 quarterly filings; a most recent filing no earlier than eight quarters before the sample close; an average of at most 100 positions per filing; and an average of at most 20 events per quarter, where the last two quantities are averaged over the fund's full filing history. Beyond these numeric cutoffs we retain only discretionary equity funds and exclude the structurally non-discretionary institutions (endowments and foundations, insurance and pension books, sovereign portfolios, corporate holding companies, funds-of-funds, single-family offices, and quant or index replicators), together with filers we cannot match to the holdings panel. The turnover cutoff was applied to the recognizable candidates; the controls were matched on breadth, size, and history, with turnover checked for balance afterwards.

\subsection{Recognizable group construction}

Table IA.I traces the recognizable group from the raw Wikipedia ``American hedge fund managers'' category to the final 24 included funds, with examples of who is dropped at each stage.

\begin{table}[htbp]
\centering
\caption{From the Wikipedia category to the recognizable group}
\setlength{\tabcolsep}{4pt}
\begin{threeparttable}
\small
\begin{tabularx}{\linewidth}{>{\raggedright\arraybackslash\hsize=0.85\hsize}Xc>{\raggedright\arraybackslash\hsize=1.15\hsize}X}
\toprule
Stage & n & Examples of dropped managers \\
\midrule
Raw Wikipedia category & 215 & --- \\
After dropping inactive and non-asset-manager entries and entries not matchable to a 13F-HR filer & 115 & Jim Cramer (inactive; fund closed 2001), Bernie Madoff (fraud), managed-futures and crypto managers whose holdings are not 13F-reportable \\
After applying the discretionary-fund filters (the filters above) & 24 & Cliff Asness (AQR, quant), Ray Dalio (Bridgewater, quant macro), Ken Griffin (Citadel, multi-strategy), Steve Cohen (Point72, too many positions and high turnover), Larry Fink (BlackRock, asset manager), Stanley Druckenmiller (Duquesne, family office) \\
Final recognizable group & 24 & --- \\
\bottomrule
\end{tabularx}
\begin{tablenotes}[flushleft]\footnotesize
\item \textit{Note.} The Wikipedia category was retrieved in May 2026. The reduction from 115 to 24 at the filters stage is large because most of the well-known managers in the category run multi-strategy, quant, or macro books whose 13F disclosures cannot be read as concentrated directional bets on individual stocks. The stage also removes a few managers who file 13Fs but cannot be matched to the holdings panel or are no longer active discretionary stock-pickers, such as Julian Robertson's Tiger Management (closed to outside capital, manager deceased 2022) and Paul Hilal's Mantle Ridge (no matchable holdings record).
\end{tablenotes}
\end{threeparttable}
\end{table}

\section{Per-fund disclosure premiums}

The main text defined the high-recognition group as the five funds with recognition z $\geq$ 1 and showed that this group produces the strongest disclosure premium (Section IV.C). Here we report the results fund by fund. The results can be read from Table IA.II to Table IA.IV. Four of the five high-recognition funds earn a premium on their own and Icahn does not, which is consistent with the interpretation in the main text and with the jackknife in Section V.A.

\begin{sidewaystable}[p]
\centering
\caption{The disclosure premium fund by fund}
\setlength{\tabcolsep}{3pt}
\begin{threeparttable}
\footnotesize
\begin{tabular*}{\linewidth}{@{\extracolsep{\fill}}rlrrrrrrr}
\toprule
\# & Fund & Rec. z & Buys & CAR[0,1] & CAR[0,3] & CAR[0,5] & CAR[0,10] & CAR[0,30] \\
\midrule
1 & \textbf{Buffett (Berkshire)} & 2.33 & 269 & 1.865$^{***}$ (<0.001) & 2.407$^{***}$ (<0.001) & 2.666$^{***}$ (<0.001) & 3.235$^{***}$ (<0.001) & 3.401$^{***}$ (<0.001) \\
2 & \textbf{Icahn} & 1.36 & 123 & 0.017 (0.889) & $-$0.444$^{***}$ (0.005) & $-$0.758$^{***}$ (0.002) & $-$1.132$^{***}$ (0.007) & $-$2.003$^{***}$ (0.005) \\
3 & \textbf{Burry (Scion)} & 1.29 & 221 & 1.589$^{***}$ (<0.001) & 2.360$^{***}$ (<0.001) & 3.643$^{***}$ (<0.001) & 4.067$^{***}$ (<0.001) & 3.060$^{***}$ (<0.001) \\
4 & \textbf{Ackman (Pershing Square)} & 1.24 & 101 & 0.462$^{**}$ (0.011) & 2.027$^{***}$ (<0.001) & 2.580$^{***}$ (<0.001) & 1.466$^{***}$ (<0.001) & $-$0.969 (0.288) \\
5 & \textbf{Peltz (Trian)} & 1.22 & 109 & 1.494$^{***}$ (<0.001) & 0.965$^{***}$ (0.002) & 0.500 (0.167) & 0.828$^{*}$ (0.081) & 3.818$^{***}$ (<0.001) \\
\midrule
6 & Tepper (Appaloosa) & 0.74 & 881 & 0.185$^{**}$ (0.025) & 0.438$^{***}$ (<0.001) & 0.618$^{***}$ (<0.001) & 0.582$^{***}$ (0.003) & 0.202 (0.551) \\
7 & Einhorn (Greenlight) & 0.50 & 718 & 0.355$^{***}$ (0.003) & 0.264$^{**}$ (0.047) & 0.857$^{***}$ (<0.001) & 1.314$^{***}$ (<0.001) & 1.914$^{***}$ (<0.001) \\
8 & Li Lu (Himalaya) & 0.42 & 18 & $-$2.176$^{**}$ (0.013) & $-$2.304$^{***}$ (0.010) & $-$1.823$^{*}$ (0.077) & $-$3.755$^{***}$ (<0.001) & $-$13.791$^{***}$ (<0.001) \\
9 & Klarman (Baupost) & 0.32 & 578 & 0.404$^{***}$ (<0.001) & 1.223$^{***}$ (<0.001) & 1.136$^{***}$ (<0.001) & 1.800$^{***}$ (<0.001) & 3.148$^{***}$ (<0.001) \\
10 & Ressler (Ares) & 0.30 & 318 & $-$0.813$^{***}$ (<0.001) & $-$1.075$^{***}$ (<0.001) & $-$1.269$^{***}$ (<0.001) & $-$0.139 (0.645) & 3.153$^{***}$ (<0.001) \\
11 & Lasry (Avenue) & 0.19 & 599 & $-$0.118 (0.289) & 0.013 (0.921) & $-$0.155 (0.396) & $-$0.168 (0.409) & $-$0.659 (0.200) \\
12 & Smith (Alden) & 0.09 & 466 & 1.081$^{***}$ (<0.001) & 0.710$^{***}$ (<0.001) & 0.677$^{***}$ (<0.001) & $-$0.045 (0.877) & $-$1.179$^{**}$ (0.022) \\
13 & Pabrai & $-$0.13 & 93 & 1.270$^{***}$ (<0.001) & 2.259$^{***}$ (<0.001) & 3.276$^{***}$ (<0.001) & 3.798$^{***}$ (<0.001) & 6.706$^{***}$ (<0.001) \\
14 & Gerstner (Altimeter) & $-$0.19 & 247 & $-$0.745$^{***}$ (<0.001) & $-$1.107$^{***}$ (<0.001) & $-$1.957$^{***}$ (<0.001) & $-$2.072$^{***}$ (<0.001) & $-$3.272$^{***}$ (<0.001) \\
15 & Ubben (ValueAct) & $-$0.41 & 252 & 0.361$^{***}$ (0.008) & 0.178 (0.237) & 0.125 (0.564) & 0.959$^{***}$ (<0.001) & $-$0.075 (0.889) \\
16 & Lichtenstein (Steel Partners) & $-$0.52 & 282 & $-$0.392$^{***}$ (0.003) & $-$0.001 (0.993) & $-$0.787$^{***}$ (0.004) & $-$0.266 (0.371) & $-$1.653$^{***}$ (0.004) \\
17 & Berkowitz (Fairholme) & $-$0.62 & 368 & $-$0.062 (0.491) & $-$0.198$^{*}$ (0.082) & $-$0.432$^{*}$ (0.059) & $-$0.610$^{**}$ (0.034) & $-$0.227 (0.701) \\
18 & Mudrick & $-$0.64 & 32 & $-$1.241$^{***}$ (0.008) & 1.389$^{**}$ (0.045) & 3.245$^{***}$ (<0.001) & 6.534$^{***}$ (<0.001) & 8.149$^{***}$ (<0.001) \\
19 & Hansen (Valiant) & $-$1.05 & 187 & $-$0.513$^{**}$ (0.017) & $-$0.038 (0.879) & 0.177 (0.485) & 2.624$^{***}$ (<0.001) & 3.631$^{***}$ (<0.001) \\
20 & Tananbaum (GoldenTree) & $-$1.13 & 895 & $-$0.189$^{*}$ (0.073) & $-$0.007 (0.947) & 0.383$^{***}$ (0.008) & $-$0.038 (0.837) & $-$0.930$^{***}$ (0.005) \\
21 & Goel (Matrix) & $-$1.14 & 481 & 0.405$^{***}$ (<0.001) & 0.868$^{***}$ (<0.001) & 0.844$^{***}$ (<0.001) & 0.608$^{**}$ (0.032) & 0.880$^{*}$ (0.055) \\
22 & Yun (Palo Alto Investors) & $-$1.21 & 683 & 0.088 (0.450) & 0.083 (0.678) & 0.817$^{***}$ (0.010) & 1.212$^{***}$ (0.003) & $-$0.080 (0.888) \\
23 & von Mueffling (Cantillon) & $-$1.33 & 506 & 0.119 (0.400) & $-$0.039 (0.836) & 0.204 (0.395) & 0.625$^{*}$ (0.075) & $-$0.080 (0.877) \\
24 & Rosenwald (Dalton) & $-$1.62 & 408 & $-$0.303$^{**}$ (0.035) & $-$0.277$^{**}$ (0.025) & $-$0.332$^{*}$ (0.072) & 0.764$^{**}$ (0.011) & 1.577$^{**}$ (0.011) \\
\bottomrule
\end{tabular*}
\begin{tablenotes}[flushleft]\footnotesize
\item \textit{Note.} Each cell is the treat $\times$ buy coefficient (\%, full specification) for the fund alone against the control group, with p-values from manager-clustered standard errors in parentheses. Funds are ordered by the Investor Recognition Index. High-recognition funds (z $\geq$ 1) in bold. With a single treated fund the clustered inference is fragile; the jackknife and the randomization inference in Section V.A of the main text are the formal tests. $^{*}p<.10$, $^{**}p<.05$, $^{***}p<.01$.
\end{tablenotes}
\end{threeparttable}
\end{sidewaystable}

Table IA.III and Table IA.IV present the mean cumulative abnormal returns on new positions for each recognizable fund and each control fund. They supplement the definition of the high-recognition group in Section IV.C of the main text.

\begin{sidewaystable}[p]
\centering
\caption{Per-fund cumulative abnormal returns: recognizable funds}
\setlength{\tabcolsep}{3pt}
\begin{threeparttable}
\small
\begin{tabular*}{\linewidth}{@{\extracolsep{\fill}}rllrrrrrrrr}
\toprule
\# & Fund & Manager & N buys & Rec. z & CAR[0,1] & CAR[0,3] & CAR[0,5] & CAR[0,10] & CAR[0,20] & CAR[0,30] \\
\midrule
1 & \textbf{Berkshire} & \textbf{Warren Buffett} & 269 & 2.33 & 1.50 & 1.71 & 1.94 & 2.18 & 2.70 & 2.77 \\
2 & \textbf{Icahn} & \textbf{Carl Icahn} & 123 & 1.36 & 0.08 & $-$0.07 & $-$0.38 & $-$0.02 & 0.97 & 0.13 \\
3 & \textbf{Scion} & \textbf{Michael Burry} & 221 & 1.29 & 0.99 & 1.27 & 1.79 & 2.49 & 3.00 & 3.68 \\
4 & \textbf{Pershing Square} & \textbf{Bill Ackman} & 101 & 1.24 & 0.67 & 1.14 & 1.69 & 1.25 & 1.07 & 0.95 \\
5 & \textbf{Trian} & \textbf{Nelson Peltz} & 109 & 1.22 & 0.71 & 0.97 & 1.27 & 1.34 & 2.30 & 2.66 \\
6 & Appaloosa & David Tepper & 881 & 0.74 & 0.05 & 0.19 & 0.22 & 0.42 & 0.16 & 0.05 \\
7 & Greenlight & David Einhorn & 718 & 0.50 & 0.15 & $-$0.07 & 0.20 & 0.85 & 1.20 & 1.15 \\
8 & Himalaya & Li Lu & 18 & 0.42 & $-$0.76 & $-$0.42 & 0.48 & $-$0.67 & $-$2.83 & $-$3.84 \\
9 & Baupost & Seth Klarman & 578 & 0.32 & 0.44 & 1.03 & 0.91 & 1.50 & 2.68 & 2.73 \\
10 & Ares & Tony Ressler & 318 & 0.30 & 0.01 & 0.31 & 0.47 & 0.43 & 0.03 & $-$0.03 \\
11 & Avenue & Marc Lasry & 599 & 0.19 & $-$0.18 & $-$0.40 & $-$0.50 & $-$0.27 & $-$0.87 & $-$1.42 \\
12 & Alden & Randall D. Smith & 466 & 0.09 & 0.67 & 0.33 & 0.22 & 0.40 & 0.19 & 0.63 \\
13 & Pabrai & Mohnish Pabrai & 93 & $-$0.13 & 0.23 & 0.21 & 0.81 & 0.92 & 1.81 & 1.69 \\
14 & Altimeter & Brad Gerstner & 247 & $-$0.19 & 0.08 & $-$0.22 & $-$0.93 & $-$0.51 & 0.37 & 1.46 \\
15 & ValueAct & Jeffrey W. Ubben & 252 & $-$0.41 & 0.41 & 0.37 & 0.34 & 0.77 & 1.10 & 1.39 \\
16 & Steel Partners & Warren Lichtenstein & 282 & $-$0.52 & $-$0.41 & $-$0.25 & $-$0.51 & $-$0.30 & $-$0.81 & $-$0.87 \\
17 & Fairholme & Bruce R. Berkowitz & 368 & $-$0.62 & 0.13 & 0.29 & 0.15 & 0.15 & 1.01 & 1.53 \\
18 & Mudrick & Jason Mudrick & 32 & $-$0.64 & $-$0.95 & $-$0.59 & $-$0.37 & 0.28 & 3.59 & 1.81 \\
19 & Valiant & Chris R. Hansen & 187 & $-$1.05 & $-$0.32 & 0.02 & 0.06 & 0.31 & 0.02 & $-$0.58 \\
20 & GoldenTree & Steven Tananbaum & 895 & $-$1.13 & 0.10 & 0.10 & 0.15 & $-$0.05 & $-$0.40 & $-$0.51 \\
21 & Matrix & David Goel & 481 & $-$1.14 & 0.12 & 0.15 & 0.16 & $-$0.18 & $-$0.35 & $-$0.67 \\
22 & Palo Alto Investors & Joon Yun & 683 & $-$1.21 & $-$0.03 & $-$0.21 & $-$0.11 & $-$0.32 & 0.29 & $-$0.82 \\
23 & Cantillon & William von Mueffling & 506 & $-$1.33 & $-$0.07 & $-$0.27 & $-$0.13 & $-$0.17 & $-$0.52 & $-$0.74 \\
24 & Dalton & James B. Rosenwald & 408 & $-$1.62 & $-$0.02 & $-$0.38 & $-$0.29 & 0.05 & $-$0.14 & 0.25 \\
\bottomrule
\end{tabular*}
\begin{tablenotes}[flushleft]\footnotesize
\item \textit{Note.} Mean cumulative abnormal return (\%, Carhart four-factor adjusted) on new-purchase events, by fund. N buys is the number of new-position events. High-recognition funds (z $\geq$ 1) in bold.
\end{tablenotes}
\end{threeparttable}
\end{sidewaystable}

\begin{sidewaystable}[p]
\centering
\caption{Per-fund cumulative abnormal returns: control funds}
\setlength{\tabcolsep}{3pt}
\begin{threeparttable}
\small
\begin{tabular*}{\linewidth}{@{\extracolsep{\fill}}rlrrrrrrrr}
\toprule
\# & Fund & N buys & AUM (\$M) & CAR[0,1] & CAR[0,3] & CAR[0,5] & CAR[0,10] & CAR[0,20] & CAR[0,30] \\
\midrule
1 & Sands & 600 & 23,608 & 0.07 & $-$0.32 & $-$0.58 & $-$0.99 & $-$1.79 & $-$2.21 \\
2 & Southeastern & 410 & 13,846 & $-$0.07 & 0.12 & 0.28 & 0.45 & 1.39 & 2.52 \\
3 & Generation Investment & 276 & 8,896 & $-$0.19 & $-$0.08 & $-$0.32 & $-$0.17 & $-$0.45 & $-$0.79 \\
4 & Independent Franchise & 103 & 8,632 & 0.60 & 1.02 & 1.43 & 1.52 & 2.13 & 2.75 \\
5 & Akre & 408 & 5,154 & 0.05 & $-$0.15 & 0.09 & $-$0.38 & $-$0.85 & $-$0.59 \\
6 & Sound Shore & 552 & 4,498 & $-$0.30 & $-$0.29 & $-$0.20 & $-$0.03 & $-$0.28 & 0.36 \\
7 & Caledonia & 83 & 4,218 & 0.08 & 0.23 & $-$0.75 & $-$0.92 & $-$0.29 & $-$0.99 \\
8 & Greenhaven Associates & 359 & 3,921 & 0.02 & $-$0.14 & 0.05 & 0.50 & 1.00 & 1.19 \\
9 & Ownership & 52 & 3,448 & $-$0.08 & 0.30 & 0.70 & 0.69 & 0.58 & 0.46 \\
10 & Steadfast & 1241 & 3,282 & $-$0.03 & $-$0.15 & $-$0.39 & $-$0.55 & $-$0.94 & $-$0.86 \\
11 & Jana & 1414 & 2,782 & 0.13 & 0.07 & 0.09 & 0.18 & $-$0.06 & $-$0.15 \\
12 & Theleme & 108 & 1,650 & $-$0.34 & $-$0.38 & $-$0.63 & $-$0.06 & $-$0.21 & 0.09 \\
13 & Cas Investment & 14 & 1,078 & 1.74 & 2.33 & 2.37 & 6.22 & 2.13 & 2.33 \\
14 & Oldfield & 106 & 1,044 & 0.13 & 0.08 & 0.43 & 0.41 & 1.54 & 2.24 \\
15 & Hawk Ridge & 582 & 1,001 & 0.02 & 0.18 & 0.18 & 0.77 & 1.34 & 2.39 \\
16 & Greenbriar & 104 & 566 & 0.02 & $-$0.04 & 0.39 & 0.70 & 0.24 & $-$0.28 \\
17 & Cyrus & 438 & 461 & 0.00 & $-$0.08 & $-$0.05 & 0.11 & $-$0.79 & $-$1.42 \\
18 & Thorntree & 170 & 432 & $-$0.26 & 0.02 & $-$0.14 & $-$0.28 & $-$0.72 & $-$0.32 \\
19 & Prime & 67 & 354 & 0.32 & 0.01 & $-$0.68 & $-$0.05 & 0.25 & 0.77 \\
20 & Mak & 214 & 319 & $-$0.27 & 0.16 & 0.06 & 0.56 & 1.99 & 3.12 \\
21 & Adw & 16 & 240 & 0.83 & $-$0.02 & $-$1.36 & $-$0.79 & $-$2.92 & $-$5.73 \\
22 & Pacifica & 71 & 205 & 0.16 & 0.57 & 0.69 & 0.60 & $-$0.18 & $-$0.88 \\
23 & Archon & 512 & 198 & $-$0.21 & $-$0.38 & $-$0.34 & $-$0.58 & $-$0.73 & $-$0.73 \\
24 & Tenzing & 185 & 175 & $-$0.06 & $-$0.11 & $-$0.13 & $-$0.47 & $-$0.61 & $-$1.53 \\
\bottomrule
\end{tabular*}
\begin{tablenotes}[flushleft]\footnotesize
\item \textit{Note.} Mean cumulative abnormal return (\%, Carhart four-factor adjusted) on new-purchase events, by control fund, ordered by AUM. N buys is the number of new-position events. Control managers are not individually identified. This table is the control-group counterpart to Table IA.III.
\end{tablenotes}
\end{threeparttable}
\end{sidewaystable}

\section{Skill versus recognition}

The main text presented the high-recognition premium with the full set of stock and fund controls (Section IV.C). Here we add a control for the fund's own past stock-picking record, which recognition correlates with (Pearson r = 0.47, p = 0.02). The results can be read from Table IA.V and Table IA.VI and Figure IA.1. The immediate premium survives every definition of the skill measure; at [0,30] it is not significant under six of the eight definitions. This is consistent with the interpretation in the main text.

\begin{table}[htbp]
\centering
\caption{The high-recognition premium with a trailing-skill control}
\setlength{\tabcolsep}{4pt}
\begin{threeparttable}
\small
\begin{tabular*}{\linewidth}{@{\extracolsep{\fill}}lcc}
\toprule
Window & Baseline & + Skill control \\
\midrule
\multicolumn{3}{l}{\textit{Panel A. High-recognition vs. Control}}\\
CAR[0,1] & 1.476$^{***}$ (<0.001) & 1.244$^{***}$ (0.001) \\
CAR[0,3] & 1.861$^{***}$ (0.001) & 1.422$^{***}$ (0.005) \\
CAR[0,5] & 2.289$^{***}$ (0.005) & 1.721$^{**}$ (0.024) \\
CAR[0,10] & 2.594$^{***}$ (0.008) & 1.808$^{**}$ (0.036) \\
CAR[0,20] & 3.125$^{***}$ (0.007) & 1.837$^{*}$ (0.061) \\
CAR[0,30] & 2.372$^{**}$ (0.024) & 0.819 (0.381) \\
\addlinespace[3pt]
\multicolumn{3}{l}{\textit{Panel B. Recognizable vs. Control}}\\
CAR[0,1] & 0.220 (0.143) & 0.154 (0.302) \\
CAR[0,3] & 0.393$^{**}$ (0.021) & 0.270$^{*}$ (0.064) \\
CAR[0,5] & 0.547$^{**}$ (0.021) & 0.384$^{*}$ (0.070) \\
CAR[0,10] & 0.730$^{**}$ (0.025) & 0.557$^{*}$ (0.064) \\
CAR[0,20] & 0.823$^{*}$ (0.089) & 0.544 (0.210) \\
CAR[0,30] & 0.456 (0.426) & 0.095 (0.856) \\
\bottomrule
\end{tabular*}
\begin{tablenotes}[flushleft]\footnotesize
\item \textit{Note.} Treat $\times$ buy coefficients (\%, full specification), estimated on the subsample of events for which the trailing skill measure exists (n = 10,395 for the high-recognition contrast, 18,775 for the recognizable contrast); p-values from manager-clustered standard errors in parentheses. Baseline re-estimates the paper's specification on this subsample; the second column adds trailing skill (the expanding mean of the CAR[0,3] of the fund's own previous new positions, current quarter excluded, minimum ten prior purchases) and its interaction with the new-position indicator. $^{*}p<.10$, $^{**}p<.05$, $^{***}p<.01$.
\end{tablenotes}
\end{threeparttable}
\end{table}

\begin{table}[htbp]
\centering
\caption{Skill-definition sweep: the high-recognition premium under every trailing-skill measure}
\setlength{\tabcolsep}{4pt}
\begin{threeparttable}
\small
\begin{tabularx}{\linewidth}{>{\raggedright\arraybackslash\hangindent=1em}Xccc}
\toprule
Skill measured from & CAR[0,1] & CAR[0,5] & CAR[0,30] \\
\midrule
No skill control (baseline) & 1.476$^{***}$ (<0.001) & 2.289$^{***}$ (0.005) & 2.372$^{**}$ (0.024) \\
CAR[0,1] of past buys & 1.035$^{***}$ (0.003) & 1.695$^{**}$ (0.022) & 1.301 (0.260) \\
CAR[0,3] of past buys (baseline measure) & 1.244$^{***}$ (0.001) & 1.721$^{**}$ (0.024) & 0.819 (0.381) \\
CAR[0,5] of past buys & 1.298$^{***}$ (0.001) & 1.731$^{**}$ (0.015) & 0.884 (0.331) \\
CAR[0,10] of past buys & 1.409$^{***}$ (0.001) & 1.994$^{***}$ (0.006) & 0.617 (0.503) \\
CAR[0,20] of past buys & 1.369$^{***}$ (<0.001) & 2.058$^{***}$ (0.003) & 0.699 (0.468) \\
CAR[0,30] of past buys & 1.383$^{***}$ (<0.001) & 2.081$^{***}$ (0.005) & 1.138 (0.129) \\
Drift only (CAR[0,30] $-$ CAR[0,3]) & 1.434$^{***}$ (<0.001) & 2.199$^{***}$ (0.004) & 1.730$^{**}$ (0.031) \\
Rolling 12 quarters (CAR[0,3]) & 1.240$^{***}$ (<0.001) & 1.973$^{***}$ (0.004) & 2.126$^{*}$ (0.057) \\
\bottomrule
\end{tabularx}
\begin{tablenotes}[flushleft]\footnotesize
\item \textit{Note.} High-recognition-vs-Control treat $\times$ buy (\%, full specification) with the trailing-skill control (skill and skill $\times$ buy) rebuilt under each definition. Rows 2--7 measure skill as the expanding mean of the stated CAR window of the fund's own previous new positions (current quarter excluded, minimum ten prior purchases with a usable outcome; n = 10,395). ``Drift only'' measures skill from the past buys' CAR[0,30] minus CAR[0,3], the abnormal return earned after the disclosure window, and is therefore the skill proxy least contaminated by the fund's own historical fame effect. ``Rolling 12 quarters'' replaces the expanding career mean with a rolling three-year record (minimum five prior purchases; n = 10,372). The baseline row re-estimates the specification without skill terms on the same subsample. p-values from manager-clustered standard errors in parentheses. $^{*}p<.10$, $^{**}p<.05$, $^{***}p<.01$.
\end{tablenotes}
\end{threeparttable}
\end{table}

\begin{figure}[htbp]
\centering
\includegraphics[width=\linewidth]{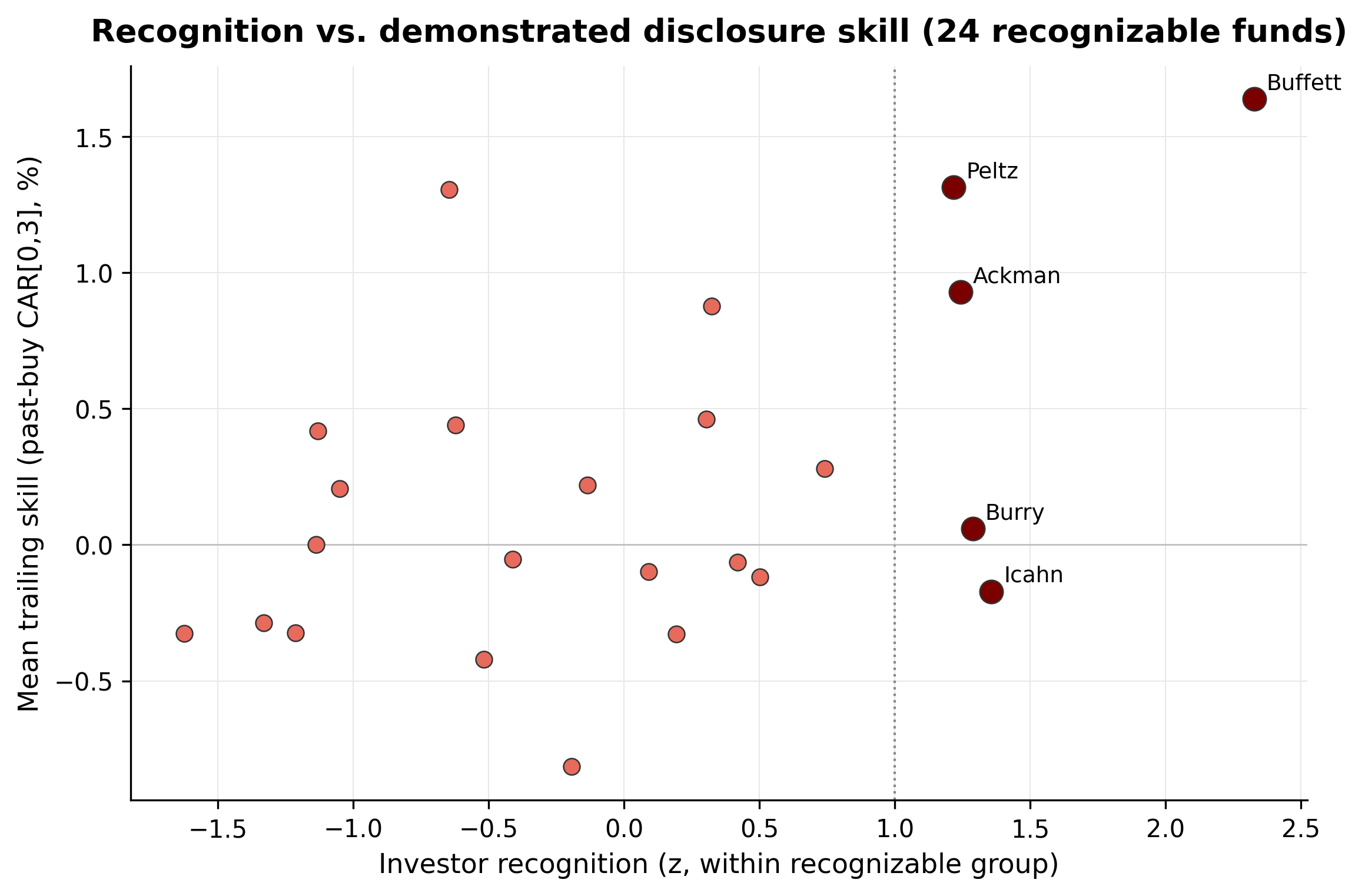}
\caption{Recognition and demonstrated disclosure skill across the 24 recognizable funds}
{\footnotesize\noindent\textit{Note.} Each point is one fund; the horizontal axis is the recognition index (z, standardized within the recognizable group) and the vertical axis is the fund's mean trailing skill, the average CAR[0,3] of its own past new positions. High-recognition funds (z $\geq$ 1, right of the dotted line) are labeled. Pearson r = 0.47 (p = 0.02).\par}
\end{figure}

\section{Recognition measured in real time}

The main text replaced the static recognition index with a rolling index built only from information available at the filing date and found the premium somewhat smaller but present at every horizon (Section V.B). Here we report the full comparison, the quarters each manager spends above the threshold, and the reaction to the rolling index. The results can be read from Table IA.VII to Table IA.IX and Figure IA.2. They show the premium under both classifications, with the rolling estimates 11 to 22\% smaller, and a positive slope of the reaction in the rolling index, which is consistent with the main text.

\begin{table}[htbp]
\centering
\caption{Static versus rolling high-recognition classification (events from 2016H2)}
\setlength{\tabcolsep}{4pt}
\begin{threeparttable}
\small
\begin{tabular*}{\linewidth}{@{\extracolsep{\fill}}lcc}
\toprule
Window & Static tiers & Rolling (as-of) tiers \\
\midrule
CAR[0,1] & 1.772$^{***}$ (<0.001) & 1.515$^{***}$ (0.004) \\
CAR[0,3] & 2.054$^{***}$ (<0.001) & 1.811$^{***}$ (0.006) \\
CAR[0,5] & 3.067$^{***}$ (0.002) & 2.650$^{**}$ (0.013) \\
CAR[0,10] & 3.392$^{**}$ (0.016) & 3.011$^{**}$ (0.036) \\
CAR[0,20] & 4.289$^{***}$ (0.002) & 3.337$^{*}$ (0.062) \\
CAR[0,30] & 3.369$^{***}$ (0.004) & 2.637$^{**}$ (0.045) \\
\bottomrule
\end{tabular*}
\begin{tablenotes}[flushleft]\footnotesize
\item \textit{Note.} High-recognition-vs-Control treat $\times$ buy (\%, full specification) on events filed from 2016H2 onward. Static tiers use the paper's full-sample recognition index (n = 4,252); rolling tiers re-classify each event with the trailing-twelve-month Wikipedia recognition index known at the filing date (n = 4,309). p-values from manager-clustered standard errors in parentheses. $^{*}p<.10$, $^{**}p<.05$, $^{***}p<.01$.
\end{tablenotes}
\end{threeparttable}
\end{table}

Table IA.VIII presents the quarters in which each manager is above the high-recognition threshold under the rolling index. It supplements Section V.B of the main text.

\begin{table}[htbp]
\centering
\caption{High-recognition-tier spells under the rolling recognition index}
\setlength{\tabcolsep}{4pt}
\begin{threeparttable}
\small
\begin{tabular*}{\linewidth}{@{\extracolsep{\fill}}llccc}
\toprule
 & & \multicolumn{3}{c}{High-recognition spell (rolling z $\geq$ 1)} \\
\cmidrule(l{3pt}){3-5}
Manager (fund) & Static tier & First quarter & Last quarter & Quarters \\
\midrule
Buffett (Berkshire) & High & 2016Q3 & 2025Q4 & 38 \\
Burry (Scion) & High & 2016Q3 & 2025Q4 & 38 \\
Ackman (Pershing Square) & High & 2018Q2 & 2025Q4 & 30 \\
Icahn & High & 2016Q3 & 2023Q4 & 25 \\
Peltz (Trian) & High & 2020Q4 & 2025Q4 & 19 \\
Ressler (Ares) & Moderate & 2023Q3 & 2025Q4 & 10 \\
Tepper (Appaloosa) & Moderate & 2018Q4 & 2025Q1 & 9 \\
\bottomrule
\end{tabular*}
\begin{tablenotes}[flushleft]\footnotesize
\item \textit{Note.} Quarters in which a manager's rolling recognition satisfies z $\geq$ 1 within the recognizable group, 2016Q3--2025Q4 (38 quarters). The remaining seventeen recognizable managers never cross the threshold. Spells need not be contiguous; first and last mark the boundaries of the high-recognition period.
\end{tablenotes}
\end{threeparttable}
\end{table}

\begin{table}[htbp]
\centering
\caption{The buy reaction and the rolling recognition index}
\setlength{\tabcolsep}{3pt}
\begin{threeparttable}
\small
\begin{tabular*}{\linewidth}{@{\extracolsep{\fill}}lcccccc}
\toprule
 & (1) & (2) & (3) & (4) & (5) & (6) \\
 & Raw & \makecell{+ Stock\\ctrls} & \makecell{+ Stock\\FE} & \makecell{+ Fund\\ctrls} & \makecell{+ Fund\\FE} & Full \\
\midrule
CAR[0,1] & 0.537$^{***}$ & 0.649$^{***}$ & 0.762$^{***}$ & 0.763$^{***}$ & 0.770$^{***}$ & 0.681$^{***}$ \\
 & (0.002) & (0.001) & (0.003) & (0.003) & (0.002) & (0.007) \\[1pt]
CAR[0,3] & 0.478$^{**}$ & 0.527$^{*}$ & 0.710$^{**}$ & 0.711$^{**}$ & 0.720$^{**}$ & 0.598$^{*}$ \\
 & (0.033) & (0.063) & (0.031) & (0.030) & (0.029) & (0.056) \\[1pt]
CAR[0,5] & 0.537$^{*}$ & 0.597 & 0.726 & 0.730 & 0.701 & 0.624 \\
 & (0.092) & (0.102) & (0.153) & (0.151) & (0.172) & (0.224) \\[1pt]
CAR[0,10] & 0.705$^{*}$ & 0.693$^{*}$ & 0.692 & 0.700 & 0.698 & 0.680 \\
 & (0.072) & (0.095) & (0.255) & (0.253) & (0.257) & (0.252) \\[1pt]
CAR[0,20] & 1.109$^{**}$ & 1.307$^{***}$ & 1.289$^{*}$ & 1.310$^{*}$ & 1.322$^{*}$ & 1.301$^{*}$ \\
 & (0.017) & (0.009) & (0.092) & (0.096) & (0.091) & (0.097) \\[1pt]
CAR[0,30] & 1.141 & 1.595$^{***}$ & 1.270 & 1.280 & 1.243 & 1.142 \\
 & (0.123) & (0.010) & (0.117) & (0.113) & (0.141) & (0.148) \\[1pt]
\bottomrule
\end{tabular*}
\begin{tablenotes}[flushleft]\footnotesize
\item \textit{Note.} Sample: new-position and complete-exit events of the 24 recognizable funds filed from 2016H2, when the rolling index is defined. Each entry is the change in the buy reaction per one standard deviation of the rolling recognition index measured at the filing date (rolling recognition z $\times$ buy, \%, Carhart four-factor adjusted), with p-values from manager-clustered standard errors in parentheses. Columns add controls cumulatively as in Table IV of the main text. $^{*}p<.10$, $^{**}p<.05$, $^{***}p<.01$.
\end{tablenotes}
\end{threeparttable}
\end{table}

\begin{figure}[htbp]
\centering
\includegraphics[width=\linewidth]{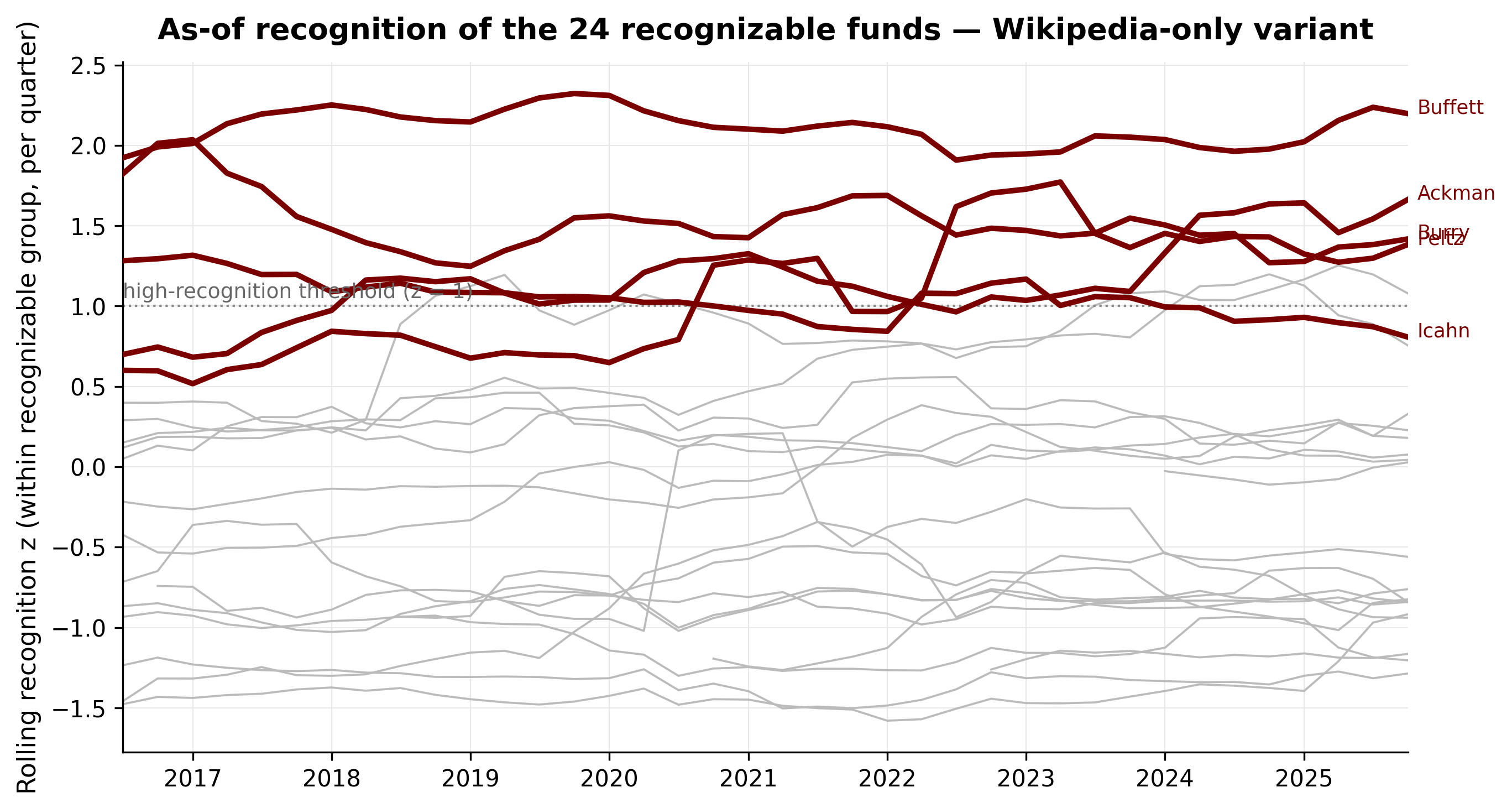}
\caption{Rolling recognition of the 24 recognizable funds}
{\footnotesize\noindent\textit{Note.} Rolling (as-of) recognition of the 24 recognizable funds, Wikipedia-only variant: trailing-twelve-month pageviews standardized within the recognizable group each quarter. Dark lines mark the five statically classified high-recognition managers; the dotted line is the high-recognition threshold (z = 1).\par}
\end{figure}

\section{Control matching}

The main text matches each recognizable fund to a control fund (Section III.C). Here we report the matching procedure, the balance of the matched sample, the controls that were removed and replaced, and the re-matching without manual exclusions. The results can be read from Table IA.X to Table IA.XV. The matched groups are balanced, and the estimates do not depend on the manual exclusions, which is consistent with the main text.

Let F denote the 24 recognizable funds with covariates X (portfolio value, concentration, and longevity), and let C denote the pool of candidate controls passing the filters in Section IA.1. We match each recognizable fund to a unique control by optimal assignment, minimizing the total Mahalanobis distance between the two groups on X, with portfolio value and concentration entered in logs given their right-skew. Because the assignment is one-to-one and has no random component, it is exact and fully reproducible. Table IA.X reports the distribution of the 24 matched-pair Mahalanobis distances, and Table IA.XI reports balance on the matched sample.

\begin{table}[htbp]
\centering
\caption{Distribution of matched-pair Mahalanobis distances}
\setlength{\tabcolsep}{4pt}
\begin{threeparttable}
\small
\begin{tabular*}{\linewidth}{@{\extracolsep{\fill}}lc}
\toprule
Statistic & Matched-pair distance \\
\midrule
Minimum & 0.13 \\
25th percentile & 0.27 \\
Median & 0.39 \\
Mean & 0.56 \\
75th percentile & 0.64 \\
Maximum & 1.87 \\
Standard deviation & 0.46 \\
\bottomrule
\end{tabular*}
\begin{tablenotes}[flushleft]\footnotesize
\item \textit{Note.} Distances are Mahalanobis on the matching covariates (log portfolio value, log concentration, longevity). The single large distance, 1.87, is the control matched to Berkshire Hathaway, whose extreme size and longevity make it the hardest fund to match; this is expected and does not affect the results.
\end{tablenotes}
\end{threeparttable}
\end{table}

\begin{table}[htbp]
\centering
\caption{Balance on the matched sample}
\setlength{\tabcolsep}{4pt}
\begin{threeparttable}
\small
\begin{tabularx}{\linewidth}{>{\raggedright\arraybackslash\hangindent=1em}Xcccc}
\toprule
Characteristic & \makecell{Recognizable\\(median)} & \makecell{Control\\(median)} & SMD & Welch p \\
\midrule
Portfolio value (\$M) & 2,644 & 1,364 & +0.09 & 0.42 \\
Concentration (positions per filing) & 18.9 & 21.0 & $-$0.11 & 0.55 \\
Longevity (quarters) & 86.5 & 64.5 & +0.46 & 0.12 \\
Turnover (events per quarter) & 8.8 & 7.2 & +0.03 & 0.91 \\
\bottomrule
\end{tabularx}
\begin{tablenotes}[flushleft]\footnotesize
\item \textit{Note.} Standardized mean differences for portfolio value and concentration are computed on the log scale; the Welch tests are on the raw scale. No matched characteristic differs significantly between the groups (Welch two-sample t-test). Turnover is reported as an out-of-sample check; it is not a matching covariate.
\end{tablenotes}
\end{threeparttable}
\end{table}

Eighteen initially matched controls were removed and replaced over successive matching rounds because, on closer inspection, they were not comparable. The largest group had wound down or gone dormant: funds that had liquidated to near-zero holdings or filed only notice forms in recent quarters (for example ESL Investments, Cedar Rock Capital, Voce Capital, and Northern Cross), and credit or distressed shops that had since dissolved or rebranded (Halcyon Asset Management). A second group had returned outside capital and converted to private or family-office vehicles (Aquamarine Capital, Meditor Group). A third were single-name control vehicles rather than stock-selection portfolios, each dominated by one position (Hill Path Capital, roughly two-thirds in SeaWorld; All-Stars Investment, roughly seven-eighths in a single name). Others were foreign or emerging-markets managers whose US 13F captures only an unrepresentative sliver of the true book (the Canadian value managers QV Investors and Beutel Goodman, the Hong Kong fund Turiya, the Brazilian manager Squadra, and the emerging-markets complex Seafarer), or were simply the wrong vehicle type (NFC Investments, a multi-asset RIA managing insurance-company portfolios). Each removed control was swapped for a US-registered, discretionary, active analogue matched on the same covariates, and every removal is logged with its reason and supporting sources in the replication package. The remainder of this section re-runs the entire matching procedure with no manual removals and re-estimates the paper on the resulting control group.

The remaining tables re-run the matching with no manual removals. The results can be read from Table IA.XII to Table IA.XV. The point estimates are unchanged or slightly larger; the pooled [0,3] and [0,5] p-values rise to 0.063 and 0.056. This is consistent with the main text.

\begin{table}[htbp]
\centering
\caption{Paper versus no-exclusion control group}
\setlength{\tabcolsep}{4pt}
\begin{threeparttable}
\small
\begin{tabular*}{\linewidth}{@{\extracolsep{\fill}}lccc}
\toprule
Window & Paper $\beta$ & No-exclusion $\beta$ & No-exclusion p \\
\midrule
\multicolumn{4}{l}{\textit{Panel A. Recognizable vs. Control}}\\
CAR[0,1] & 0.200 & 0.210 & 0.159 \\
CAR[0,3] & 0.341 & 0.316 & 0.063 \\
CAR[0,5] & 0.499 & 0.474 & 0.056 \\
CAR[0,10] & 0.684 & 0.682 & 0.041 \\
CAR[0,20] & 0.772 & 0.931 & 0.049 \\
CAR[0,30] & 0.435 & 0.712 & 0.171 \\
CAR[-10,-1] & $-$0.431 & $-$0.139 & 0.653 \\
\addlinespace[3pt]
\multicolumn{4}{l}{\textit{Panel B. High-recognition vs. Control}}\\
CAR[0,1] & 1.422 & 1.458 & <0.001 \\
CAR[0,3] & 1.869 & 1.793 & <0.001 \\
CAR[0,5] & 2.309 & 2.189 & 0.001 \\
CAR[0,10] & 2.547 & 2.612 & 0.002 \\
CAR[0,20] & 3.091 & 3.212 & 0.001 \\
CAR[0,30] & 2.319 & 2.565 & 0.005 \\
CAR[-10,-1] & 0.392 & 0.774 & 0.302 \\
\bottomrule
\end{tabular*}
\begin{tablenotes}[flushleft]\footnotesize
\item \textit{Note.} Treat $\times$ buy coefficients (\%, full specification). ``Paper'' is the published estimate; ``No-exclusion'' re-estimates it with the control group produced by the matcher with zero manual removals (n = 17,737 recognizable, 8,962 high-recognition at CAR[0,1]), with the winsorization and within-recognizable standardization rebuilt on the new sample. p-values from manager-clustered standard errors.
\end{tablenotes}
\end{threeparttable}
\end{table}

\begin{table}[htbp]
\centering
\caption{No-exclusion high-recognition contrast: wild-cluster bootstrap}
\setlength{\tabcolsep}{4pt}
\begin{threeparttable}
\small
\begin{tabular*}{\linewidth}{@{\extracolsep{\fill}}lccc}
\toprule
Window & $\beta$ & WCR p & WCU p \\
\midrule
CAR[0,1] & 1.458 & 0.001 & <0.001 \\
CAR[0,3] & 1.793 & 0.033 & 0.001 \\
CAR[0,5] & 2.189 & 0.057 & 0.003 \\
CAR[0,30] & 2.565 & 0.098 & 0.006 \\
\bottomrule
\end{tabular*}
\begin{tablenotes}[flushleft]\footnotesize
\item \textit{Note.} Restricted (WCR) and unrestricted (WCU) wild-cluster bootstrap p-values (Webb six-point weights, B = 9,999, G = 29 clusters) for the no-exclusion High-recognition-vs-Control contrast, mirroring the paper's small-cluster inference.
\end{tablenotes}
\end{threeparttable}
\end{table}

Table IA.XIV and Table IA.XV present the balance of the re-matched sample and the nine control funds that return to it. They supplement the matching procedure above.

\begin{table}[htbp]
\centering
\caption{Balance on the no-exclusion matched sample}
\setlength{\tabcolsep}{4pt}
\begin{threeparttable}
\small
\begin{tabularx}{\linewidth}{>{\raggedright\arraybackslash\hangindent=1em}Xcccc}
\toprule
Characteristic & \makecell{Recognizable\\(median)} & \makecell{Control\\(median)} & SMD & Welch p \\
\midrule
Portfolio value (\$M) & 2,316 & 1,786 & +0.06 & 0.42 \\
Concentration (positions per filing) & 17.9 & 19.2 & $-$0.11 & 0.42 \\
Longevity (quarters) & 65.0 & 66.5 & +0.04 & 0.89 \\
\bottomrule
\end{tabularx}
\begin{tablenotes}[flushleft]\footnotesize
\item \textit{Note.} Balance between the 24 recognizable funds and the 24 no-exclusion controls on the matching covariates; standardized mean differences for portfolio value and concentration are computed on the log scale. Fund characteristics here are computed from the holdings panel rebuilt for the re-matching exercise, so the recognizable-fund medians differ from Table IA.XI, which uses the paper's fund metrics.
\end{tablenotes}
\end{threeparttable}
\end{table}

\begin{table}[htbp]
\centering
\caption{The nine returning control funds}
\setlength{\tabcolsep}{4pt}
\begin{threeparttable}
\small
\begin{tabularx}{\linewidth}{ll>{\raggedright\arraybackslash}X}
\toprule
Control fund & Matched recognizable fund & Reason for the paper's removal \\
\midrule
ESL Investments & Pershing Square (Ackman) & Wound down; last holdings 13F in 2022; became a control vehicle for Sears. \\
Meditor Group & Steel Partners (Lichtenstein) & Returned outside capital in 2013; stopped filing after 2024Q3. \\
Voce Capital & Scion (Burry) & Dormant; no filings since mid-2023. \\
Cedar Rock Capital & Trian (Peltz) & Liquidated to zero in 2022Q1; dissolved in 2025. \\
Genesis Fund Managers & Baupost (Klarman) & Near-zero recent 13Fs; emerging-markets manager whose 13F captured only a US ADR sliver. \\
QV Investors & Palo Alto Investors (Yun) & Canadian manager; the US 13F is an unrepresentative slice of the book. \\
NFC Investments & Dalton (Rosenwald) & Multi-asset insurance RIA, not a long-equity stock picker. \\
All-Stars Investment & Mudrick & Roughly 88\% in a single name; China-focused hybrid vehicle. \\
Hill Path Capital & Himalaya (Li Lu) & Single-company activist control position (roughly 65\% SeaWorld). \\
\bottomrule
\end{tabularx}
\begin{tablenotes}[flushleft]\footnotesize
\item \textit{Note.} The nine previously removed controls that the no-exclusion matcher returns, the recognizable fund each is matched to, and a condensed version of the documented removal reason (full documentation in the replication package). All nine are retained in the no-exclusion estimation of Table IA.XII and Table IA.XIII.
\end{tablenotes}
\end{threeparttable}
\end{table}

\section{Pre-disclosure dynamics and information leakage}

The main text reported no pre-disclosure run-up in returns (Section IV.A). Here we test the pre-filing window day by day, look at trading volume, and exclude the purchases that were public before the 13F through the fund's own 13D or 13G filing. The results can be read from Table IA.XVI and Table IA.XVII and Figure IA.3 and Figure IA.4. There is no run-up in returns or volume before the filing, and excluding the telegraphed purchases raises the premium, which is consistent with the main text.

\begin{table}[htbp]
\centering
\caption{Daily abnormal-return differences around the filing date}
\setlength{\tabcolsep}{4pt}
\begin{threeparttable}
\small
\begin{tabular*}{\linewidth}{@{\extracolsep{\fill}}lcccc}
\toprule
& \multicolumn{2}{c}{High-recognition $-$ Control} & \multicolumn{2}{c}{Recognizable $-$ Control} \\
Day & Diff. (bps) & p & Diff. (bps) & p \\
\midrule
$-$10 & $-$1.9 & 0.899 & $-$1.9 & 0.713 \\
$-$9 & $-$5.4 & 0.398 & $-$8.8$^{**}$ & 0.044 \\
$-$8 & 4.3 & 0.809 & 2.6 & 0.648 \\
$-$7 & 3.2 & 0.758 & $-$1.0 & 0.872 \\
$-$6 & 7.1 & 0.755 & $-$0.2 & 0.977 \\
$-$5 & $-$7.2 & 0.481 & $-$12.6$^{**}$ & 0.037 \\
$-$4 & 25.1$^{***}$ & 0.006 & 14.5$^{**}$ & 0.046 \\
$-$3 & 6.4 & 0.468 & $-$2.6 & 0.606 \\
$-$2 & $-$20.3 & 0.122 & $-$0.2 & 0.976 \\
$-$1 & 21.5 & 0.188 & 1.7 & 0.757 \\
0 & 43.2$^{***}$ & 0.002 & 3.8 & 0.553 \\
+1 & 67.5$^{***}$ & 0.001 & 16.6$^{***}$ & 0.010 \\
+2 & 24.5$^{***}$ & 0.006 & 4.9 & 0.361 \\
+3 & 2.4 & 0.655 & 3.3 & 0.601 \\
+4 & 10.4 & 0.353 & 5.4 & 0.269 \\
+5 & 26.1$^{***}$ & 0.005 & 0.7 & 0.858 \\
+6 & 23.8$^{**}$ & 0.028 & 2.4 & 0.624 \\
+7 & $-$3.2 & 0.547 & $-$2.2 & 0.693 \\
+8 & 6.6 & 0.357 & $-$5.6 & 0.318 \\
+9 & 2.6 & 0.740 & 2.0 & 0.753 \\
+10 & 9.3 & 0.322 & 8.9$^{*}$ & 0.071 \\
\bottomrule
\end{tabular*}
\begin{tablenotes}[flushleft]\footnotesize
\item \textit{Note.} Difference in mean daily abnormal returns (Carhart four-factor, basis points) between treated and control buy events, by trading day relative to the filing date; p-values from two-sample tests with manager-clustered standard errors. One of the ten high-recognition pre-filing days is significant at 5\%, against three consecutive days from day 0. $^{*}p<.10$, $^{**}p<.05$, $^{***}p<.01$.
\end{tablenotes}
\end{threeparttable}
\end{table}

\begin{table}[htbp]
\centering
\caption{Excluding purchases telegraphed by the fund's own 13D/13G}
\setlength{\tabcolsep}{4pt}
\begin{threeparttable}
\small
\begin{tabular*}{\linewidth}{@{\extracolsep{\fill}}lcc}
\toprule
Window & All buys & Excluding telegraphed buys \\
\midrule
\multicolumn{3}{l}{\textit{Panel A. High-recognition vs. Control}}\\
CAR[0,1] & 1.422$^{***}$ (<0.001) & 1.518$^{***}$ (<0.001) \\
CAR[0,3] & 1.869$^{***}$ (<0.001) & 2.028$^{***}$ (<0.001) \\
CAR[0,5] & 2.309$^{***}$ (0.002) & 2.478$^{***}$ (<0.001) \\
CAR[0,10] & 2.547$^{***}$ (0.004) & 2.749$^{***}$ (<0.001) \\
CAR[0,20] & 3.091$^{***}$ (0.003) & 3.427$^{***}$ (<0.001) \\
CAR[0,30] & 2.319$^{**}$ (0.016) & 2.646$^{***}$ (0.003) \\
\addlinespace[3pt]
\multicolumn{3}{l}{\textit{Panel B. Recognizable vs. Control}}\\
CAR[0,1] & 0.200 (0.159) & 0.198 (0.168) \\
CAR[0,3] & 0.341$^{**}$ (0.043) & 0.326$^{*}$ (0.060) \\
CAR[0,5] & 0.499$^{**}$ (0.031) & 0.486$^{**}$ (0.037) \\
CAR[0,10] & 0.684$^{**}$ (0.027) & 0.672$^{**}$ (0.034) \\
CAR[0,20] & 0.772$^{*}$ (0.099) & 0.756 (0.108) \\
CAR[0,30] & 0.435 (0.434) & 0.438 (0.431) \\
\bottomrule
\end{tabular*}
\begin{tablenotes}[flushleft]\footnotesize
\item \textit{Note.} Treat $\times$ buy coefficients (\%, full specification) with and without the buy events preceded ($\leq$ 90 days) by the same fund's own Schedule 13D or 13G on the same stock. There are 982 such events in the full event set; 211 of them (152 from high-recognition funds) fall in the full-specification estimation sample (n = 10,820 high-recognition, 19,536 recognizable after exclusion). Excluding the telegraphed buys strengthens the high-recognition premium in every window, so pre-13F visibility biases the paper's estimates toward zero. p-values from manager-clustered standard errors in parentheses. $^{*}p<.10$, $^{**}p<.05$, $^{***}p<.01$.
\end{tablenotes}
\end{threeparttable}
\end{table}

\begin{figure}[htbp]
\centering
\includegraphics[width=\linewidth]{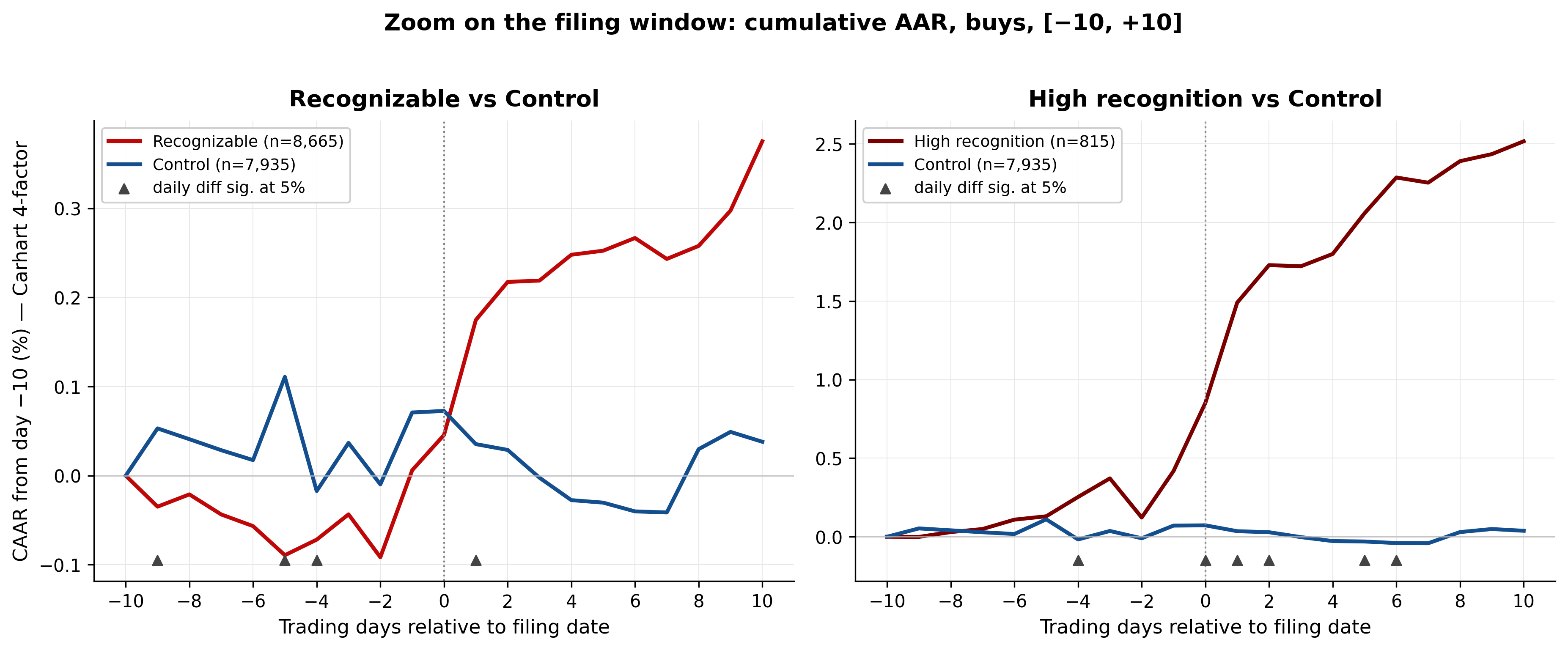}
\caption{Cumulative average abnormal return around the filing date}
{\footnotesize\noindent\textit{Note.} Cumulative average abnormal return from day $-$10 around the filing date, buy events, Carhart four-factor: recognizable versus control (left) and high-recognition versus control (right). Triangles mark days on which the daily treated-control difference is significant at 5\%. Both treated paths are flat before the filing and jump on and just after it. The event counts (8,665 recognizable and 7,935 control purchases) differ slightly from Figure 1 of the main text (8,668 and 7,940), which is built from the full [$-$10,+30] panel.\par}
\end{figure}

\begin{figure}[htbp]
\centering
\includegraphics[width=\linewidth]{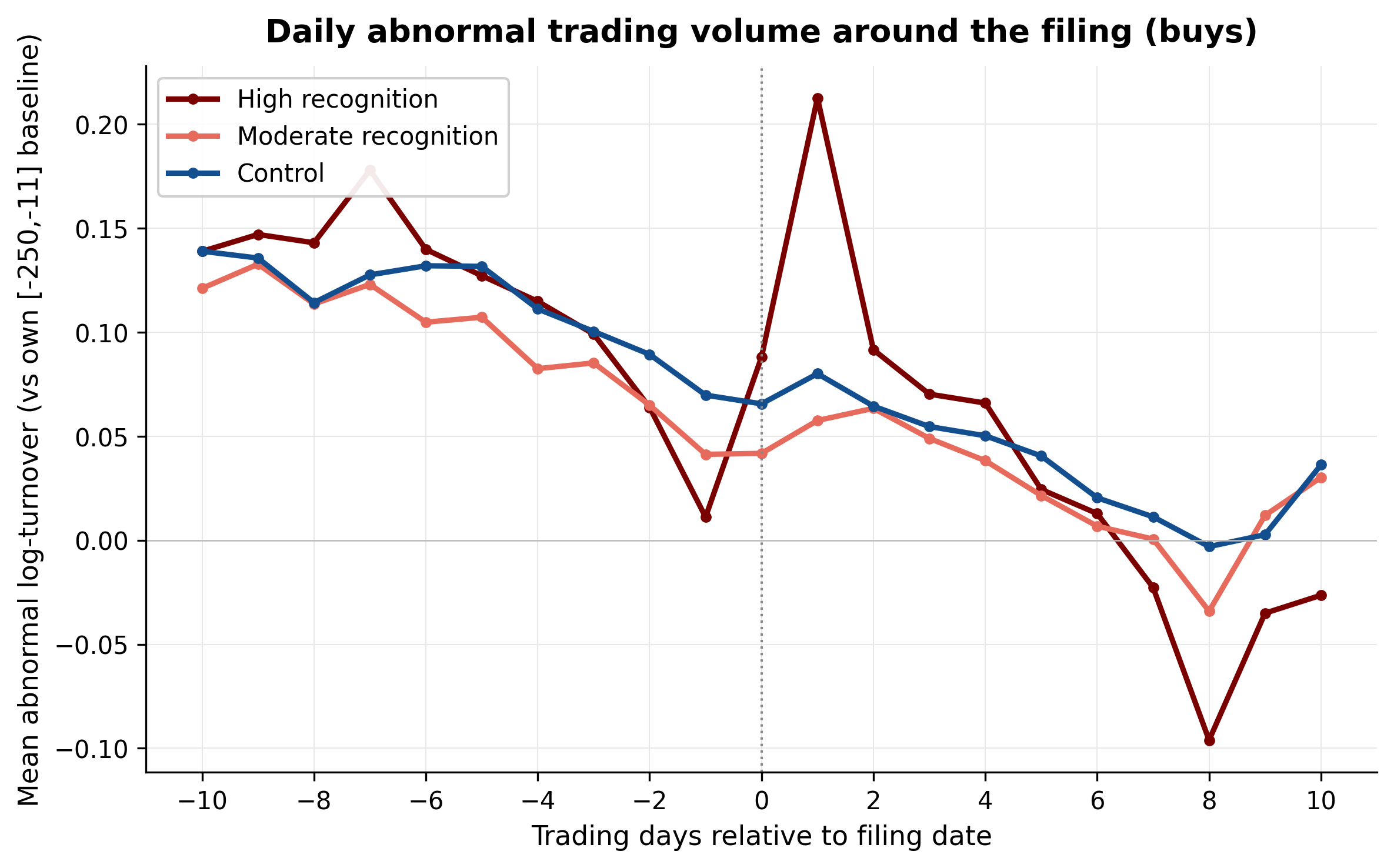}
\caption{Daily abnormal trading volume around the filing date}
{\footnotesize\noindent\textit{Note.} Daily abnormal trading volume around the filing date, buy events: mean abnormal log-turnover relative to the stock's own [$-$250,$-$11] baseline, by tier. High-recognition volume is indistinguishable from the controls before the filing and spikes on the first post-filing day (p = 0.005).\par}
\end{figure}

\section{Calendar-time portfolios: additional results}

The main text presented the calendar-time alphas of portfolios that copy the disclosed purchases (Section VI.D). Here we report weighting and entry-day variants, long-short portfolios, and exit portfolios. The results can be read from Table IA.XVIII to Table IA.XX and Figure IA.5. The alphas do not depend on daily rebalancing, a copier who enters one day later keeps about half of the 21-day alpha and three-quarters of the 63-day alpha, and exits earn none, which is consistent with the main text.

\begin{table}[htbp]
\centering
\caption{Weighting and entry-day variants: rebalanced versus buy-and-hold}
\setlength{\tabcolsep}{4pt}
\begin{threeparttable}
\small
\begin{tabularx}{\linewidth}{>{\raggedright\arraybackslash\hangindent=1em}Xccc}
\toprule
Variant & H = 21 & H = 63 & H = 126 \\
\midrule
\multicolumn{4}{l}{\textit{Panel A. High-recognition buy portfolio}}\\
Daily-rebalanced, entry day +1 (baseline) & 9.8$^{***}$ (<0.001) & 4.4$^{***}$ (0.003) & 1.4 (0.164) \\
Buy-and-hold, entry day +1 & 9.4$^{***}$ (0.001) & 3.8$^{***}$ (0.010) & 1.1 (0.274) \\
Daily-rebalanced, entry day +2 & 4.9$^{*}$ (0.064) & 3.2$^{**}$ (0.029) & 1.0 (0.312) \\
Buy-and-hold, entry day +2 & 4.8$^{*}$ (0.075) & 2.7$^{*}$ (0.068) & 0.7 (0.493) \\
\addlinespace[3pt]
\multicolumn{4}{l}{\textit{Panel B. High-recognition $-$ Control long-short}}\\
Daily-rebalanced, entry day +1 (baseline) & 10.4$^{***}$ (<0.001) & 4.8$^{***}$ (0.003) & 0.8 (0.438) \\
Buy-and-hold, entry day +1 & 10.2$^{***}$ (<0.001) & 4.3$^{***}$ (0.006) & 0.9 (0.412) \\
Daily-rebalanced, entry day +2 & 5.3$^{*}$ (0.058) & 3.7$^{**}$ (0.019) & 0.4 (0.692) \\
Buy-and-hold, entry day +2 & 5.3$^{*}$ (0.063) & 3.4$^{**}$ (0.030) & 0.4 (0.680) \\
\bottomrule
\end{tabularx}
\begin{tablenotes}[flushleft]\footnotesize
\item \textit{Note.} Daily Carhart alphas (basis points per day). ``Daily-rebalanced'' resets weights to equal each day (the convention of the paper's Table); ``buy-and-hold'' lets each position enter at equal size and drift untouched for its H trading days, so the portfolio never trades inside a holding window. Entry day +1 starts each spell at the close of the filing day; entry day +2 starts it one trading day later, at a price any follower could obtain. The buy-and-hold, entry +2 row is the closest approximation to an implementable copying strategy, gross of transaction costs. HAC standard errors (five lags); p-values in parentheses. $^{*}p<.10$, $^{**}p<.05$, $^{***}p<.01$.
\end{tablenotes}
\end{threeparttable}
\end{table}

\begin{table}[htbp]
\centering
\caption{Long-short calendar-time alphas}
\setlength{\tabcolsep}{4pt}
\begin{threeparttable}
\small
\begin{tabularx}{\linewidth}{>{\raggedright\arraybackslash\hangindent=1em}Xccc}
\toprule
Portfolio & H = 21 & H = 63 & H = 126 \\
\midrule
High-recognition $-$ Control & 10.4$^{***}$ (<0.001) & 4.8$^{***}$ (0.003) & 0.8 (0.438) \\
Recognizable $-$ Control & 2.3 (0.186) & 2.4$^{***}$ (0.005) & 1.4$^{**}$ (0.027) \\
High-recognition $-$ Moderate-recognition & 7.8$^{**}$ (0.017) & 1.9 (0.272) & $-$0.3 (0.797) \\
\bottomrule
\end{tabularx}
\begin{tablenotes}[flushleft]\footnotesize
\item \textit{Note.} Daily Carhart alphas (basis points per day) of long-short calendar-time portfolios, estimated on days where both legs have at least five open positions; HAC standard errors (five lags), p-values in parentheses. The high-recognition-minus-moderate-recognition contrast shows that the short-horizon alpha is specific to the high-recognition tier even within the recognizable group. $^{*}p<.10$, $^{**}p<.05$, $^{***}p<.01$.
\end{tablenotes}
\end{threeparttable}
\end{table}

\begin{table}[htbp]
\centering
\caption{Exit portfolios and delayed entry (H = 63)}
\setlength{\tabcolsep}{4pt}
\begin{threeparttable}
\small
\begin{tabular*}{\linewidth}{@{\extracolsep{\fill}}lcc}
\toprule
Portfolio & Alpha (bps/day) & p \\
\midrule
\multicolumn{3}{l}{\textit{Panel A. Complete-exit portfolios}}\\
High-recognition exits & 1.9 & 0.288 \\
Recognizable exits & 1.2 & 0.100 \\
Control exits & 1.5$^{**}$ & 0.020 \\
\addlinespace[3pt]
\multicolumn{3}{l}{\textit{Panel B. Buy portfolios, entry at the second post-filing day}}\\
High-recognition buys & 3.2$^{**}$ & 0.029 \\
Moderate-recognition buys & 2.6$^{***}$ & 0.006 \\
Recognizable buys & 2.7$^{***}$ & 0.002 \\
Control buys & 0.6 & 0.369 \\
\bottomrule
\end{tabular*}
\begin{tablenotes}[flushleft]\footnotesize
\item \textit{Note.} Daily Carhart alphas at the 63-day horizon. Panel A forms portfolios from complete-exit disclosures in place of new positions: no group's exits carry an alpha economically distinguishable from the controls', so the buy effect has no mirror image. Panel B delays entry by one further trading day, to a price any follower could obtain; the high-recognition alpha remains positive and significant at about three-quarters of the paper's 63-day headline figure (4.4 basis points). $^{*}p<.10$, $^{**}p<.05$, $^{***}p<.01$.
\end{tablenotes}
\end{threeparttable}
\end{table}

\begin{figure}[htbp]
\centering
\includegraphics[width=\linewidth]{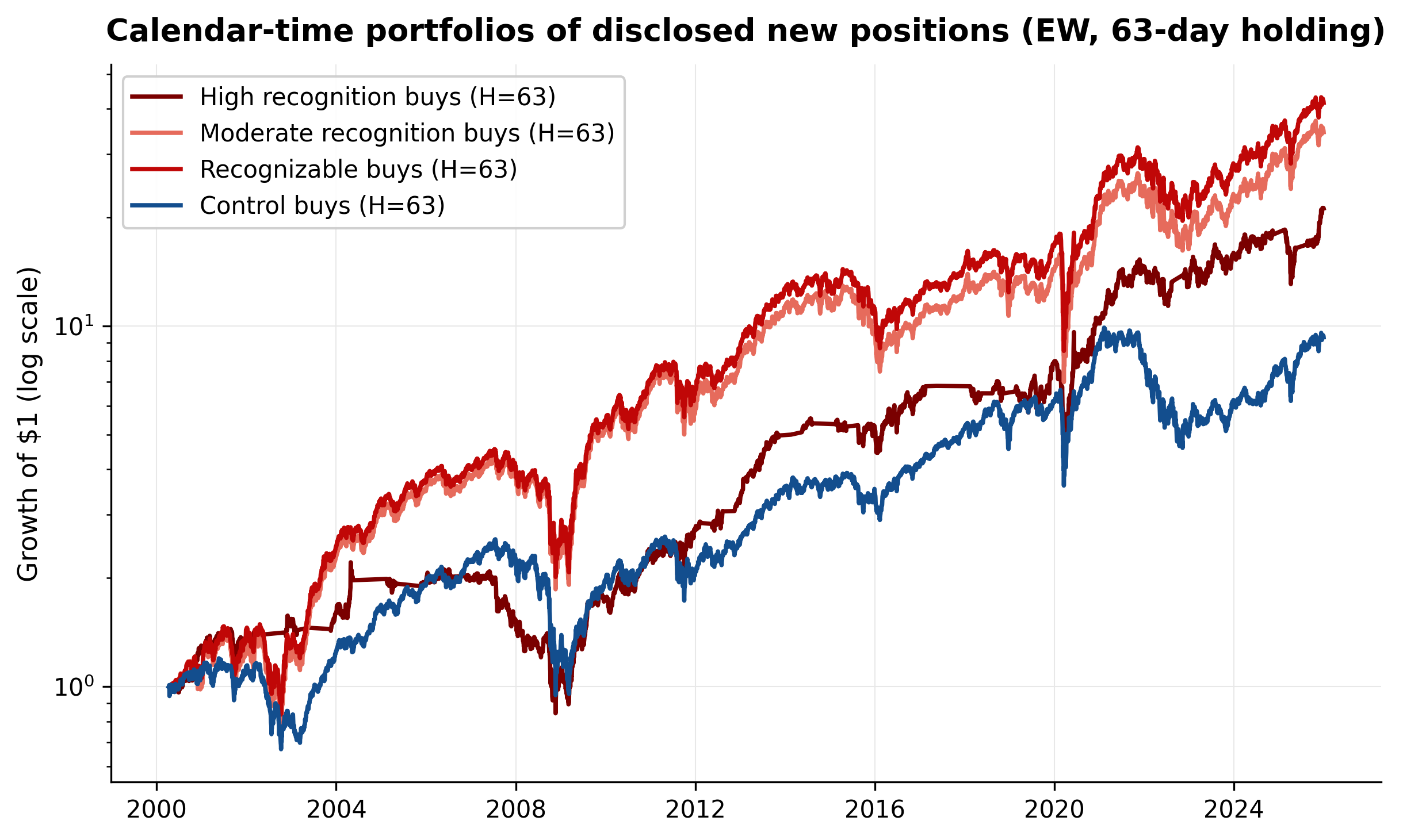}
\caption{Growth of one dollar in the calendar-time buy portfolios}
{\footnotesize\noindent\textit{Note.} Growth of one dollar in the calendar-time buy portfolios, 63-day holding period, equal-weighted, log scale. The high-recognition portfolio's flat stretches are periods with fewer than five open positions, when the portfolio is out of the market.\par}
\end{figure}

\section{Trade-level results}

The main text presented the number and size of trades after high-recognition disclosures and the split of the trading response by stock characteristic (Section VI). Here we report the same outcomes for every recognition tier, the specification ladder, an alternative baseline window, the wild-cluster bootstrap, the retail order imbalance, and the splits. The results can be read from Table IA.XXI to Table IA.XXVII and Figure IA.6 to Figure IA.8. The increase in trading is largest in the high-recognition tier; the retail count does not depend on the specification or the baseline, while the total count is significant only once the pre-window control enters; and the response does not concentrate where information is costly, which is consistent with the main text.

\begin{table}[htbp]
\centering
\caption{Number of trades by recognition tier}
\setlength{\tabcolsep}{4pt}
\begin{threeparttable}
\small
\begin{tabular*}{\linewidth}{@{\extracolsep{\fill}}lccccc}
\toprule
 & \multicolumn{5}{c}{Window} \\
\cmidrule(l{3pt}){2-6}
 & [$-$5,$-$1] & [0,1] & [0,3] & [0,5] & [0,10] \\
\midrule
\multicolumn{6}{l}{\textit{Panel A. High-recognition purchases vs. control purchases}}\\
\addlinespace[3pt]
\multicolumn{6}{l}{\textit{Total trades}}\\
\quad Recognizable & $-$3.66$^{**}$ & 2.43 & 2.69$^{*}$ & 2.03 & 2.07 \\
\quad  & (0.037) & (0.158) & (0.089) & (0.237) & (0.197) \\[1pt]
\quad Top half & $-$3.45$^{*}$ & 3.09$^{*}$ & 2.76$^{**}$ & 2.13 & 2.10 \\
\quad  & (0.080) & (0.083) & (0.039) & (0.146) & (0.194) \\[1pt]
\quad Bottom half & $-$3.75 & $-$0.21 & 1.55 & 1.16 & 1.96 \\
\quad  & (0.105) & (0.941) & (0.560) & (0.681) & (0.436) \\[1pt]
\quad High-recognition & $-$5.97$^{**}$ & 7.24$^{**}$ & 4.50 & 2.64 & 1.76 \\
\quad  & (0.028) & (0.027) & (0.139) & (0.324) & (0.547) \\[1pt]
\quad Moderate-recognition & $-$3.00 & 0.97 & 1.95 & 1.52 & 1.94 \\
\quad  & (0.112) & (0.557) & (0.219) & (0.402) & (0.256) \\[1pt]
\addlinespace[3pt]
\multicolumn{6}{l}{\textit{Non-retail trades}}\\
\quad Recognizable & $-$3.64$^{**}$ & 2.27 & 2.57$^{*}$ & 1.88 & 1.93 \\
\quad  & (0.039) & (0.181) & (0.096) & (0.262) & (0.221) \\[1pt]
\quad Top half & $-$3.42$^{*}$ & 2.86$^{*}$ & 2.60$^{**}$ & 1.97 & 1.92 \\
\quad  & (0.081) & (0.100) & (0.045) & (0.171) & (0.231) \\[1pt]
\quad Bottom half & $-$3.74 & $-$0.30 & 1.49 & 1.03 & 1.87 \\
\quad  & (0.108) & (0.914) & (0.575) & (0.715) & (0.457) \\[1pt]
\quad High-recognition & $-$5.97$^{**}$ & 6.71$^{**}$ & 4.00 & 2.22 & 1.42 \\
\quad  & (0.025) & (0.038) & (0.188) & (0.406) & (0.625) \\[1pt]
\quad Moderate-recognition & $-$3.00 & 0.88 & 1.92 & 1.44 & 1.86 \\
\quad  & (0.114) & (0.593) & (0.222) & (0.423) & (0.271) \\[1pt]
\addlinespace[3pt]
\multicolumn{6}{l}{\textit{Retail trades}}\\
\quad Recognizable & $-$2.86 & 4.16$^{*}$ & 3.80$^{*}$ & 3.24 & 3.34$^{*}$ \\
\quad  & (0.112) & (0.071) & (0.089) & (0.148) & (0.100) \\[1pt]
\quad Top half & $-$3.13 & 5.60$^{**}$ & 4.78$^{**}$ & 4.14$^{**}$ & 4.39$^{**}$ \\
\quad  & (0.140) & (0.041) & (0.029) & (0.042) & (0.017) \\[1pt]
\quad Bottom half & $-$2.66 & 0.15 & 1.11 & 1.29 & 1.92 \\
\quad  & (0.232) & (0.956) & (0.683) & (0.656) & (0.474) \\[1pt]
\quad High-recognition & $-$4.41 & 15.12$^{***}$ & 12.30$^{***}$ & 8.98$^{**}$ & 6.55$^{*}$ \\
\quad  & (0.209) & (0.001) & (0.002) & (0.012) & (0.080) \\[1pt]
\quad Moderate-recognition & $-$2.20 & 1.29 & 1.62 & 1.60 & 2.32 \\
\quad  & (0.259) & (0.483) & (0.389) & (0.442) & (0.234) \\[1pt]
\bottomrule
\end{tabular*}
\end{threeparttable}
\end{table}

\begin{table}[htbp]
\ContinuedFloat
\centering
\caption[]{Number of trades by recognition tier (continued)}
\setlength{\tabcolsep}{4pt}
\begin{threeparttable}
\small
\begin{tabular*}{\linewidth}{@{\extracolsep{\fill}}lccccc}
\toprule
 & \multicolumn{5}{c}{Window} \\
\cmidrule(l{3pt}){2-6}
 & [$-$5,$-$1] & [0,1] & [0,3] & [0,5] & [0,10] \\
\midrule
\multicolumn{6}{l}{\textit{Panel B. Exits vs. control exits}}\\
\addlinespace[3pt]
\multicolumn{6}{l}{\textit{Total trades}}\\
\quad Recognizable & $-$2.39 & $-$0.07 & $-$1.34 & $-$1.57 & $-$1.28 \\
\quad  & (0.183) & (0.967) & (0.397) & (0.324) & (0.383) \\[1pt]
\quad Top half & $-$1.43 & 0.47 & $-$1.03 & $-$2.01 & $-$2.19 \\
\quad  & (0.510) & (0.828) & (0.591) & (0.275) & (0.216) \\[1pt]
\quad Bottom half & $-$2.68 & $-$0.80 & $-$1.04 & $-$0.90 & $-$0.44 \\
\quad  & (0.256) & (0.697) & (0.548) & (0.638) & (0.815) \\[1pt]
\quad High-recognition & $-$1.55 & 6.36$^{*}$ & 4.23 & 2.19 & 3.73$^{*}$ \\
\quad  & (0.663) & (0.089) & (0.145) & (0.420) & (0.091) \\[1pt]
\quad Moderate-recognition & $-$2.51 & $-$1.33 & $-$2.28 & $-$2.42 & $-$2.23 \\
\quad  & (0.141) & (0.462) & (0.173) & (0.170) & (0.146) \\[1pt]
\addlinespace[3pt]
\multicolumn{6}{l}{\textit{Non-retail trades}}\\
\quad Recognizable & $-$2.40 & $-$0.13 & $-$1.37 & $-$1.55 & $-$1.27 \\
\quad  & (0.178) & (0.942) & (0.386) & (0.330) & (0.392) \\[1pt]
\quad Top half & $-$1.43 & 0.37 & $-$1.11 & $-$2.06 & $-$2.23 \\
\quad  & (0.513) & (0.863) & (0.558) & (0.262) & (0.211) \\[1pt]
\quad Bottom half & $-$2.67 & $-$0.78 & $-$0.98 & $-$0.79 & $-$0.36 \\
\quad  & (0.253) & (0.706) & (0.576) & (0.682) & (0.849) \\[1pt]
\quad High-recognition & $-$1.33 & 6.02 & 3.95 & 1.97 & 3.63 \\
\quad  & (0.707) & (0.106) & (0.174) & (0.470) & (0.102) \\[1pt]
\quad Moderate-recognition & $-$2.57 & $-$1.33 & $-$2.27 & $-$2.37 & $-$2.20 \\
\quad  & (0.130) & (0.469) & (0.179) & (0.183) & (0.157) \\[1pt]
\addlinespace[3pt]
\multicolumn{6}{l}{\textit{Retail trades}}\\
\quad Recognizable & $-$1.29 & 1.12 & $-$0.12 & $-$0.38 & $-$0.59 \\
\quad  & (0.569) & (0.554) & (0.944) & (0.821) & (0.692) \\[1pt]
\quad Top half & $-$0.64 & 2.19 & 0.89 & 0.19 & $-$1.06 \\
\quad  & (0.801) & (0.347) & (0.664) & (0.922) & (0.541) \\[1pt]
\quad Bottom half & $-$1.47 & $-$0.34 & $-$0.78 & $-$0.98 & $-$0.27 \\
\quad  & (0.595) & (0.873) & (0.716) & (0.663) & (0.901) \\[1pt]
\quad High-recognition & $-$4.84 & 10.37$^{***}$ & 7.08$^{**}$ & 5.24 & 5.31$^{*}$ \\
\quad  & (0.313) & (0.010) & (0.040) & (0.110) & (0.059) \\[1pt]
\quad Moderate-recognition & $-$0.56 & $-$0.57 & $-$1.30 & $-$1.49 & $-$1.59 \\
\quad  & (0.793) & (0.728) & (0.433) & (0.380) & (0.287) \\[1pt]
\bottomrule
\end{tabular*}
\begin{tablenotes}[flushleft]\footnotesize
\item \textit{Note.} Treated tier minus control, estimated separately for purchases (Panel A) and complete exits (Panel B): the coefficient on the tier indicator in a regression of the outcome on the paper's stock and fund controls, stock and year-quarter fixed effects, and the event's own pre-window value of the outcome, with p-values from manager-clustered standard errors in parentheses. Each outcome is net of the event's own [-15,-6] baseline, averaged over the window, and scaled by 100. Sample 2015 to 2025. $^{*}p<.10$, $^{**}p<.05$, $^{***}p<.01$. Trade counts are in log points (approximately percent).
\end{tablenotes}
\end{threeparttable}
\end{table}

\begin{table}[htbp]
\centering
\caption{Size of trades by recognition tier}
\setlength{\tabcolsep}{4pt}
\begin{threeparttable}
\small
\begin{tabular*}{\linewidth}{@{\extracolsep{\fill}}lccccc}
\toprule
 & \multicolumn{5}{c}{Window} \\
\cmidrule(l{3pt}){2-6}
 & [$-$5,$-$1] & [0,1] & [0,3] & [0,5] & [0,10] \\
\midrule
\multicolumn{6}{l}{\textit{Panel A. High-recognition purchases vs. control purchases}}\\
\addlinespace[3pt]
\multicolumn{6}{l}{\textit{Average trade size}}\\
\quad Recognizable & $-$0.47 & 0.08 & 0.76 & 0.92 & 1.00 \\
\quad  & (0.450) & (0.937) & (0.385) & (0.257) & (0.220) \\[1pt]
\quad Top half & $-$0.63 & 0.80 & 1.55 & 1.83$^{**}$ & 2.01$^{**}$ \\
\quad  & (0.325) & (0.552) & (0.167) & (0.043) & (0.035) \\[1pt]
\quad Bottom half & 0.02 & $-$0.88 & 0.07 & 0.22 & 0.46 \\
\quad  & (0.987) & (0.494) & (0.949) & (0.826) & (0.636) \\[1pt]
\quad High-recognition & $-$0.37 & $-$1.37 & $-$0.60 & 0.38 & 0.65 \\
\quad  & (0.721) & (0.547) & (0.758) & (0.813) & (0.688) \\[1pt]
\quad Moderate-recognition & $-$0.43 & 0.32 & 1.17 & 1.17 & 1.30 \\
\quad  & (0.510) & (0.760) & (0.206) & (0.199) & (0.138) \\[1pt]
\addlinespace[3pt]
\multicolumn{6}{l}{\textit{Average retail trade size}}\\
\quad Recognizable & 0.03 & 0.96 & 0.57 & 0.15 & $-$0.18 \\
\quad  & (0.977) & (0.464) & (0.601) & (0.867) & (0.875) \\[1pt]
\quad Top half & $-$0.75 & 1.42 & 1.07 & 0.31 & 0.21 \\
\quad  & (0.490) & (0.384) & (0.435) & (0.766) & (0.889) \\[1pt]
\quad Bottom half & 1.35 & 0.51 & 0.04 & 0.46 & 0.44 \\
\quad  & (0.471) & (0.763) & (0.980) & (0.731) & (0.744) \\[1pt]
\quad High-recognition & $-$2.65 & $-$1.26 & $-$2.59 & $-$2.68 & $-$0.63 \\
\quad  & (0.218) & (0.697) & (0.353) & (0.246) & (0.772) \\[1pt]
\quad Moderate-recognition & 0.07 & 1.14 & 1.00 & 0.64 & 0.07 \\
\quad  & (0.952) & (0.368) & (0.370) & (0.522) & (0.957) \\[1pt]
\addlinespace[3pt]
\multicolumn{6}{l}{\textit{$\geq$\$50k dollar share}}\\
\quad Recognizable & $-$0.16 & $-$0.17 & 0.20 & 0.24 & 0.13 \\
\quad  & (0.458) & (0.703) & (0.593) & (0.516) & (0.707) \\[1pt]
\quad Top half & $-$0.19 & 0.27 & 0.63 & 0.62 & 0.47 \\
\quad  & (0.502) & (0.584) & (0.163) & (0.109) & (0.223) \\[1pt]
\quad Bottom half & $-$0.08 & $-$0.58 & $-$0.10 & $-$0.07 & $-$0.11 \\
\quad  & (0.840) & (0.362) & (0.851) & (0.890) & (0.787) \\[1pt]
\quad High-recognition & $-$0.16 & 0.02 & 0.31 & 0.53 & 0.35 \\
\quad  & (0.687) & (0.971) & (0.532) & (0.110) & (0.424) \\[1pt]
\quad Moderate-recognition & $-$0.14 & $-$0.21 & 0.27 & 0.28 & 0.17 \\
\quad  & (0.549) & (0.674) & (0.525) & (0.510) & (0.666) \\[1pt]
\bottomrule
\end{tabular*}
\end{threeparttable}
\end{table}

\begin{table}[htbp]
\ContinuedFloat
\centering
\caption[]{Size of trades by recognition tier (continued)}
\setlength{\tabcolsep}{4pt}
\begin{threeparttable}
\small
\begin{tabular*}{\linewidth}{@{\extracolsep{\fill}}lccccc}
\toprule
 & \multicolumn{5}{c}{Window} \\
\cmidrule(l{3pt}){2-6}
 & [$-$5,$-$1] & [0,1] & [0,3] & [0,5] & [0,10] \\
\midrule
\multicolumn{6}{l}{\textit{Panel B. Exits vs. control exits}}\\
\addlinespace[3pt]
\multicolumn{6}{l}{\textit{Average trade size}}\\
\quad Recognizable & $-$0.17 & 0.62 & 0.44 & 0.72 & 1.25 \\
\quad  & (0.797) & (0.532) & (0.566) & (0.334) & (0.115) \\[1pt]
\quad Top half & $-$0.49 & $-$0.50 & $-$0.39 & 0.02 & 0.47 \\
\quad  & (0.508) & (0.656) & (0.656) & (0.981) & (0.573) \\[1pt]
\quad Bottom half & $-$0.19 & 1.94$^{*}$ & 1.59$^{*}$ & 1.64$^{*}$ & 2.47$^{***}$ \\
\quad  & (0.829) & (0.074) & (0.094) & (0.063) & (0.010) \\[1pt]
\quad High-recognition & $-$0.59 & $-$0.59 & $-$0.77 & $-$0.37 & 0.72 \\
\quad  & (0.683) & (0.852) & (0.734) & (0.859) & (0.707) \\[1pt]
\quad Moderate-recognition & $-$0.06 & 0.83 & 0.70 & 0.90 & 1.47$^{*}$ \\
\quad  & (0.923) & (0.369) & (0.341) & (0.228) & (0.073) \\[1pt]
\addlinespace[3pt]
\multicolumn{6}{l}{\textit{Average retail trade size}}\\
\quad Recognizable & $-$0.74 & $-$0.28 & $-$0.42 & $-$0.35 & 0.24 \\
\quad  & (0.472) & (0.856) & (0.718) & (0.767) & (0.840) \\[1pt]
\quad Top half & $-$1.02 & $-$0.86 & $-$0.90 & $-$1.28 & $-$0.67 \\
\quad  & (0.445) & (0.598) & (0.442) & (0.256) & (0.529) \\[1pt]
\quad Bottom half & 0.02 & $-$1.44 & $-$1.00 & $-$0.64 & $-$0.15 \\
\quad  & (0.988) & (0.568) & (0.586) & (0.725) & (0.941) \\[1pt]
\quad High-recognition & 1.10 & $-$4.83 & $-$1.81 & $-$2.73 & $-$0.50 \\
\quad  & (0.630) & (0.163) & (0.444) & (0.161) & (0.698) \\[1pt]
\quad Moderate-recognition & $-$0.90 & 0.26 & $-$0.24 & $-$0.26 & 0.14 \\
\quad  & (0.367) & (0.871) & (0.848) & (0.854) & (0.920) \\[1pt]
\addlinespace[3pt]
\multicolumn{6}{l}{\textit{$\geq$\$50k dollar share}}\\
\quad Recognizable & 0.09 & 0.23 & 0.17 & 0.32 & 0.46 \\
\quad  & (0.751) & (0.571) & (0.609) & (0.278) & (0.136) \\[1pt]
\quad Top half & $-$0.21 & 0.07 & 0.15 & 0.40 & 0.50 \\
\quad  & (0.505) & (0.897) & (0.704) & (0.278) & (0.194) \\[1pt]
\quad Bottom half & 0.29 & 0.46 & 0.24 & 0.33 & 0.60$^{*}$ \\
\quad  & (0.417) & (0.353) & (0.555) & (0.332) & (0.079) \\[1pt]
\quad High-recognition & $-$0.53 & 0.15 & $-$0.12 & 0.09 & 0.10 \\
\quad  & (0.411) & (0.912) & (0.903) & (0.922) & (0.905) \\[1pt]
\quad Moderate-recognition & 0.19 & 0.25 & 0.25 & 0.38 & 0.57$^{*}$ \\
\quad  & (0.519) & (0.536) & (0.423) & (0.204) & (0.060) \\[1pt]
\bottomrule
\end{tabular*}
\begin{tablenotes}[flushleft]\footnotesize
\item \textit{Note.} Treated tier minus control, estimated separately for purchases (Panel A) and complete exits (Panel B): the coefficient on the tier indicator in a regression of the outcome on the paper's stock and fund controls, stock and year-quarter fixed effects, and the event's own pre-window value of the outcome, with p-values from manager-clustered standard errors in parentheses. Each outcome is net of the event's own [-15,-6] baseline, averaged over the window, and scaled by 100. Sample 2015 to 2025. $^{*}p<.10$, $^{**}p<.05$, $^{***}p<.01$. Trade sizes are in log points; the dollar share of trades of at least \$50,000 is in percentage points.
\end{tablenotes}
\end{threeparttable}
\end{table}

\begin{table}[htbp]
\centering
\caption{Retail order imbalance by recognition tier}
\setlength{\tabcolsep}{4pt}
\begin{threeparttable}
\small
\begin{tabular*}{\linewidth}{@{\extracolsep{\fill}}lccccc}
\toprule
 & \multicolumn{5}{c}{Window} \\
\cmidrule(l{3pt}){2-6}
 & [$-$5,$-$1] & [0,1] & [0,3] & [0,5] & [0,10] \\
\midrule
\multicolumn{6}{l}{\textit{Panel A. High-recognition purchases vs. control purchases}}\\
\addlinespace[3pt]
\multicolumn{6}{l}{\textit{Retail order imbalance}}\\
\quad Recognizable & $-$0.84 & 0.81 & 0.47 & 0.54 & 0.60 \\
\quad  & (0.100) & (0.297) & (0.467) & (0.386) & (0.247) \\[1pt]
\quad Top half & $-$0.70 & 1.63 & 1.38$^{*}$ & 1.38$^{**}$ & 1.17$^{*}$ \\
\quad  & (0.216) & (0.104) & (0.056) & (0.039) & (0.060) \\[1pt]
\quad Bottom half & $-$1.03 & $-$0.04 & $-$0.49 & $-$0.26 & $-$0.02 \\
\quad  & (0.124) & (0.964) & (0.593) & (0.721) & (0.970) \\[1pt]
\quad High-recognition & $-$0.47 & 3.95$^{**}$ & 2.61$^{***}$ & 2.57$^{***}$ & 2.61$^{***}$ \\
\quad  & (0.642) & (0.013) & (0.003) & (0.001) & (<0.001) \\[1pt]
\quad Moderate-recognition & $-$0.99$^{*}$ & 0.17 & 0.02 & 0.10 & 0.17 \\
\quad  & (0.054) & (0.829) & (0.977) & (0.871) & (0.724) \\[1pt]
\addlinespace[3pt]
\multicolumn{6}{l}{\textit{Panel B. Exits vs. control exits}}\\
\addlinespace[3pt]
\multicolumn{6}{l}{\textit{Retail order imbalance}}\\
\quad Recognizable & $-$0.01 & $-$0.66 & $-$0.75 & $-$0.84 & $-$0.34 \\
\quad  & (0.983) & (0.272) & (0.130) & (0.101) & (0.445) \\[1pt]
\quad Top half & $-$0.16 & $-$0.43 & $-$0.83 & $-$1.21$^{*}$ & $-$0.74 \\
\quad  & (0.845) & (0.635) & (0.257) & (0.089) & (0.187) \\[1pt]
\quad Bottom half & 0.08 & $-$1.05 & $-$0.89 & $-$0.81 & $-$0.09 \\
\quad  & (0.918) & (0.263) & (0.136) & (0.189) & (0.878) \\[1pt]
\quad High-recognition & $-$1.67$^{**}$ & $-$1.75 & $-$1.37 & $-$1.94$^{**}$ & $-$1.85$^{**}$ \\
\quad  & (0.036) & (0.141) & (0.203) & (0.037) & (0.024) \\[1pt]
\quad Moderate-recognition & 0.14 & $-$0.25 & $-$0.49 & $-$0.56 & 0.01 \\
\quad  & (0.853) & (0.683) & (0.333) & (0.256) & (0.974) \\[1pt]
\addlinespace[3pt]
\multicolumn{6}{l}{\textit{Panel C. Small-cluster inference, High-recognition $-$ Control}}\\
\quad WCR p, purchases & 0.531 & 0.003 & 0.004 & 0.003 & 0.002 \\
\quad WCU p, purchases & 0.513 & <0.001 & <0.001 & <0.001 & <0.001 \\
\quad WCR p, exits & 0.016 & 0.085 & 0.114 & 0.032 & 0.013 \\
\quad WCU p, exits & <0.001 & 0.031 & 0.044 & 0.001 & <0.001 \\
\bottomrule
\end{tabular*}
\begin{tablenotes}[flushleft]\footnotesize
\item \textit{Note.} Treated tier minus control, estimated separately for purchases (Panel A) and complete exits (Panel B): the coefficient on the tier indicator in a regression of the outcome on the paper's stock and fund controls, stock and year-quarter fixed effects, and the event's own pre-window value of the outcome, with p-values from manager-clustered standard errors in parentheses. Each outcome is net of the event's own [-15,-6] baseline, averaged over the window, and scaled by 100. Sample 2015 to 2025. $^{*}p<.10$, $^{**}p<.05$, $^{***}p<.01$. The retail order imbalance is retail buy minus retail sell dollar volume over their sum, in percentage points. Panel C re-tests the high-recognition contrast with the restricted (WCR) and unrestricted (WCU) wild-cluster bootstrap (Webb six-point weights, B = 9,999).
\end{tablenotes}
\end{threeparttable}
\end{table}

\begin{table}[htbp]
\centering
\caption{Specification ladder for the trading response to high-recognition purchases}
\setlength{\tabcolsep}{4pt}
\begin{threeparttable}
\small
\begin{tabular*}{\linewidth}{@{\extracolsep{\fill}}lccccc}
\toprule
 & \multicolumn{5}{c}{Window} \\
\cmidrule(l{3pt}){2-6}
 & [$-$5,$-$1] & [0,1] & [0,3] & [0,5] & [0,10] \\
\midrule
\multicolumn{6}{l}{\textit{Total trades}}\\
\quad (1) Raw & $-$6.44$^{***}$ & 5.60$^{*}$ & 4.26 & 2.44 & $-$1.00 \\
\quad  & (<0.001) & (0.054) & (0.149) & (0.353) & (0.667) \\[1pt]
\quad (2) +Stock ctrls & $-$4.25$^{***}$ & 9.52$^{***}$ & 7.46$^{***}$ & 5.67$^{**}$ & 2.18 \\
\quad  & (0.009) & (<0.001) & (0.006) & (0.016) & (0.328) \\[1pt]
\quad (3) +Stock FE & $-$5.08$^{*}$ & 2.58 & 0.39 & $-$1.34 & $-$3.52 \\
\quad  & (0.079) & (0.510) & (0.913) & (0.679) & (0.266) \\[1pt]
\quad (4) +Time FE & $-$5.97$^{**}$ & 3.18 & 0.92 & $-$0.70 & $-$1.52 \\
\quad  & (0.028) & (0.394) & (0.795) & (0.828) & (0.639) \\[1pt]
\quad (5) +Pre-window & $-$5.97$^{**}$ & 7.24$^{**}$ & 4.50 & 2.64 & 1.76 \\
\quad  & (0.028) & (0.027) & (0.139) & (0.324) & (0.547) \\[1pt]
\addlinespace[3pt]
\multicolumn{6}{l}{\textit{Non-retail trades}}\\
\quad (1) Raw & $-$6.52$^{***}$ & 4.91$^{*}$ & 3.55 & 1.75 & $-$1.62 \\
\quad  & (<0.001) & (0.078) & (0.211) & (0.491) & (0.480) \\[1pt]
\quad (2) +Stock ctrls & $-$4.31$^{***}$ & 8.92$^{***}$ & 6.74$^{***}$ & 5.00$^{**}$ & 1.69 \\
\quad  & (0.008) & (<0.001) & (0.008) & (0.025) & (0.437) \\[1pt]
\quad (3) +Stock FE & $-$5.06$^{*}$ & 2.04 & $-$0.11 & $-$1.78 & $-$3.93 \\
\quad  & (0.076) & (0.595) & (0.974) & (0.580) & (0.215) \\[1pt]
\quad (4) +Time FE & $-$5.97$^{**}$ & 2.66 & 0.44 & $-$1.10 & $-$1.86 \\
\quad  & (0.025) & (0.467) & (0.902) & (0.731) & (0.562) \\[1pt]
\quad (5) +Pre-window & $-$5.97$^{**}$ & 6.71$^{**}$ & 4.00 & 2.22 & 1.42 \\
\quad  & (0.025) & (0.038) & (0.188) & (0.406) & (0.625) \\[1pt]
\bottomrule
\end{tabular*}
\end{threeparttable}
\end{table}

\begin{table}[htbp]
\ContinuedFloat
\centering
\caption[]{Specification ladder for the trading response to high-recognition purchases (continued)}
\setlength{\tabcolsep}{4pt}
\begin{threeparttable}
\small
\begin{tabular*}{\linewidth}{@{\extracolsep{\fill}}lccccc}
\toprule
 & \multicolumn{5}{c}{Window} \\
\cmidrule(l{3pt}){2-6}
 & [$-$5,$-$1] & [0,1] & [0,3] & [0,5] & [0,10] \\
\midrule
\multicolumn{6}{l}{\textit{Retail trades}}\\
\quad (1) Raw & $-$5.55$^{***}$ & 16.10$^{***}$ & 14.25$^{***}$ & 11.95$^{***}$ & 7.63$^{**}$ \\
\quad  & (0.003) & (0.002) & (0.002) & (0.003) & (0.011) \\[1pt]
\quad (2) +Stock ctrls & $-$3.22 & 19.56$^{***}$ & 16.34$^{***}$ & 13.36$^{***}$ & 8.12$^{***}$ \\
\quad  & (0.117) & (<0.001) & (<0.001) & (0.002) & (0.007) \\[1pt]
\quad (3) +Stock FE & $-$3.81 & 12.56$^{**}$ & 9.90$^{**}$ & 6.36 & 3.03 \\
\quad  & (0.310) & (0.028) & (0.044) & (0.171) & (0.471) \\[1pt]
\quad (4) +Time FE & $-$4.41 & 12.06$^{**}$ & 9.58$^{**}$ & 6.42 & 4.17 \\
\quad  & (0.209) & (0.022) & (0.031) & (0.128) & (0.324) \\[1pt]
\quad (5) +Pre-window & $-$4.41 & 15.12$^{***}$ & 12.30$^{***}$ & 8.98$^{**}$ & 6.55$^{*}$ \\
\quad  & (0.209) & (0.001) & (0.002) & (0.012) & (0.080) \\[1pt]
\addlinespace[3pt]
\multicolumn{6}{l}{\textit{Retail order imbalance}}\\
\quad (1) Raw & $-$0.04 & 3.99$^{***}$ & 3.34$^{***}$ & 3.15$^{***}$ & 2.47$^{***}$ \\
\quad  & (0.910) & (<0.001) & (<0.001) & (<0.001) & (<0.001) \\[1pt]
\quad (2) +Stock ctrls & $-$0.62 & 2.97$^{***}$ & 2.74$^{***}$ & 2.63$^{***}$ & 2.37$^{***}$ \\
\quad  & (0.291) & (0.001) & (0.005) & (0.006) & (<0.001) \\[1pt]
\quad (3) +Stock FE & $-$0.58 & 4.33$^{***}$ & 2.18$^{**}$ & 2.24$^{**}$ & 2.04$^{**}$ \\
\quad  & (0.606) & (0.009) & (0.047) & (0.017) & (0.018) \\[1pt]
\quad (4) +Time FE & $-$0.47 & 3.78$^{**}$ & 2.44$^{**}$ & 2.43$^{***}$ & 2.50$^{***}$ \\
\quad  & (0.642) & (0.039) & (0.022) & (0.009) & (0.003) \\[1pt]
\quad (5) +Pre-window & $-$0.47 & 3.95$^{**}$ & 2.61$^{***}$ & 2.57$^{***}$ & 2.61$^{***}$ \\
\quad  & (0.642) & (0.013) & (0.003) & (0.001) & (<0.001) \\[1pt]
\bottomrule
\end{tabular*}
\begin{tablenotes}[flushleft]\footnotesize
\item \textit{Note.} High-recognition purchases against control purchases. Column (1) is the raw difference; (2) adds the stock controls; (3) adds stock fixed effects and the fund controls; (4) adds year-quarter fixed effects; (5) adds the event's own pre-window value of the outcome, which is the specification used in the main text. p-values from manager-clustered standard errors in parentheses. $^{*}p<.10$, $^{**}p<.05$, $^{***}p<.01$.
\end{tablenotes}
\end{threeparttable}
\end{table}

\begin{table}[htbp]
\centering
\caption{Alternative baseline window}
\setlength{\tabcolsep}{4pt}
\begin{threeparttable}
\small
\begin{tabular*}{\linewidth}{@{\extracolsep{\fill}}lccccc}
\toprule
 & \multicolumn{5}{c}{Window} \\
\cmidrule(l{3pt}){2-6}
 & [$-$5,$-$1] & [0,1] & [0,3] & [0,5] & [0,10] \\
\midrule
\multicolumn{6}{l}{\textit{Panel A. Purchases}}\\
\addlinespace[3pt]
\multicolumn{6}{l}{\textit{Total trades}}\\
\quad Baseline [$-$15,$-$6] & $-$5.97$^{**}$ & 7.24$^{**}$ & 4.50 & 2.64 & 1.76 \\
\quad  & (0.028) & (0.027) & (0.139) & (0.324) & (0.547) \\[1pt]
\quad Baseline [$-$15,$-$11] & $-$2.63 & 8.20$^{**}$ & 5.72$^{*}$ & 4.03 & 3.18 \\
\quad  & (0.480) & (0.013) & (0.065) & (0.143) & (0.279) \\[1pt]
\addlinespace[3pt]
\multicolumn{6}{l}{\textit{Non-retail trades}}\\
\quad Baseline [$-$15,$-$6] & $-$5.97$^{**}$ & 6.71$^{**}$ & 4.00 & 2.22 & 1.42 \\
\quad  & (0.025) & (0.038) & (0.188) & (0.406) & (0.625) \\[1pt]
\quad Baseline [$-$15,$-$11] & $-$2.55 & 7.72$^{**}$ & 5.28$^{*}$ & 3.67 & 2.88 \\
\quad  & (0.493) & (0.017) & (0.087) & (0.181) & (0.326) \\[1pt]
\addlinespace[3pt]
\multicolumn{6}{l}{\textit{Retail trades}}\\
\quad Baseline [$-$15,$-$6] & $-$4.41 & 15.12$^{***}$ & 12.30$^{***}$ & 8.98$^{**}$ & 6.55$^{*}$ \\
\quad  & (0.209) & (0.001) & (0.002) & (0.012) & (0.080) \\[1pt]
\quad Baseline [$-$15,$-$11] & $-$1.84 & 15.51$^{***}$ & 12.87$^{***}$ & 9.55$^{***}$ & 7.13$^{*}$ \\
\quad  & (0.653) & (0.001) & (0.001) & (0.009) & (0.051) \\[1pt]
\addlinespace[3pt]
\multicolumn{6}{l}{\textit{Average trade size}}\\
\quad Baseline [$-$15,$-$6] & $-$0.37 & $-$1.37 & $-$0.60 & 0.38 & 0.65 \\
\quad  & (0.721) & (0.547) & (0.758) & (0.813) & (0.688) \\[1pt]
\quad Baseline [$-$15,$-$11] & $-$0.25 & $-$0.66 & 0.01 & 0.90 & 1.10 \\
\quad  & (0.867) & (0.777) & (0.995) & (0.580) & (0.483) \\[1pt]
\addlinespace[3pt]
\multicolumn{6}{l}{\textit{$\geq$\$50k dollar share}}\\
\quad Baseline [$-$15,$-$6] & $-$0.16 & 0.02 & 0.31 & 0.53 & 0.35 \\
\quad  & (0.687) & (0.971) & (0.532) & (0.110) & (0.424) \\[1pt]
\quad Baseline [$-$15,$-$11] & $-$0.07 & 0.33 & 0.63 & 0.77$^{**}$ & 0.59 \\
\quad  & (0.899) & (0.609) & (0.218) & (0.023) & (0.171) \\[1pt]
\addlinespace[3pt]
\multicolumn{6}{l}{\textit{Retail order imbalance}}\\
\quad Baseline [$-$15,$-$6] & $-$0.47 & 3.95$^{**}$ & 2.61$^{***}$ & 2.57$^{***}$ & 2.61$^{***}$ \\
\quad  & (0.642) & (0.013) & (0.003) & (0.001) & (<0.001) \\[1pt]
\quad Baseline [$-$15,$-$11] & 1.37 & 4.68$^{***}$ & 3.28$^{***}$ & 3.10$^{***}$ & 3.11$^{***}$ \\
\quad  & (0.167) & (0.003) & (0.002) & (0.002) & (<0.001) \\[1pt]
\bottomrule
\end{tabular*}
\end{threeparttable}
\end{table}

\begin{table}[htbp]
\ContinuedFloat
\centering
\caption[]{Alternative baseline window (continued)}
\setlength{\tabcolsep}{4pt}
\begin{threeparttable}
\small
\begin{tabular*}{\linewidth}{@{\extracolsep{\fill}}lccccc}
\toprule
 & \multicolumn{5}{c}{Window} \\
\cmidrule(l{3pt}){2-6}
 & [$-$5,$-$1] & [0,1] & [0,3] & [0,5] & [0,10] \\
\midrule
\multicolumn{6}{l}{\textit{Panel B. Exits}}\\
\addlinespace[3pt]
\multicolumn{6}{l}{\textit{Total trades}}\\
\quad Baseline [$-$15,$-$6] & $-$1.55 & 6.36$^{*}$ & 4.23 & 2.19 & 3.73$^{*}$ \\
\quad  & (0.663) & (0.089) & (0.145) & (0.420) & (0.091) \\[1pt]
\quad Baseline [$-$15,$-$11] & $-$2.54 & 5.92$^{*}$ & 3.59 & 1.52 & 2.89 \\
\quad  & (0.560) & (0.076) & (0.171) & (0.527) & (0.136) \\[1pt]
\addlinespace[3pt]
\multicolumn{6}{l}{\textit{Non-retail trades}}\\
\quad Baseline [$-$15,$-$6] & $-$1.33 & 6.02 & 3.95 & 1.97 & 3.63 \\
\quad  & (0.707) & (0.106) & (0.174) & (0.470) & (0.102) \\[1pt]
\quad Baseline [$-$15,$-$11] & $-$2.42 & 5.56$^{*}$ & 3.32 & 1.28 & 2.77 \\
\quad  & (0.572) & (0.096) & (0.207) & (0.594) & (0.155) \\[1pt]
\addlinespace[3pt]
\multicolumn{6}{l}{\textit{Retail trades}}\\
\quad Baseline [$-$15,$-$6] & $-$4.84 & 10.37$^{***}$ & 7.08$^{**}$ & 5.24 & 5.31$^{*}$ \\
\quad  & (0.313) & (0.010) & (0.040) & (0.110) & (0.059) \\[1pt]
\quad Baseline [$-$15,$-$11] & $-$5.20 & 10.03$^{***}$ & 6.41$^{*}$ & 4.68 & 4.61$^{*}$ \\
\quad  & (0.366) & (0.006) & (0.054) & (0.134) & (0.097) \\[1pt]
\addlinespace[3pt]
\multicolumn{6}{l}{\textit{Average trade size}}\\
\quad Baseline [$-$15,$-$6] & $-$0.59 & $-$0.59 & $-$0.77 & $-$0.37 & 0.72 \\
\quad  & (0.683) & (0.852) & (0.734) & (0.859) & (0.707) \\[1pt]
\quad Baseline [$-$15,$-$11] & $-$0.72 & $-$0.33 & $-$0.48 & $-$0.18 & 0.64 \\
\quad  & (0.720) & (0.918) & (0.829) & (0.928) & (0.711) \\[1pt]
\addlinespace[3pt]
\multicolumn{6}{l}{\textit{$\geq$\$50k dollar share}}\\
\quad Baseline [$-$15,$-$6] & $-$0.53 & 0.15 & $-$0.12 & 0.09 & 0.10 \\
\quad  & (0.411) & (0.912) & (0.903) & (0.922) & (0.905) \\[1pt]
\quad Baseline [$-$15,$-$11] & 0.03 & 0.42 & 0.17 & 0.34 & 0.31 \\
\quad  & (0.976) & (0.758) & (0.855) & (0.679) & (0.671) \\[1pt]
\addlinespace[3pt]
\multicolumn{6}{l}{\textit{Retail order imbalance}}\\
\quad Baseline [$-$15,$-$6] & $-$1.67$^{**}$ & $-$1.75 & $-$1.37 & $-$1.94$^{**}$ & $-$1.85$^{**}$ \\
\quad  & (0.036) & (0.141) & (0.203) & (0.037) & (0.024) \\[1pt]
\quad Baseline [$-$15,$-$11] & $-$1.36 & $-$1.29 & $-$0.87 & $-$1.37 & $-$1.29$^{*}$ \\
\quad  & (0.180) & (0.257) & (0.402) & (0.119) & (0.094) \\[1pt]
\bottomrule
\end{tabular*}
\begin{tablenotes}[flushleft]\footnotesize
\item \textit{Note.} High-recognition minus control under the baseline used in the main text, days [-15,-6] before the filing, and under a shorter baseline that ends further from the filing, days [-15,-11]. Full specification; p-values from manager-clustered standard errors in parentheses. $^{*}p<.10$, $^{**}p<.05$, $^{***}p<.01$.
\end{tablenotes}
\end{threeparttable}
\end{table}

\begin{table}[htbp]
\centering
\caption{Wild-cluster bootstrap for the trading response}
\setlength{\tabcolsep}{4pt}
\begin{threeparttable}
\small
\begin{tabular*}{\linewidth}{@{\extracolsep{\fill}}lccccc}
\toprule
 & \multicolumn{5}{c}{Window} \\
\cmidrule(l{3pt}){2-6}
 & [$-$5,$-$1] & [0,1] & [0,3] & [0,5] & [0,10] \\
\midrule
\multicolumn{6}{l}{\textit{Panel A. Purchases}}\\
\addlinespace[3pt]
\multicolumn{6}{l}{\textit{Total trades}}\\
\quad WCR p & 0.024 & 0.024 & 0.091 & 0.198 & 0.378 \\
\quad WCU p & 0.002 & 0.004 & 0.045 & 0.151 & 0.377 \\
\addlinespace[3pt]
\multicolumn{6}{l}{\textit{Non-retail trades}}\\
\quad WCR p & 0.022 & 0.033 & 0.124 & 0.266 & 0.474 \\
\quad WCU p & <0.001 & 0.006 & 0.078 & 0.224 & 0.474 \\
\addlinespace[3pt]
\multicolumn{6}{l}{\textit{Retail trades}}\\
\quad WCR p & 0.207 & 0.001 & 0.002 & 0.010 & 0.048 \\
\quad WCU p & 0.142 & <0.001 & <0.001 & 0.001 & 0.022 \\
\addlinespace[3pt]
\multicolumn{6}{l}{\textit{Retail order imbalance}}\\
\quad WCR p & 0.531 & 0.003 & 0.004 & 0.003 & 0.002 \\
\quad WCU p & 0.513 & <0.001 & <0.001 & <0.001 & <0.001 \\
\addlinespace[3pt]
\multicolumn{6}{l}{\textit{Average trade size}}\\
\quad WCR p & 0.610 & 0.504 & 0.745 & 0.776 & 0.640 \\
\quad WCU p & 0.605 & 0.457 & 0.724 & 0.780 & 0.641 \\
\addlinespace[3pt]
\multicolumn{6}{l}{\textit{Panel B. Exits}}\\
\addlinespace[3pt]
\multicolumn{6}{l}{\textit{Total trades}}\\
\quad WCR p & 0.562 & 0.073 & 0.118 & 0.326 & 0.093 \\
\quad WCU p & 0.559 & 0.027 & 0.040 & 0.264 & 0.024 \\
\addlinespace[3pt]
\multicolumn{6}{l}{\textit{Non-retail trades}}\\
\quad WCR p & 0.615 & 0.086 & 0.138 & 0.368 & 0.101 \\
\quad WCU p & 0.608 & 0.034 & 0.059 & 0.318 & 0.029 \\
\addlinespace[3pt]
\multicolumn{6}{l}{\textit{Retail trades}}\\
\quad WCR p & 0.209 & 0.010 & 0.031 & 0.068 & 0.049 \\
\quad WCU p & 0.182 & <0.001 & 0.001 & 0.013 & 0.005 \\
\addlinespace[3pt]
\multicolumn{6}{l}{\textit{Retail order imbalance}}\\
\quad WCR p & 0.016 & 0.085 & 0.114 & 0.032 & 0.013 \\
\quad WCU p & <0.001 & 0.031 & 0.044 & 0.001 & <0.001 \\
\addlinespace[3pt]
\multicolumn{6}{l}{\textit{Average trade size}}\\
\quad WCR p & 0.557 & 0.830 & 0.673 & 0.819 & 0.644 \\
\quad WCU p & 0.558 & 0.832 & 0.667 & 0.821 & 0.640 \\
\bottomrule
\end{tabular*}
\begin{tablenotes}[flushleft]\footnotesize
\item \textit{Note.} Restricted (WCR) and unrestricted (WCU) wild-cluster bootstrap p-values (Webb six-point weights, B = 9,999, 29 clusters) for the high-recognition minus control coefficient of the full specification, purchases and exits separately.
\end{tablenotes}
\end{threeparttable}
\end{table}

\begin{table}[htbp]
\centering
\caption{The trading response to high-recognition purchases by stock characteristic}
\setlength{\tabcolsep}{3pt}
\begin{threeparttable}
\footnotesize
\begin{tabularx}{\linewidth}{>{\raggedright\arraybackslash\hangindent=1em}Xcccccc}
\toprule
 & \multicolumn{3}{c}{[0,1]} & \multicolumn{3}{c}{[0,5]} \\
\cmidrule(l{3pt}r{3pt}){2-4}\cmidrule(l{3pt}){5-7}
Characteristic (low / high) & Low & High & Diff. (p) & Low & High & Diff. (p) \\
\midrule
\multicolumn{7}{l}{\textit{Retail trades}}\\
Market capitalization (small / large) & 23.6 & 11.6$^{**}$ & $-$10.5 (0.262) & 12.7 & 6.9$^{*}$ & $-$2.5 (0.747) \\
Analyst coverage (scarce / broad) & 7.2 & 14.6$^{***}$ & 8.6 (0.388) & 1.7 & 9.7$^{**}$ & 12.0 (0.046) \\
Illiquidity (liquid / illiquid) & 10.9$^{*}$ & 21.4$^{*}$ & 12.5 (0.190) & 5.8 & 14.4 & 7.2 (0.362) \\
Institutional ownership (under-held / widely held) & 10.0$^{**}$ & 21.8$^{***}$ & 8.1 (0.319) & 6.1 & 11.4 & 4.3 (0.546) \\
Momentum (weak / strong) & 29.5$^{***}$ & 11.1$^{*}$ & $-$3.4 (0.605) & 16.9$^{*}$ & 7.9 & 1.1 (0.796) \\
Realized drift after day 5 (losers / winners) & 10.7$^{**}$ & 15.2$^{*}$ & 6.9 (0.230) & 7.4 & 7.2 & 4.6 (0.433) \\
\addlinespace[3pt]
\multicolumn{7}{l}{\textit{Total trades}}\\
Market capitalization (small / large) & 6.7 & 5.1 & $-$5.6 (0.362) & $-$0.7 & 1.9 & 0.3 (0.970) \\
Analyst coverage (scarce / broad) & 0.6 & 7.2 & 9.3 (0.191) & $-$4.9 & 3.6 & 11.7 (0.020) \\
Illiquidity (liquid / illiquid) & 4.8 & 2.2 & 2.6 (0.682) & 1.4 & $-$3.7 & $-$0.5 (0.943) \\
Institutional ownership (under-held / widely held) & 4.2 & 13.6$^{**}$ & 9.4 (0.054) & 2.4 & 4.0 & 3.0 (0.559) \\
Momentum (weak / strong) & 16.0$^{**}$ & 6.1 & $-$0.6 (0.917) & 8.1 & 5.6 & 3.4 (0.525) \\
Realized drift after day 5 (losers / winners) & 5.1 & 6.8 & 5.7 (0.294) & 3.4 & 1.9 & 5.4 (0.384) \\
\addlinespace[3pt]
\multicolumn{7}{l}{\textit{Retail order imbalance}}\\
Market capitalization (small / large) & 8.8 & 2.8$^{**}$ & $-$4.1 (0.359) & 4.2 & 2.2$^{**}$ & $-$3.6 (0.132) \\
Analyst coverage (scarce / broad) & 5.0 & 3.7$^{***}$ & $-$4.3 (0.166) & 1.1 & 3.1$^{***}$ & 0.2 (0.894) \\
Illiquidity (liquid / illiquid) & 2.9$^{**}$ & 1.1 & 3.0 (0.525) & 2.4$^{***}$ & $-$0.6 & 1.0 (0.732) \\
Institutional ownership (under-held / widely held) & 2.2 & 5.4$^{*}$ & 2.8 (0.205) & 0.4 & 5.1$^{***}$ & 4.2 (0.045) \\
Momentum (weak / strong) & 6.7$^{**}$ & 4.4$^{*}$ & 1.9 (0.528) & 4.5$^{**}$ & 2.6 & 0.3 (0.844) \\
Realized drift after day 5 (losers / winners) & 4.4 & 2.4 & 1.6 (0.624) & 2.5 & 1.8 & 0.4 (0.834) \\
\bottomrule
\end{tabularx}
\begin{tablenotes}[flushleft]\footnotesize
\item \textit{Note.} High-recognition purchases against control purchases, full specification, estimated on the low and high half of each characteristic (median split; the splitting characteristic is excluded from the controls). The first label in parentheses is the low half and the second the high half. "diff." is the coefficient on the interaction of the high-recognition indicator with the high-half indicator in the pooled regression, with its p-value in parentheses; because the controls are pooled it need not equal the difference between the two half estimates. Trade counts in log points; the retail order imbalance in percentage points. $^{*}p<.10$, $^{**}p<.05$, $^{***}p<.01$.
\end{tablenotes}
\end{threeparttable}
\end{table}

\begin{figure}[htbp]
\centering
\includegraphics[width=\linewidth]{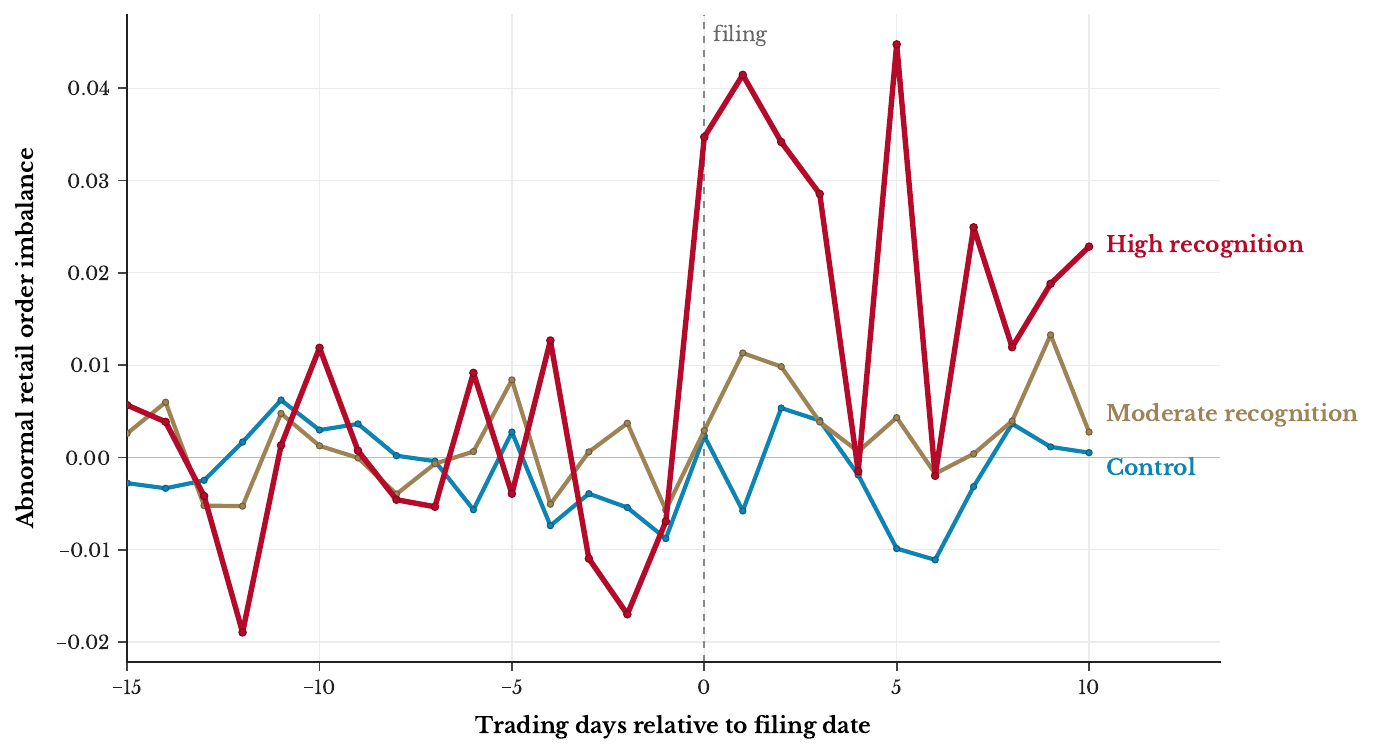}
\caption{Daily abnormal retail order imbalance around the filing date, purchases, by recognition tier}
{\footnotesize\noindent\textit{Note.} Retail buy minus retail sell dollar volume over their sum, net of the stock's own [-15,-6] baseline, averaged across purchase events by tier, from day $-$15 to day +10 relative to the filing date.\par}
\end{figure}

\begin{figure}[htbp]
\centering
\includegraphics[width=\linewidth]{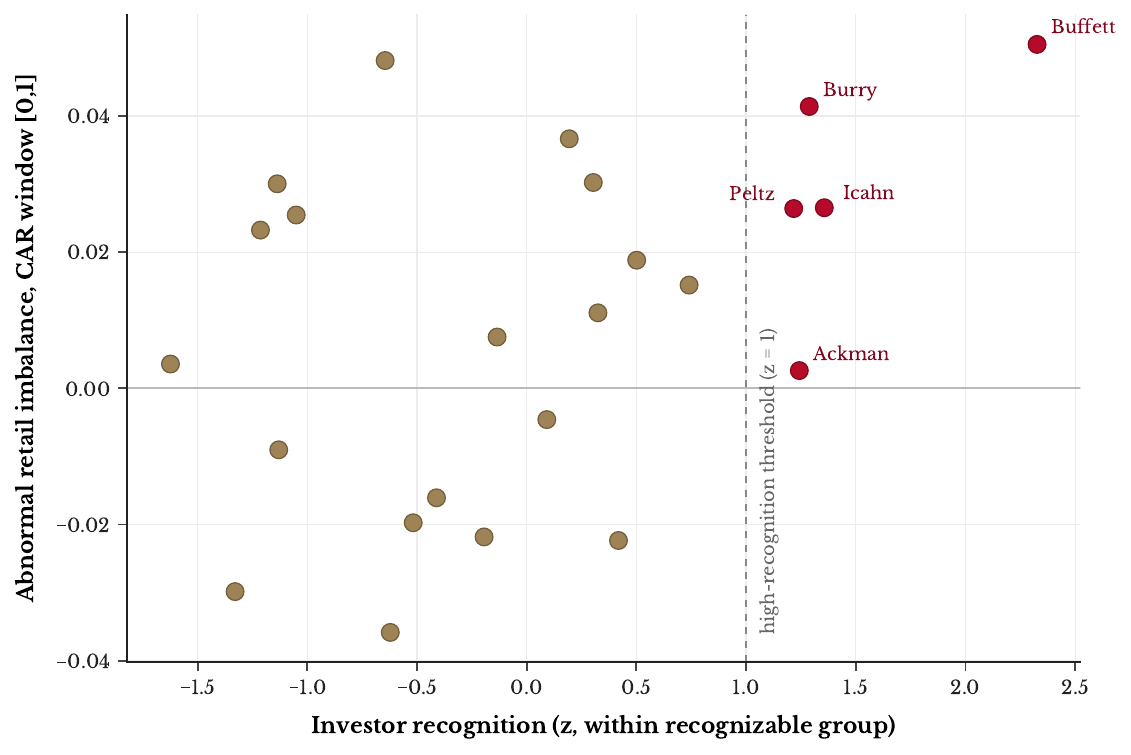}
\caption{Investor recognition and the retail buy tilt within the recognizable group}
{\footnotesize\noindent\textit{Note.} Each point is one recognizable fund: the Investor Recognition Index (z, standardized within the recognizable group) against the fund's mean abnormal retail order imbalance over [0,1] after its purchases (percentage points). High-recognition funds (z $\geq$ 1) are labeled.\par}
\end{figure}

\begin{figure}[htbp]
\centering
\includegraphics[width=\linewidth]{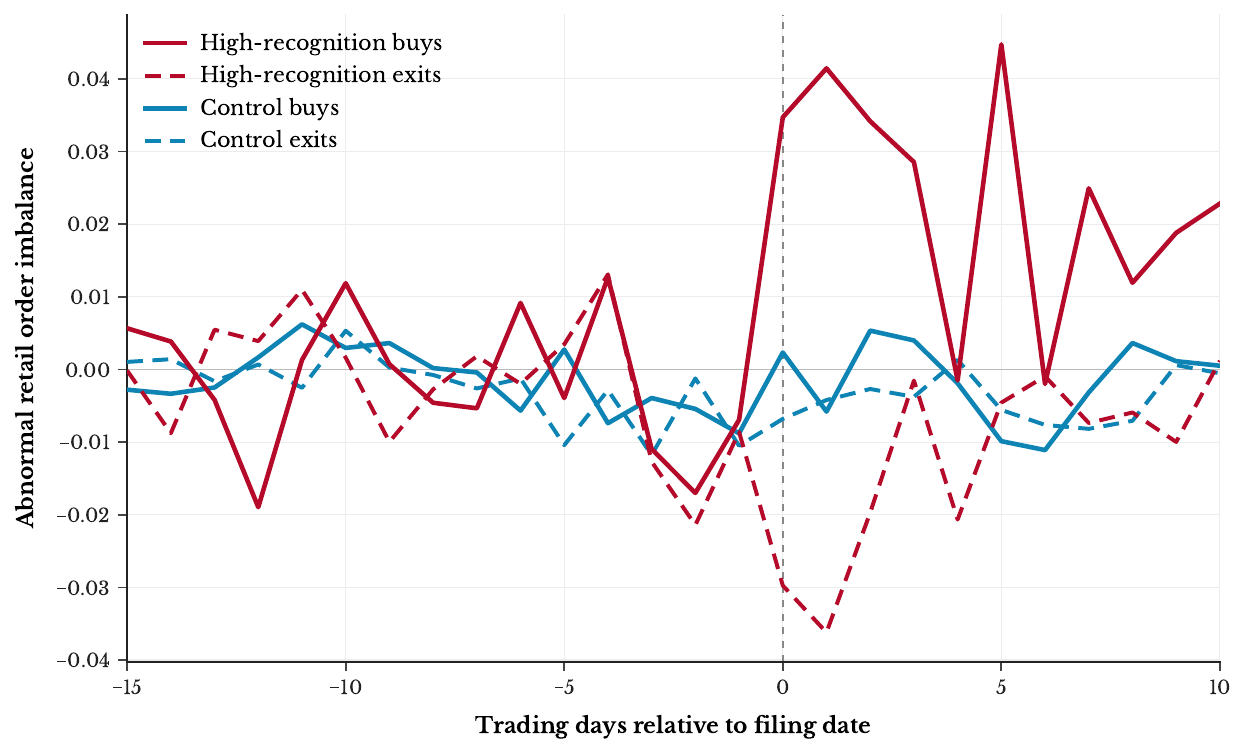}
\caption{Retail order flow around high-recognition and control disclosures: purchases and complete exits}
{\footnotesize\noindent\textit{Note.} Daily abnormal retail order imbalance, averaged across events by tier and disclosure type. Retail flow tilts toward buying after high-recognition purchases and toward selling after high-recognition exits, and does neither around control disclosures.\par}
\end{figure}

\end{document}